      [1999/12/01 v1.4c Il Nuovo Cimento]
\documentclass[varenna]{cimento}
\usepackage{graphicx}  % got figures? uncomment this
\title{X-ray observations of Clusters of Galaxies}
\author{M.~ARNAUD}
\institute{
CEA/DSM/DAPNIA Service d'Astrophysique L'Orme des Merisiers, Bat. 709, 91191 Gif sur Yvette Cedex, France}

\def\etal{et  al.\ }
\def\araa{{Ann.\ Rev.\ Astron.\ Ap.}}

\def\aj{{Astron.\ J.}}
\def\apj{ApJ}

\def\apjs{{ApJ\ Suppl.}}
\def\aas{{A\&A Suppl. Ser.}}
\def\aa{{A\&A}}
\def\mnras{{MNRAS}}

\def\pasp{{PASP}}
\def\pasj{{PASJ}}

\def\keV {\rm keV}
\def\msun{{M_{\odot}}}

\def\rc{{R_{\rm c}}}

\def\ne{{n_{\rm e}}}
\def\nH{{n_{\rm H}}}

\def\kT{{\rm k}T}

\def \fgas  {f_{\rm gas}}
\def \fb {f_{\rm b}}
\def \fgal  {f_{\rm gal}}
\def \Mgas  {M_{\rm gas}}
\def \Mv  {M_{\rm 200}}
\def \Rv  {R_{200}}
\def \MV  {M_{\rm V}}
\def \RV  {R_{V}}
\def \rhoDM {\rho_{\rm DM}}
\def \rhog {\rho_{\rm gas}}

\def \dc {\delta_{\rm c}}
\def \rhocz {\rho_{\rm c}(z)}
\def \ne {n_{\rm e}}
\def \Lx {L_{\rm X}}

\def \LxT {\hbox{$\Lx$--$T$}}
\def \LxM {\hbox{$\Lx$--$M$}}
\def \MgT {\hbox{$\Mgas$--$T$}} 
\def \MT {\hbox{$M$--$T$}} 
  
\def \EMT {\hbox{$EM$--$T$}} 
\def \ST {\hbox{$S$--$T$}} 
\def\fgT  {\hbox{$\fgas$--$T$}} 

\def\Omm{{\Omega_{\rm m}}}
\def\Omb{{\Omega_{\rm b}}}
\def\Oml{{\Omega_{\Lambda}}}

\def \xmm {\hbox{\it XMM-Newton}}
\def \ginga {\hbox{\it GINGA}}
\def \asca {\hbox {\it ASCA}}
\def \rosat  {\hbox {\it ROSAT}}
\def \chandra  {\hbox {\it Chandra}}
\def \sax {\hbox {\it Beppo-SAX}}
\def \rxte {\hbox {\it RXTE}}
\def \exosat {\hbox {\it EXOSAT}}
\def \etal {et al.}
\def\lesssim{\mathrel{\hbox{\rlap{\hbox{\lower4pt\hbox{$\sim$}}}\hbox{$< 
$}}}}
\def\gtrsim{\mathrel{\hbox{\rlap{\hbox{\lower4pt\hbox{$\sim$}}}\hbox{$>$ 
}}}}

\begin{document}

\maketitle

\begin{abstract}
X-ray observations of clusters provide key information on the dark matter, on the formation of structures in the Universe, and can be used to constrain the cosmological parameters. I review our current knowledge, with emphasis on recent XMM and Chandra results.
\end{abstract}

\section{Introduction}

In the standard CDM (Cold Dark Matter)  cosmological scenario,  initial density fluctuations, generated in the early Universe,  grow under the influence of gravity. The Universe gets more and more structured with time. At large scale, the matter density distribution exhibits a web-like topology with  expanding voids surrounded by contracting sheets and filaments.   Massive clusters of galaxies, located at the crossing of filaments, define the nodes of this cosmic web.  Clusters of galaxies are the largest collapsed structures,  with masses ranging from  $10^{13}~\msun$ for small groups  to $10^{15}~\msun$ for the richest clusters.   Because of their size, their mass content  reflects that of the Universe: $\sim 85\%$ of the mass is made of Dark Matter. The main baryonic cluster component is a hot X-ray emitting intracluster gas, as shown by the first X-ray images obtained with the Einstein satellite  \cite{jones84}.  Only a few percent of the mass in clusters lies in the optical galaxies.  

In the CDM  scenario, the amplitude of the initial density fluctuations decreases with increasing scale. As a result,  low-mass objects form first and  then merge together to form more massive objects. Clusters of galaxies are the last manifestation of this hierarchical clustering.  They started to form in the recent cosmological epoch ($z\sim 2$) and the cluster population is continuously evolving.  Individual clusters are not immutable objects.  A cluster continuously accretes matter and smaller groups along the  filaments, and, occasionally, merges  with another cluster of similar mass.   

Clusters of galaxies are key  objects for cosmological studies (see the review \cite{voit04}).  By studying the properties of clusters of galaxies we can test the scenario of structure formation and better understand the gravitational collapse of the Dark Matter and the baryon specific physics. Furthermore, because the history of structure formation depends strongly on the cosmology, studies of cluster samples can also constrain cosmological parameters.  In this course, I will focus on the X-ray observations relevant for these issues and  on properties which can also been studied with SZ observations.  In particular, I  will not discuss  abundance measurements,  although they are important for our understanding of the history of galaxy and star formation and of the heating and enrichment of the intra-cluster medium. This course cannot be exhaustive. More information on X-ray clusters can be found in the reference book ``X--ray emission from clusters of galaxies`` \cite{sarazin88} and in more recent books \cite{feretti02,mulchaey04} and reviews quoted in the text.

\section{Observing clusters in X-ray}

\subsection{X-ray emission of clusters}

The Intra-Cluster medium (ICM) is a hot, tenuous and optically thin plasma. The gas density varies from $\sim 10^{-4}\rm{cm^{-3}} $ in the outer regions of clusters to a few $\sim 10^{-2}\rm{cm^{-3}} $ in the center.  The ICM has mean temperatures in the range  $\kT=0.5-15\keV$, reflecting the depth of the potential well ($\kT  \propto {\rm GM}/R$). Note that the ICM is not an isothermal plasma, although temperature variations are usually small (see below). The ICM  is enriched in heavy elements, with typical abundances  of $1/3$ the solar value.  At the temperature of the ICM, H and He are fully ionized. Most electrons come from these two elements. The electron density is nearly independent  of the ionization state and is given by $\ne \sim 1.2 \nH$, where $\nH$ is the hydrogen density. The ionization stage of the other elements depends on the temperature.  

The X-ray emission is that of a coronal plasma at ionization equilibrium \cite[for detailed description]{sarazin88}.  For a volume element $dV$ of electronic density $\ne$, temperature $T$, abundances $[Z/H]$, the number of photons emitted by unit time in the energy range $[E, E+dE]$ can be written as
\begin{equation}
	dN(E)  =  \ne^2 \epsilon (E, T, [Z/H])  dE dV
\end{equation}
where $\epsilon (E, T, [Z/H])$ is the photon emissivity at energy $E$.   The intensity scales  as the {\it square} of the density, because all emission processes (like the Bremsstrahlung emission) result from collisions between electron and ions.   

\begin{figure}[t]
\begin{minipage}[b]{0.50\textwidth}
\centering 
\includegraphics[width=\textwidth]{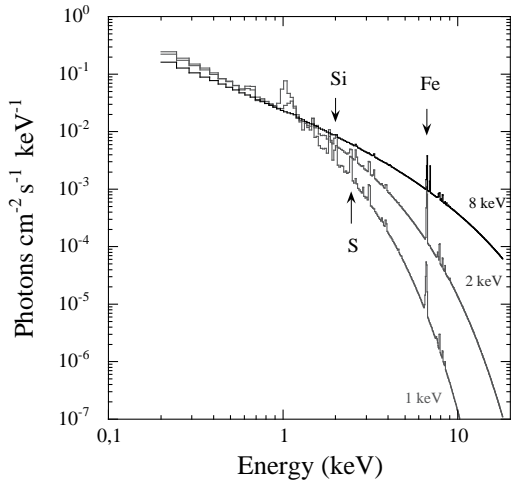}     
 \end{minipage}
\hspace{+0.02\textwidth}
\begin{minipage}[b]{0.48\textwidth}
 \caption[ ]{The X-ray emission from a thin plasma with 0.35 solar abundance at different temperatures,  $\kT= 1, 2$ and  $8~\keV$.}
\label{fig:spectre}
\end{minipage}
\end{figure}

Examples of X-ray spectra for typical cluster temperatures are shown in Fig.~\ref{fig:spectre}.   Due to the high temperature, the continuum emission is dominated by thermal Bremsstrahlung, the main species by far contributing to the emission being H and He.  The emissivity of this continuum is very sensitive to temperature for energies greater than $\kT$ and rather insensitive to it below. This is due to the exponential cut-off of the Bremsstrahlung emission. Indeed, it scales as $g(E,T)T^{-1/2} exp(-E/kT)$, where $g(E,T)$, the Gaunt factor, is a gradual function of $kT$.  The only line that clearly stands out at all  temperatures is the Iron K line complex around $6.7~keV$ (see Fig.~\ref{fig:spectre}).  We can also observe the K lines of other elements ($Z>8$,  H and He-like ionization states), as well as the L-shell complex of lower ionization states of Iron. However the intensity of these lines rapidly decreases with increasing temperature.  Except for the cool clusters ($\kT \lesssim  4 \keV$) or in the cooling core present in some clusters,  one cannot expect to measure the abundance of elements other than Iron because they are completely ionized.  

%%%%%%%%%%%%%%%%
%%%%%%%%%%%%%%%%

\subsection{Extracting Physical information from X-ray observations}
\label{sec:obsphys}

From above, it is clear that X-ray observations give access to the two characteristics of the ICM, which are the density and the temperature.  The shape of the spectrum determines the temperature \footnote{and also the heavy element  abundances from the line equivalent widths and possibly the redshift from the line positions}, whereas the normalization provides the emission measure $EM = \int \ne^2dV$. 

\subsubsection{Gas temperature}
Its determination requires spectroscopic data.  The temperature  is derived by fitting the observed spectrum (as in Fig.~\ref{fig:RDCSspec})
 with a thermal emission model convolved with the instrument response (i.e.,  taking into account how the effective area and spectral resolution vary over the energy range).  The model spectrum at Earth is computed from the emitted spectrum, taking into account the cluster redshift and the galactic absorption (which mostly affects the emission below $1-2~\keV$). It is important to keep in mind that:

\noindent $\bullet$  The temperature is constrained by the position of the exponential cut-off in the spectrum. In order to have a proper determination of the temperature, we need spectroscopic instruments  sensitive up to energies greater than $\kT$, i.e.,  typically $10~\keV$.

\noindent $\bullet$   The ICM is not strictly isothermal. This means that a temperature inferred from an isothermal fit to the data is actually a 'mean' value along the line of sight and in the considered cluster region. This temperature is not simply, as often thought, the 'emission weighted' temperature \cite{mazzotta04}. 

\subsubsection{Gas density}
The emissivity is not very sensitive to $\kT$ at low energy.   Therefore X-ray  images or surface brightness profile (as in Fig.~\ref{fig:RDCSsx})
extracted in a soft energy band ($E \lesssim 2~\keV$) are used to determine the gas density distribution. X-ray images reflect the ICM morphology. Note that one must not forget projection effects and that density contrast are enhanced because the X-ray emission increases as the square of the density.  The emission measure along the line of sight at projected radius $r$,
$EM(r)=\int\ne^2~dl$, can be deduced from the surface brightness, $S(\theta)$:
\begin{eqnarray}
EM(r) = \frac{4~\pi~(1+z)^{4}~S(\theta)}{\Lambda(T,z)}~~~~~{\rm with } ~~~~~~~~ r =
d_{\rm A}(z)~\theta 
\label{eq:sx2em}
\end{eqnarray}
where $d_{\rm A}(z)$ is the angular distance at the cluster redshift $z$.
$\Lambda(T,z)$ is the emissivity in the considered energy band, taking into account the absorption by our galaxy, the redshift, and
the instrumental spectral response. Because $\Lambda(T,z)$ depends only weakly on the temperature in the soft band, it is essentially insensitive to temperature gradients (except for cool systems, \cite{pratt03}) and one can generally use the average cluster temperature. 

The gas density radial profile $n_{\rm g}(R)$ is usually derived from Eq.~\ref{eq:sx2em}, assuming spherical symmetry. In that case $EM(r) = \int_{r}^{\infty} \ne^2(R)~R dR/\sqrt{R^{2}-r^{2}}$. One can use deprojection techniques  or parametric models fitted to the data. A  popular model is the so-called isothermal $\beta$-model: $n(R) = n_{0} \left[1+\left(R/\rc\right)^2\right]^{-3\beta/2}$, which gives $S(\theta) = S_{0} \left[1+\left(\theta/\theta_{\rm c}\right)^2\right]^{-3\beta+1/2}$. This model fits reasonably well cluster profiles at large radii, but it underestimates the density in central cooling core of clusters (see also Sect.~\ref{sec:statgas}). 

\subsubsection{Cluster properties}
Depending on the quality of the data, various physical cluster properties can then be derived. The easiest to determine is the X--ray luminosity, which is computed  from the observed flux  in the instrument energy  band \footnote{Conversion from count rates to bolometric luminosity depends on the temperature. The usual way is to use a $\LxT$ relation. }  A higher statistical quality is required to measure a surface brightness profile  or a spectrum. From these two quantities  one derives the global temperature, the gas density radial  profile and therefore the gas mass.  If spatially resolved spectroscopic measurements are available, one can also derive the temperature profile. This gives access to the entropy profile, key information on the history of the ICM (see Sec.~\ref{sec:statgas}). One can also deduce the total mass profile from the Hydrostatic Equilibrium Equation: 
\begin{equation}
M(r) = - \frac{kT\ r}{{\rm G} \mu m_p}   \left[ \frac{d \ln{n}}{d \ln{r}} + \frac{d \ln{T}}{d \ln{r}} \right]
\label{eq:HE}
\end{equation}
\noindent where G and $m_p$ are the gravitational constant and proton
mass and $\mu \sim 0.61$. Note that an approximation of the total mass can be obtained by simply assuming isothermality. 

%%%%%%%%%%%%%%%%
\begin{figure}[t]
\includegraphics[width=\textwidth]{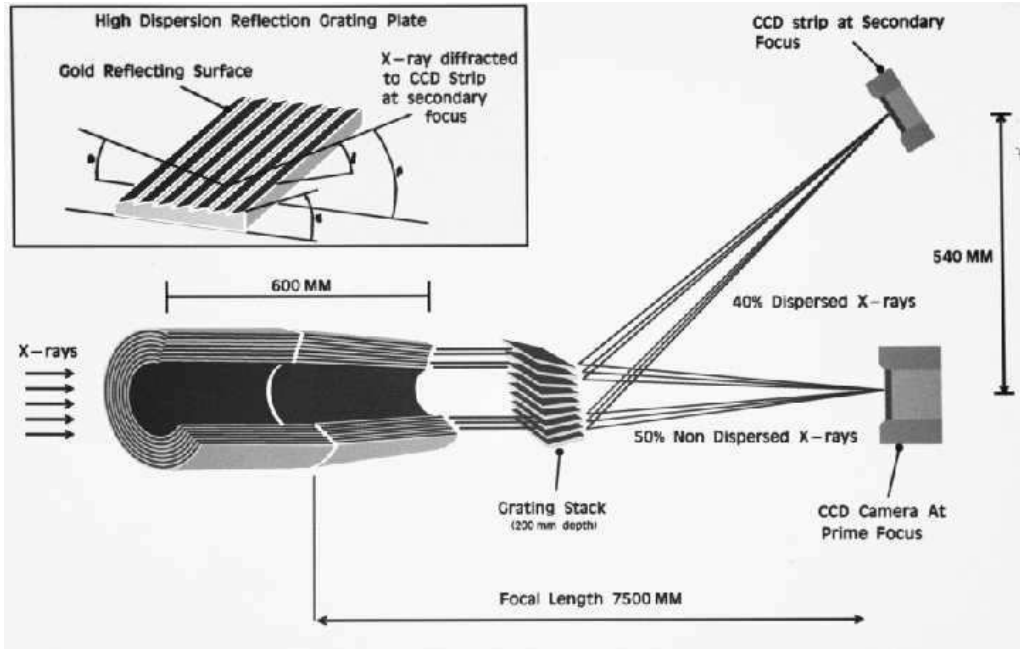}   
 \caption[ ]{The light pass in the two XMM telescopes equipped with the EPIC and RGS experiments.
After collection  by the mirror module (58 Wolter shells)  about 40\% of the
light is focused to the EPIC CCD camera, while a similar amount  is dispersed by a Reflection Grating Array. The diffracted beam is imaged by a CCD camera on the
Rowland circle for high resolution spectroscopy. In the third telescope, all
the light is collected by the EPIC CCD experiment. The Figure  was
reproduced from ESA SP-1097.}
\label{fig:xmm}
\end{figure}
%%%%%%%%%%%%%%%%

\subsection{Modern X--ray observatories}
X-rays are absorbed by the Earth's atmosphere. Therefore, X-ray observatories are put on board satellites. Two X--ray satellites are now in operation, \xmm\ and \chandra.  They are  fully described in \cite{jansen01,denherder01,struder01,turner01}  and \cite{weisskopf02}.  The general concept in the same. The X-ray photons are collected and focused by grazing incidence telescopes (Fig.~\ref{fig:xmm}). The focal plane is equipped with  CCD cameras, allowing for the measurement of the position and energy of each individual incoming photon (Fig.~\ref{fig:xmm}).  This permits spatially resolved spectroscopy at medium resolution ($\Delta E=60-140$ eV), in the energy range between $0.3$ and $10~\keV$.  Spectroscopy at higher resolution is performed using gratings, the 'dispersed' spectrum being read by CCDs (Fig.~\ref{fig:xmm}).  Because this is dispersive spectroscopy, the sensitivity is degraded, the spatial information is essentially lost and the spectral resolution rapidly deteriorates as the extent of the source increases. The spectral resolution of the XMM grating instruments (XMM/RGS) is $\Delta(E)/E=0.1-0.3\%$ in the $0.3-2~\keV$ energy band for point sources.  Such spectroscopy is limited to clusters with sharply peaked emission profiles. It is restricted to the central $\theta \lesssim 2' $ region. The sensitivity of \chandra\  grating instruments is too low for cluster studies. 

The \xmm\ and \chandra\ observatories are complementary.   \chandra\ has an extremely good spatial resolution of  $\Delta\theta= 0.5"$ (compared to $8"$ for \xmm). The strength of \xmm\ is its  exceptional collecting area and thus sensitivity: three high-throughput telescopes are operating in parallel. The  Field of view  is $30'$ in diameter, well adapted to cluster studies.   \chandra\ has only one telescope, a smaller field of view of $17'X17'$  (for the ACIS-I instrument) and an effective area typically $3$($5$) times lower than \xmm\ at $1.5(8)~\keV$.

As compared to the previous generation of satellites, \xmm\ and \chandra\ represent a giant step forward in term of sensitivity and spatial resolution. The \rosat\  satellite \cite{trumper90}  had good imagery capability ($\Delta\theta=15"$) but much lower effective area  and very poor spectroscopic capability. The high energy cut-off of the telescope was $E\sim 2~\keV$, so that accurate temperature measurement was limited to very cool clusters. \asca\ was the first X--ray observatory \cite{tanaka94} with telescopes working up to $10~\keV$ and a CCD camera at the focal plane at one of the telescope (the other telescopes were equipped with proportional counters). As compared to spectroscopy made before with collimated spectrometers (like with the \exosat\ or \ginga\ satellite), the gain in sensitivity  was very important.  It was also the first time one could do spatially resolved spectroscopy of clusters. However, this was limited by the relatively large and energy dependent Point Spread Function.  The spatial resolution of \sax\  was better, but above all  it had the capability of observing sources over more than three decades of energy, from $0.1$ to $200~\keV$ \cite{dicocco97}.  \rxte\ \cite{bradt93} has no imaging capability but was also used to study hard X-ray emission from clusters. 

%%%%%%%%%%%%%%%%
\begin{figure}[t]
\begin{minipage}[c]{0.48\textwidth}
\centering 
\includegraphics[width=\textwidth]{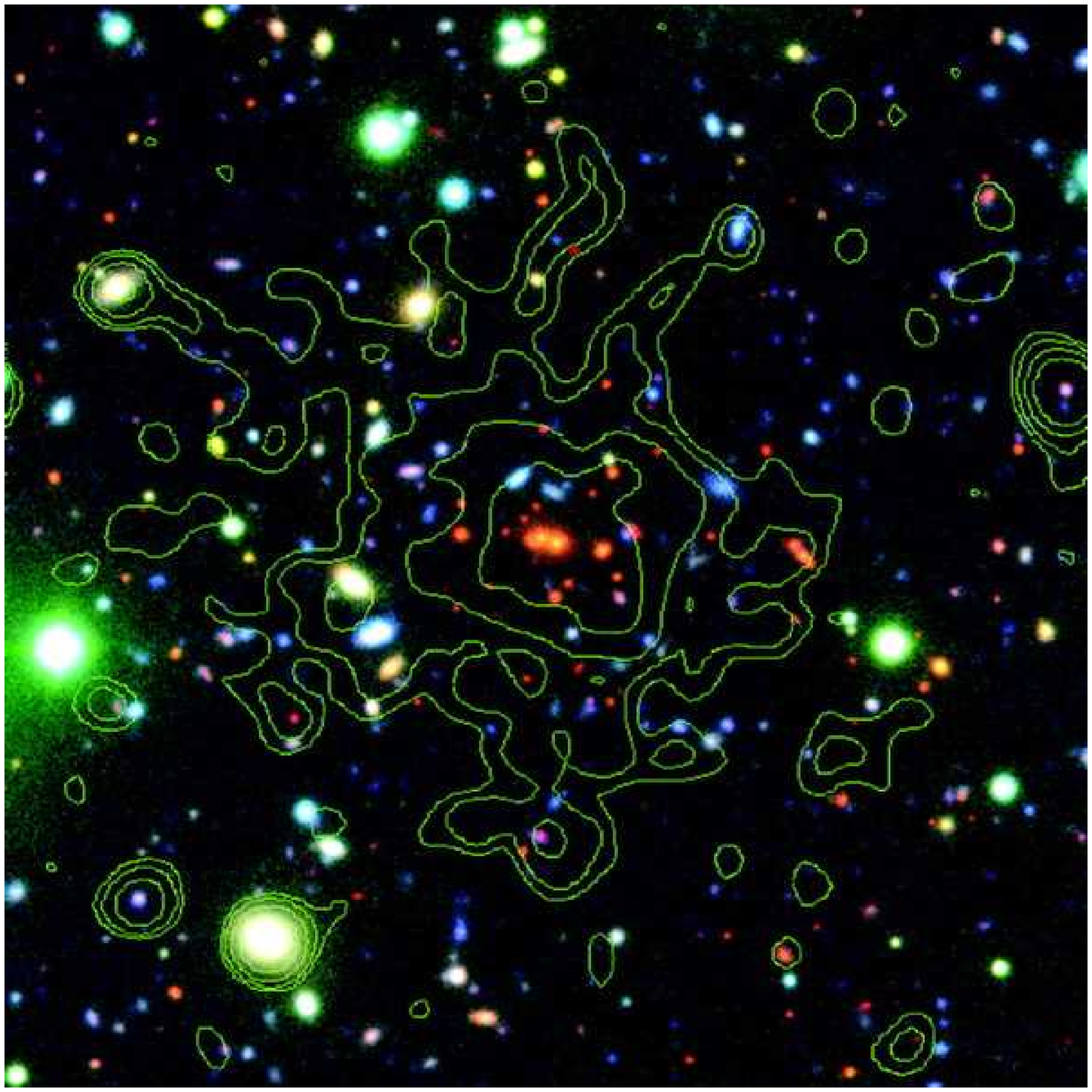}     
\end{minipage}
\hspace{+0.02\textwidth}
\begin{minipage}[c]{0.48\textwidth}
\centering  \caption[ ]{Image of the cluster RDCS 1252.9-2927 at $z=1.24$.  Contours of the \chandra\ emission  are  overlaid on a composite VLT image. Figure from \cite{rosati04}}
 \label{fig:RDCSim}
\end{minipage}
\begin{minipage}[c]{0.48\textwidth}
\centering 
\includegraphics[width=\textwidth]{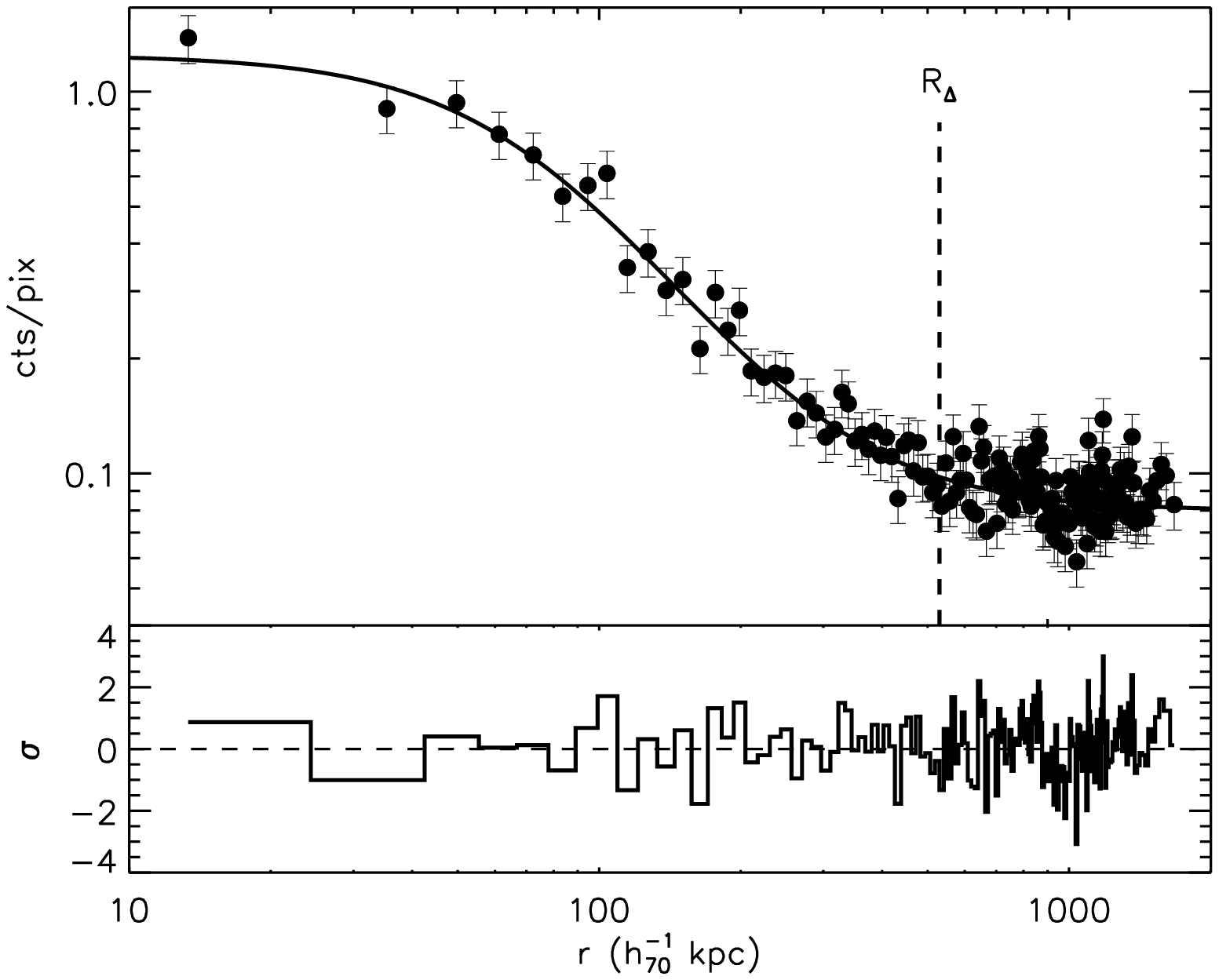}     
\end{minipage}
\hspace{+0.02\textwidth}
\begin{minipage}[c]{0.48\textwidth}
\centering 
\includegraphics[width=\textwidth]{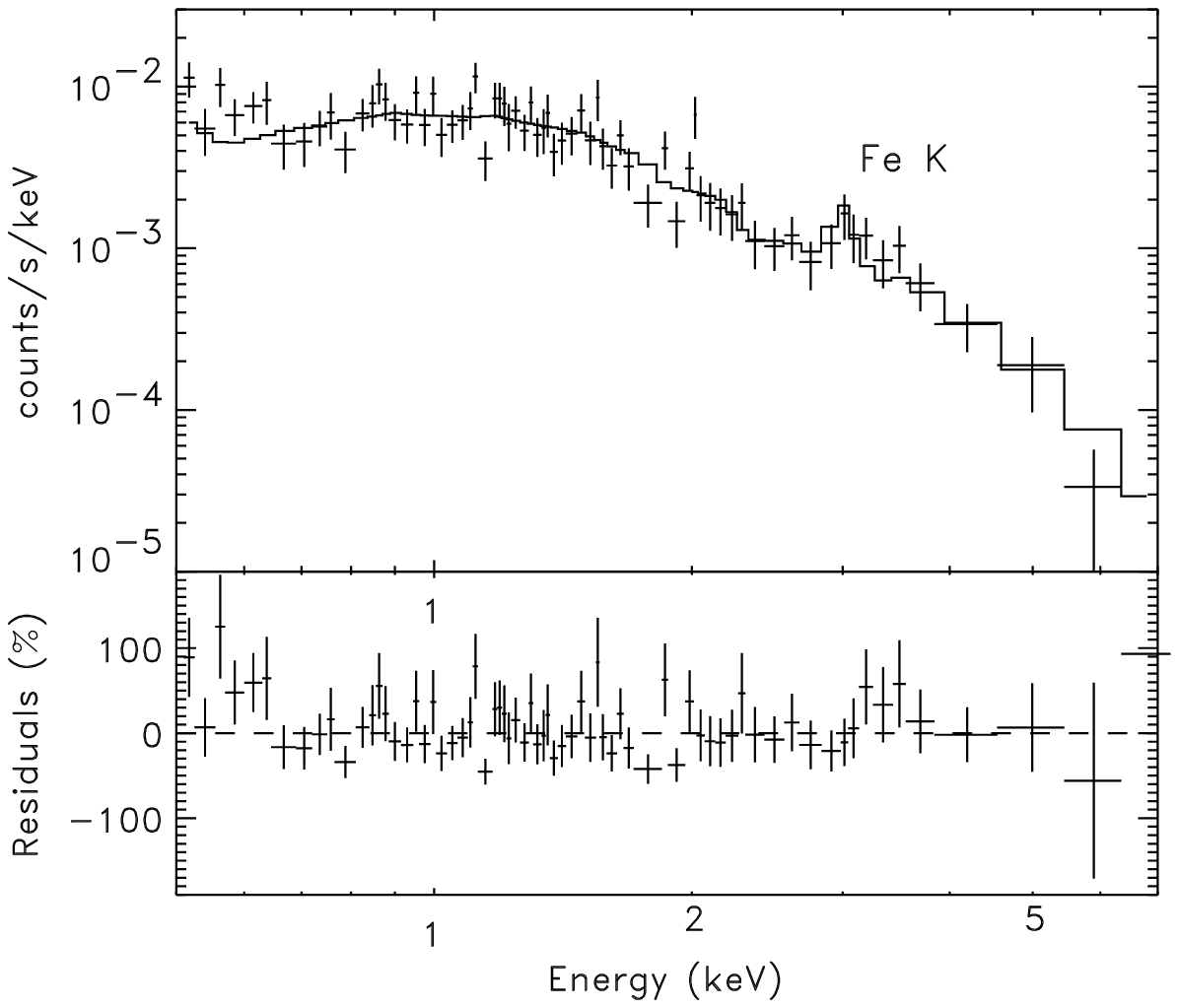}     
\end{minipage}
\begin{minipage}[t]{0.48\textwidth}
 \caption[ ]{Surface brightness profile of RDCS 1252.9-2927 measured with \chandra\ (data points), with best-fit  model (solid line) and residuals.  Figure from \cite{rosati04}}
 \label{fig:RDCSsx}
\end{minipage}
\hspace{+0.02\textwidth}
\begin{minipage}[t]{0.48\textwidth}
 \caption[ ]{X-ray spectrum  of RDCS 1252.9-2927 (data points) and best-fit thermal model (solid line) from  \xmm\ observation.  Note the redshifted Iron line. Figure from \cite{rosati04}}
 \label{fig:RDCSspec}
\end{minipage}
 \end{figure}
%%%%%%%%%%%%%%%%

	With \xmm\ and \chandra:

\noindent $\bullet$  We can map the gas distribution in nearby clusters from very deep inside the core,  at the scale of a few $kpc$ with \chandra\ \cite{fabian01},  up to very close to the virial radius with \xmm. Temperatures profiles (and thus mass profiles) can be measured over a wide radial range, down to $0.001\Rv$ with \chandra\  \cite{lewis03} up to close the virial radius with \xmm, even in low mass systems \cite{pratt05}. Last but not least, we have now precise temperature maps for unrelaxed objects \cite{sun02,belsole04} and we can resolve very sharp density features \cite{markevitch00}. 

\noindent $\bullet$ We can measure basic cluster properties up to high z ($z\sim 1.3$) and down to the \rosat\ detection limit (with \xmm).  This includes morphology from images, gas density radial profile, global temperature and gas mass  (e.g. \cite{arnaud02b,rosati04}).  As an example,   Fig.~\ref{fig:RDCSim},~\ref{fig:RDCSsx},~\ref{fig:RDCSspec} show the \chandra\ and \xmm\ observations of RDCS 1252.9-2927 at $z=1.24$ \cite{rosati04}. Total mass  and entropy can be derived assuming isothermality \cite{ettori04}.  For the brighter distant clusters, crude temperature profiles  \cite{arnaud02b} or maps \cite{maughan03} can be obtained.  

\noindent $\bullet$ We can find and identify new clusters \cite{valtchanov04}. Clusters at all redshifts appear as extended sources in \xmm\ and \chandra\ images. However, only \chandra\ has the capability to perfectly remove point source  contamination. The typical  flux limit for \xmm\ serendipitous surveys is $10^{-14} {\rm ergs/s/cm^{2}}$ in the $0.5-2~\keV$ band \cite{land04}, about  $3$ times  lower than with \rosat. 

\section{Hierarchical Cluster Formation}

\subsection{Substructures and merger events in the local Universe}
The early Einstein images showed that clusters present a large variety of morphology, from regular clusters to very complex systems with multiple substructures  \cite{jones92}.  Since substructures in the ICM are erased on a typical time scale of
a few $Gyr$, their detection is the signature
of recent dynamical evolution. This manifold morphology indicates a variety of dynamical states in the local cluster population.  These observations were consistent with the idea that clusters are still forming to-day, as expected in
the hierarchical formation model.  

The presence of substructures by itself does not tell much about the cluster formation process.   For instance, we expect substructures both if clusters form through mergers of smaller systems (CDM scenario), or by divisions of larger systems (top-down scenario).  A direct support of the hierarchical CDM model was the observation of specific signatures of merger events in density and temperature maps, as predicted from numerical simulations (see the detailed reviews  \cite{sarazin02}  on the physics of merger events and  \cite{buote02} on  X--ray observations of these events with \asca\ and \rosat). The first unambiguous signature  of a merger event was provided by the \rosat\ observation  of gas compression in the subcluster of A2256,  as expected if this subcluster was falling towards the main cluster \cite{briel91}.  The \asca\ observation of a temperature increase in the interaction region between the sub-clusters of Cygn-A, was even clearer evidence that  two sub-clusters were colliding \cite{markevitch99}. It also showed  that shocks induced by
mergers do contribute to the ICM heating.  This is  is a strong  prediction of numerical simulations of cluster mergers \cite{schindler93,ricker01}.   When two sub-clusters collide, the interaction region is heated and compressed.  As the relative motion
in mergers is moderately supersonic ($v \sim 2000~{\rm kms/s}$),  shocks are driven in the ICM.  As a result, the gas of the final cluster  is heated to  its virial temperature, which is higher than the virial temperatures of the initial sub-clusters.  

%%%%%%%%%%%%%%%%
\begin{figure}[t]
\begin{minipage}[c]{0.48\textwidth}
\centering 
\includegraphics[width=\textwidth]{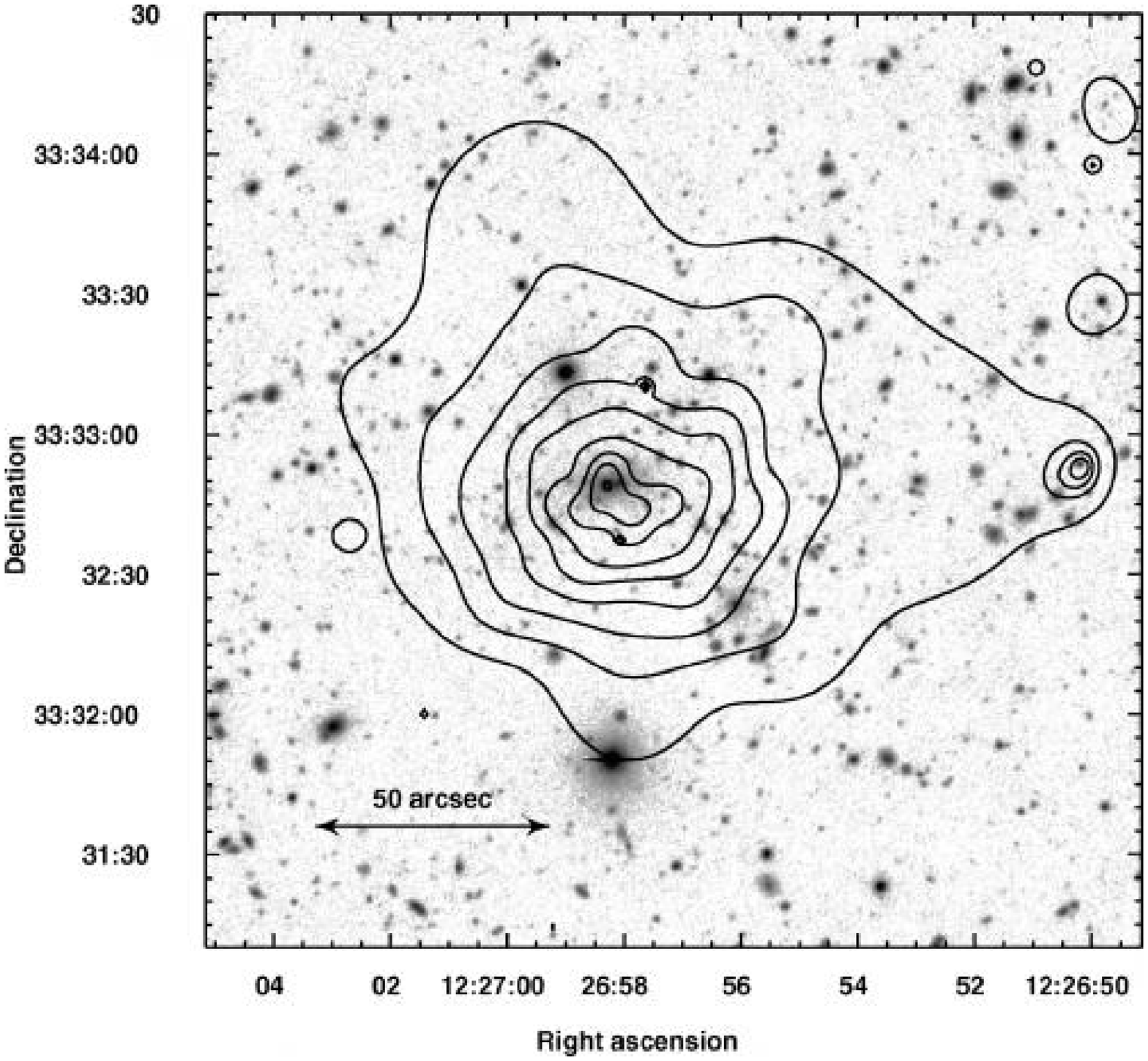}     % includes figure foo.eps
 \end{minipage}
\hspace{+0.02\textwidth}
\begin{minipage}[c]{0.48\textwidth}
\centering
 \includegraphics[width=\textwidth]{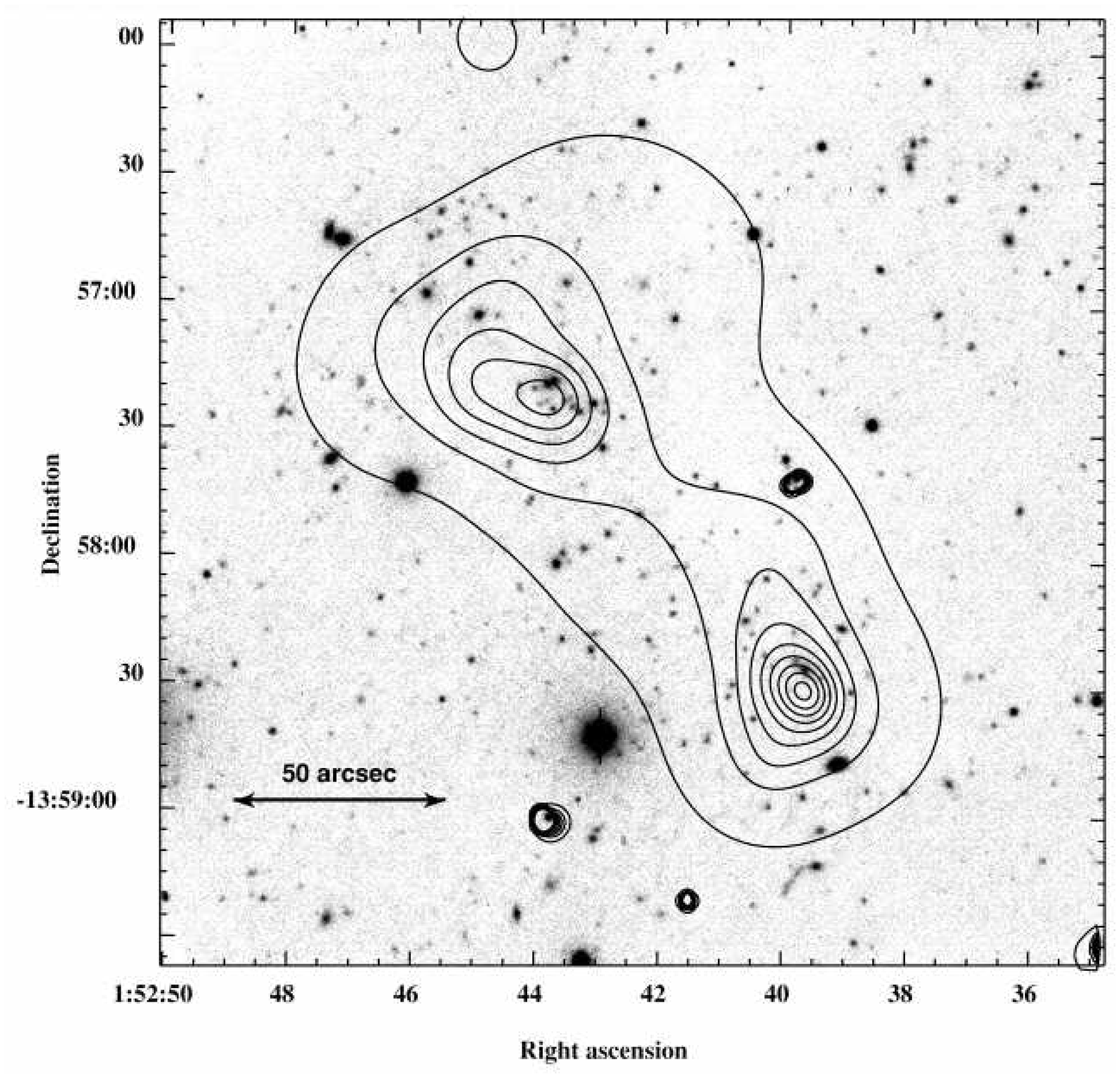}     % includes figure foo.eps
\end{minipage}
\begin{minipage}[t]{0.48\textwidth}
 \caption[ ]{RXJ1226.9+3332, a relaxed massive cluster  at $z=0.89$.   The  contours of X-ray emission detected by \xmm\  is overlaid on a Subaru I-band image. Figure from \cite{maughan04}}
 \label{fig:RXJ1226}
\end{minipage}
\hspace{+0.02\textwidth}
\begin{minipage}[t]{0.48\textwidth}
 \caption[ ]{The merging cluster RX J0152.7-1357  at $z=0.83$.  The \chandra\  X-ray contours are overlaid on a Keck II I-band image. Figure from \cite{maughan03}. The \chandra\ temperature map shows a temperature increase between the two subclusters  indicating that they have started to merge }
 \label{fig:RXJ0152}
\end{minipage}
 \end{figure}
%%%%%%%%%%%%%%%%
%%%%%%%%%%%%%%%%%%%%%%%%
\begin{figure}[t]
\begin{minipage}[c]{0.48\textwidth}
\centering 
\includegraphics[width=\textwidth]{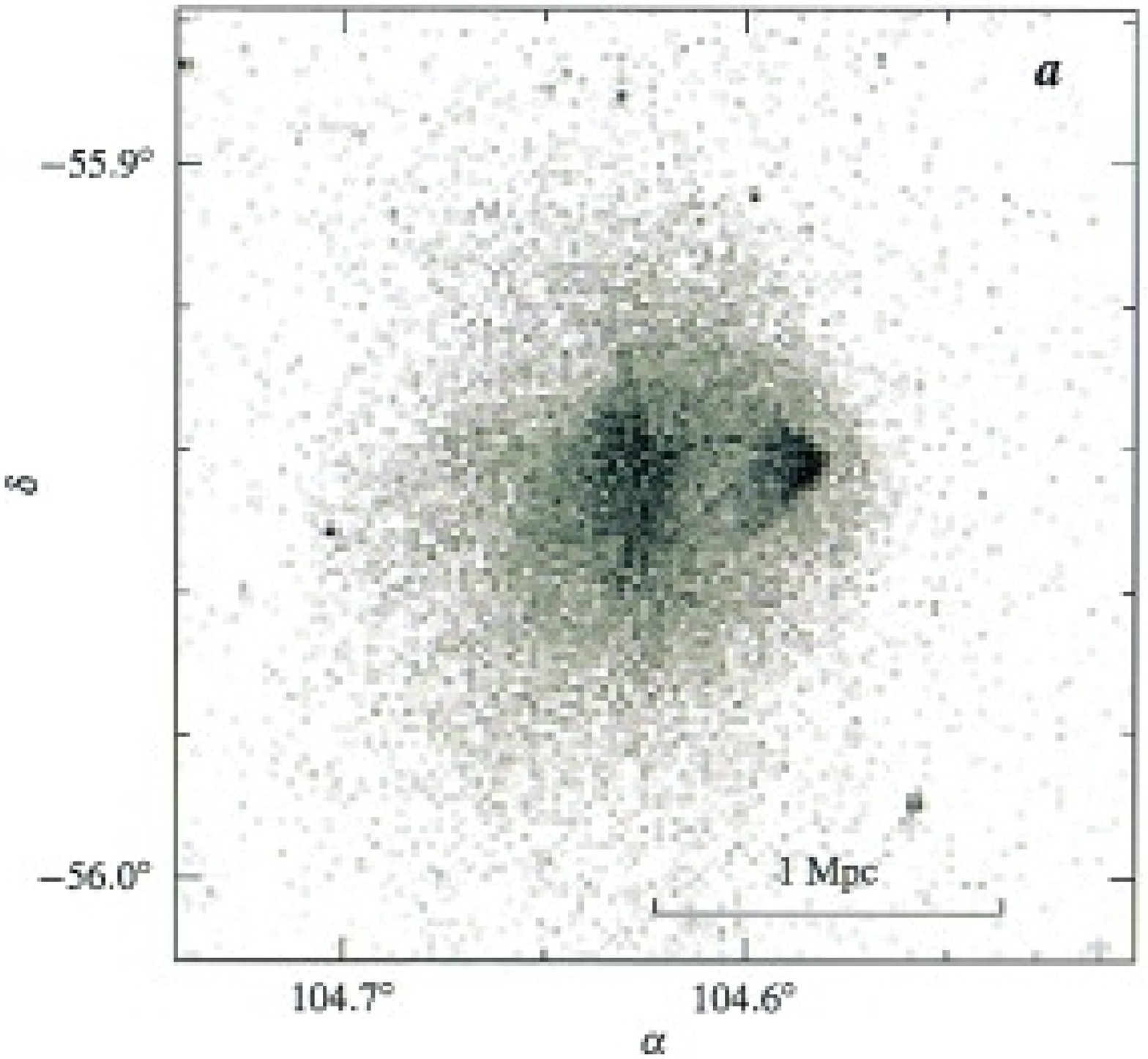}    
 \end{minipage}
\hspace{+0.02\textwidth}
\begin{minipage}[c]{0.48\textwidth}
\caption[ ]{\chandra\ image of 1E0657-56. Note the "bullet" apparently just exiting the cluster core and moving westward. The bullet is preceded by an X-ray brightness edge that resembles a bow shock. Figure from \cite{markevitch02} }
 \label{fig:1Eim}
\end{minipage}
\includegraphics[width=\textwidth]{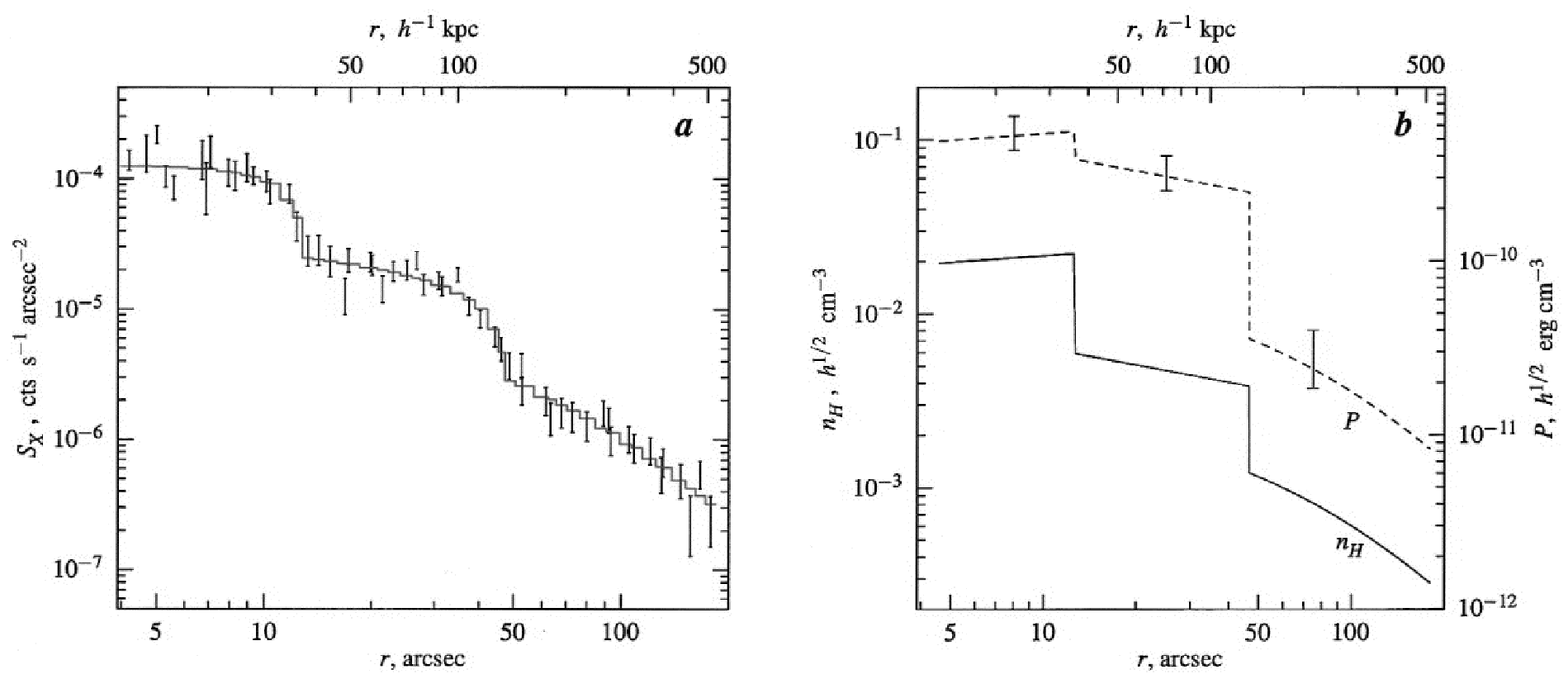}     
\caption[ ]{Surface brightness profile in a $120\,^{\circ}$ sector centered on the bullet and directed westward (Left) and corresponding temperature and pressure profile (right).  The first edge is a cold front while the second is a shock. Figure from \cite{markevitch02}}
  \label{fig:1Eprof}
\end{figure}
%%%%%%%%%%%%%%%%%%%%%%%%
 
 \subsection{Cluster formation at high $z$}

With \xmm\ and \chandra\ we now extend the study of  substructures to the distant Universe. As expected from the hierarchical formation models, we observe a variety of morphology (and thus dynamical state) up to very high z.  A clear case of a double cluster is RX J1053.7+5735, observed with \xmm\  at $z=1.14$ \cite{hashimoto04}. This cluster is very probably a  merger between two nearly equal mass systems.  In contrast, RXJ1226.9+3332 ($z=0.9$) observed with \xmm\ is a massive cluster (the temperature is $\kT=11\pm1~\keV$), with a very regular morphology (Fig.~\ref{fig:RXJ1226}) indicative of a relaxed state \cite{maughan04}. Note that the existence of massive and relaxed clusters at so high $z$ is expected in low $\Omega$ Universe but is very unlikely in a critical $\Omega=1$ Universe.  We also see unambiguous evidence of merging activity up to $z\sim0.8$ (Fig.~\ref{fig:RXJ0152}). The crude \chandra\ temperature map shows a temperature increase between the two subclusters of  RX J0152.7-1357, indicating that they have started to merge \cite{maughan03}. Interestingly, when the two subunits will have completely merged the cluster will have the mass of Coma. 

\subsection{The detailed physics of merger events and the effect of the Large Scale environment}
 With \chandra\ and \xmm, the
temperature structure in nearby clusters  can be mapped with unprecedented
accuracy. This permits much  deeper investigations of the dynamical process of cluster formation.  

\subsubsection{Shocks and Cold Fronts}
We now understand better what happens during a merger event. The \chandra\  observation of the merging cluster 1E0657-56  provides a clear 'text-book' example of a shock  \cite{markevitch02} as shown in Fig.~\ref{fig:1Eim} and Fig.~\ref{fig:1Eprof}. The derived Mach number is about $M=2$ \cite{markevitch02}. Evidence of strong shocks is rare, but they might be difficult to detect due to projection effects \cite{mazzotta04}.  More surprisingly,  \chandra\  revealed  the presence of  "Cold fronts" in merging clusters, first discovered in A2142 \cite{markevitch00} and A3667 \cite{vikhlinin01a}. Cold fronts are contact discontinuities between the cool core of a subcluster moving at near sonic velocity and the surrounding main cluster gas. Across the discontinuity, there is an abrupt jump of the gas density and temperature (Fig.~\ref{fig:1Eprof}).  The pressure is approximately continuous (Fig.~\ref{fig:1Eprof}), which shows that the discontinuity is  not a shock.  1E0657-56 exhibits both a cold front and a shock, located ahead of the cold front (Fig.~\ref{fig:1Eim}, \ref{fig:1Eprof}). The observations of cold fronts show  that the cool core of infalling subclusters can survive the passage through the main cluster core,  whereas  the gas in the outer region of  the subcluster  is  stripped by ram pressure \cite{markevitch02}. Evidence of gas stripping during mergers is also provided by the observation of cool trails behind some merging subclusters as, for instance, in the \xmm\ observation of A1644 \cite{reiprich04}. Ultimately, the cool core is destroyed: for instance the shape of the subcluster remnant in 1E0657-56  shows that it is being actively destroyed by gas-dynamic instabilities \cite{markevitch02}. 
 
The observation of cold fronts has further implication for the physics at play in clusters. The very steep temperature gradient and smooth surface of the cold fronts  imply that thermal conduction and Kelvin-Helmotz instabilities are suppressed,  probably by magnetic fields \cite{ettori00,vikhlinin01a,vikhlinin01b}.  

%%%%%%%%%%%%%%%%%%%%%%%%%%%% 
 \begin{figure}[t]
\begin{minipage}[c]{0.48\textwidth}
\centering 
\includegraphics[width=\textwidth]{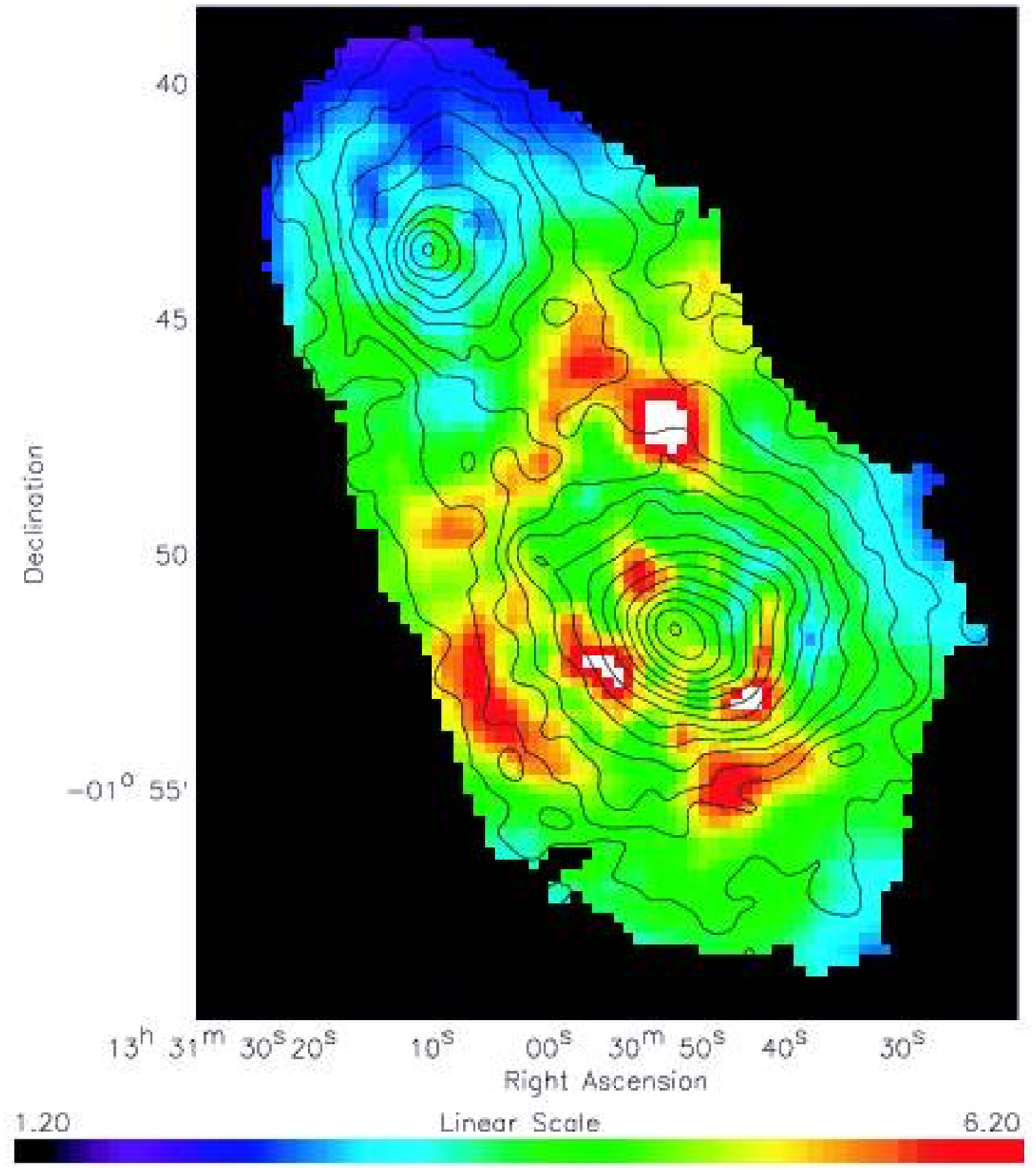}     
 \end{minipage}
\hspace{+0.02\textwidth}
\begin{minipage}[c]{0.48\textwidth}
\centering
\includegraphics[width=\textwidth]{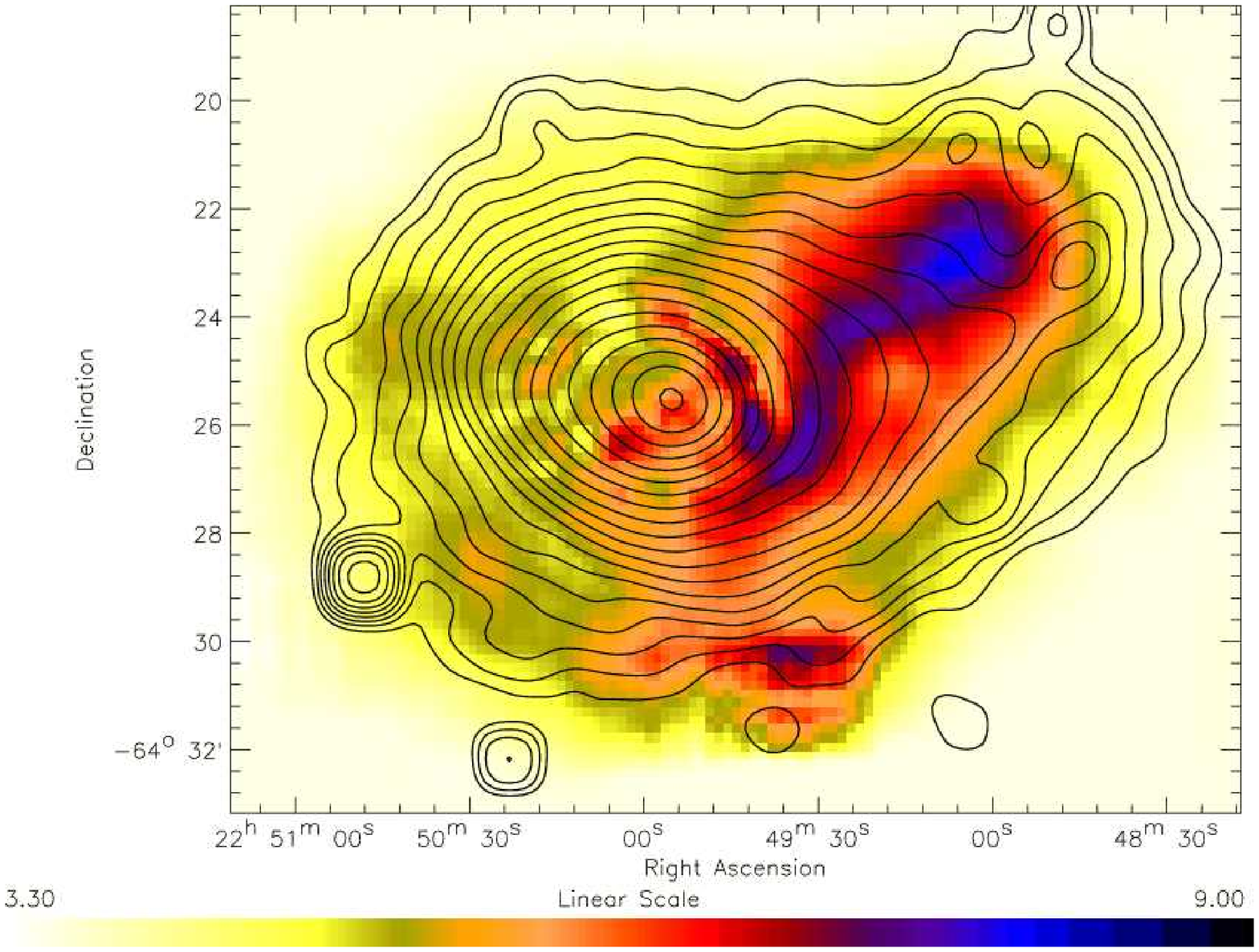}    
\end{minipage}
\begin{minipage}[t]{0.48\textwidth}
 \caption[ ]{The double cluster A1750. The contours of the \xmm\  emission are overlaid on the temperature map. An arc-like hot region can be seen between the two sub-clusters, A1750N in the north and A1750C in the South. The two clusters have started to collide and the gas is being shocked and compressed. The hot regions in A1750 C are regions of shocked gas from an older merger which occurred about 1-2 billion years ago. Figure from \cite{belsole04}. }
 \label{fig:A1750}
\end{minipage}
\hspace{+0.02\textwidth}
\begin{minipage}[t]{0.48\textwidth}
 \caption[ ]{The merging cluster A3921. The contours of the \xmm\  emission  are overlaid on the temperature map. In contrast to A1750, the hot shocked region is not perpendicular to the line joining the centre of the two subclusters, but nearly parallel to it. This is interpreted in terms of an off-axis merger:  the less massive sub-cluster infalling from the SE has already passed the core of the main cluster with a non-null impact parameter.  Figure from \cite{belsole05}}
 \label{fig:A3921}
\end{minipage}
 \end{figure} 
%%%%%%%%%%%%%%%%%%%%%%%%%%%% 

\subsubsection{Formation history  and  dynamical state}

It appears clearly now  that the dynamical state of clusters  does not always simply depend  on the most recent merger event. Multiple on-going merger events  are seen in several clusters, for instance in Coma \cite{neumann03} and A2744 \cite{kempner04}.  The gas relaxation time can be longer than the interval between successive merger events,  especially in dense environments  \cite{belsole04}. For instance, the \xmm\ observation of the double cluster A1750  \cite{belsole04}, located in a super-cluster,  revealed an increase of temperature in the region between the two sub-clusters A1750 C and A1750N (Fig.~\ref{fig:A1750}). This indicates that they have started to merge. However, the interaction is too recent to explain the temperature substructures observed in A1750C  (Fig.~\ref{fig:A1750}). They are likely to result from a more ancient merger.  

Various observations confirm  the importance of the cluster large-scale environment on its formation history. We have now unambiguous evidence that sub-clusters are accreted  preferentially along the direction of filament(s) connecting to the cluster, as expected in the standard scenario of cluster formation. It is provided, for instance, by the observations of A85 \cite{durret03}, A1367 \cite{forman03} and Coma \cite{arnaud01a,neumann03}.   The merger of sub-clusters can occur with an non-zero impact parameter,  e.g. in A3921 (Fig.~\ref{fig:A3921}, \cite{belsole05}) and Coma \cite{neumann03}, probably as a result of large-scale tidal torques. 

\subsection{Statistical studies of cluster morphology}
Beyond the detailed study of specific clusters, statistical analysis  of cluster morphology provide important
information on structure formation.   Till now, such studies have been restricted to X-ray images. They will certainly be extended in the future to temperature maps, now available with \xmm\ and \chandra.  These studies first  require to design quantitative measurements of substructures.  This is not trivial and many estimators can be used:  centre-shifts  \cite{mohr95}, power-ratios \cite{buote96}, $\beta$ and Lee statistics \cite{schucker01} \ldots.  The power-ratio method is of particular interest, because  power ratios  are closely related to the cluster dynamical state \cite{buote96}. 

Substructures are common in nearby clusters. A systematic study of substructures in $470$ clusters detected in the Rosat All Sky Survey show that $\sim 50\%$ of clusters have significant substructures \cite{schucker01}. Similar numbers are obtained from earlier Einstein observations \cite{mohr95,jones92,jones99} or  \rosat\ pointed observations \cite{buote96}.  A positive correlation is observed between the presence of substructures and the density of the cluster environment (the number density of surrounding clusters), as expected in the hierarchical scenario \cite{schucker01}. An anti-correlation is observed with the presence of a cooling core \cite{buote96,schucker01}, suggesting that cooling cores are destroyed during mergers. 

The power ratio technique was used to quantify the evolution of morphology with redshift, using a sample of $40$ distant clusters observed by \chandra. As expected from hierarchical models of structure formation, high-redshift clusters have more substructures and are dynamically more active than low-redshift clusters \cite{jeltema05}. 

\subsection{Mergers and non thermal emission}
The ICM is not only filled with hot thermal gas.  The presence of a large
scale magnetic field and relativistic electrons is revealed by radio
observations of diffuse synchrotron emission: regular, centrally located,
 radio halos and/or irregular, peripherally located, radio relics
\cite[for a review]{giovannini02}.  Further evidence is provided by the detection of
non thermal hard X--ray emission, interpreted as coming from the inverse
Compton scattering of the cosmic microwave background by the
relativistic electrons (see, e.g.,  the observation of A2256 with \sax\ \cite{fuscofemiano00} and~\rxte\ \cite{rephaeli03}). Note that the signal to noise ratio of such observations is still low and their interpretation ambiguous \cite{rephaeli03}. 

It has been known for some time that
radio halos are associated with non relaxed clusters, as expected, e.g.,    if 
relativistic electrons are (re)-accelerated by merger shocks. For instance,  there is  a correlation between the dipole power ratio (a measure of the departure from the virialised state) and the power of the radio halo \cite{buote01}. 
 However, the origin and acceleration mechanism of the relativistic electrons is uncertain \cite{brunetti02}.   
A recent comparison of the radio and \chandra\ temperature maps of merging clusters \cite{govoni04} 
suggests that radio halo electrons are mostly accelerated by the turbulence induced by mergers, rather than directly by shocks. However, when strongly supersonic shocks are present, they can also contribute to the acceleration \cite{govoni04}. First direct evidence of turbulence in the ICM was obtained  recently from  the \xmm\ pressure spectrum of Coma \cite{schucker04}.  

\section{Structural and scaling properties of the cluster population}

\subsection{The cluster population}
A wide range of morphological and physical properties are observed in
the galaxy cluster population.  As mentioned above, substructures are common in galaxy clusters, with
a rich variety in the type and scale of subclustering.  The total mass of clusters ranges
from $10^{13}~\msun$ for groups up to a few $10^{15}~\msun$ for very
rich systems, while the X-ray luminosity of the hot intra-cluster
medium varies by orders of magnitude, from $10^{41}$ to a few
$10^{45}$ ergs/s, and the observed temperature from typically $0.3$ up
to $\sim 15~\keV$.

However, clusters do not occupy the whole parameter space of physical
properties.  We do see strong correlations between observed physical
characteristics of nearby clusters, like luminosity, gas mass, total mass, size and
temperature  \cite{arnaud99,mohr99,finoguenov01,mohr97}.  Furthermore there is also some
regularity in shape.  If one excludes major merger events or very
complex systems ($\sim 20\%$ of systems), the cluster morphology is
usually dominated by a regular centrally peaked main component.  In
that case, the surface brightness, outside the cooling flow region, is
reasonably fitted by a $\beta$-model, once minor substructures (like small
secondary subclusters falling onto a main virialised component) are
excised \cite{neumann99}.

\subsection{The self-similar model of cluster formation}
\label{sec:statstandard}

Regularity in the cluster population is expected on
theoretical grounds.  The simplest models of structure formation, purely based on gravitation, predict that galaxy clusters constitute a self-similar population.  In this section, I briefly summarize  the main characteristics of such models.  Further enlightening discussions on cluster formation can be found in \cite{evrard02r} and in \cite{voit04} and in other reviews in this book.

As clusters are dark matter dominated objects, their formation and evolution is driven by gravity.  The hierarchical  collapse of initial density fluctuations of dark matter, which produces the cluster population,  is a complex dynamical phenomena. It was first modeled using simple spherical collapse models, while a full treatment of the 3-D hierarchical clustering  is now made with  N-body simulations.  From these theoretical works, key characteristics of  the cluster population emerge:

\noindent $\bullet$Major merger events are rare.  At a given time, a cluster can be seen as a collapsed  halo of dark matter, which is accreting surrounding matter and  small groups. The virialized part of the cluster corresponds roughly to a fixed density contrast $\delta \sim 200$ as compared  to the critical density of the Universe, $\rhocz $ at the considered redshift:
\begin{equation}
        \frac{\MV}{\frac{4\pi}{3}\ \RV^{3}} =   \delta  \rhocz
\label{eq:rho}
\end{equation}
where $\MV$ and $\RV$  are the 'virial' mass and radius.  $\rhocz= h^{2}(z) 3 H_0^2/( 8 \pi G)$, where $h(z)$ is the Hubble constant normalized to its local value: $h^{2}(z)=\Omm(1+z)^{3} +\Oml$, where $\Omm$ is the cosmological density parameter and $\Oml$ the cosmological constant

\noindent $\bullet$There is a strong similarity in  the internal structure of virialised dark matter halos.  This reflects the fact that there is no characteristic scale in the problem.

The gas properties directly follow from the dark matter properties, assuming that the gas evolution is purely driven by gravitation, i.e., by the evolution of the potential of the dark matter.

\begin{figure}[t]
\begin{minipage}[c]{0.48\textwidth}
\centering
\includegraphics[width=\textwidth]{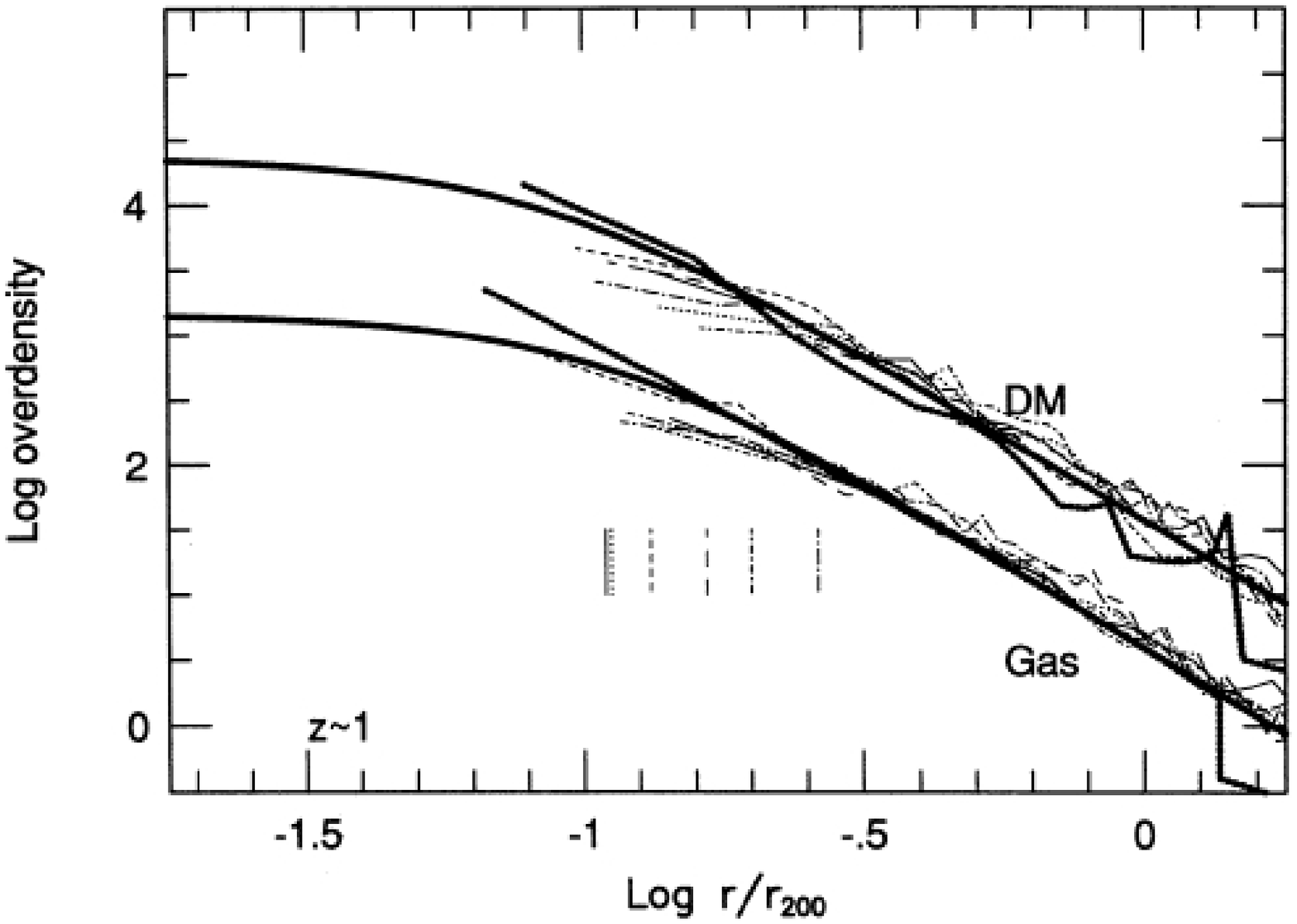}      
 \end{minipage}
\hspace{+0.02\textwidth}
\begin{minipage}[c]{0.48\textwidth}
\centering \includegraphics[width=\textwidth]{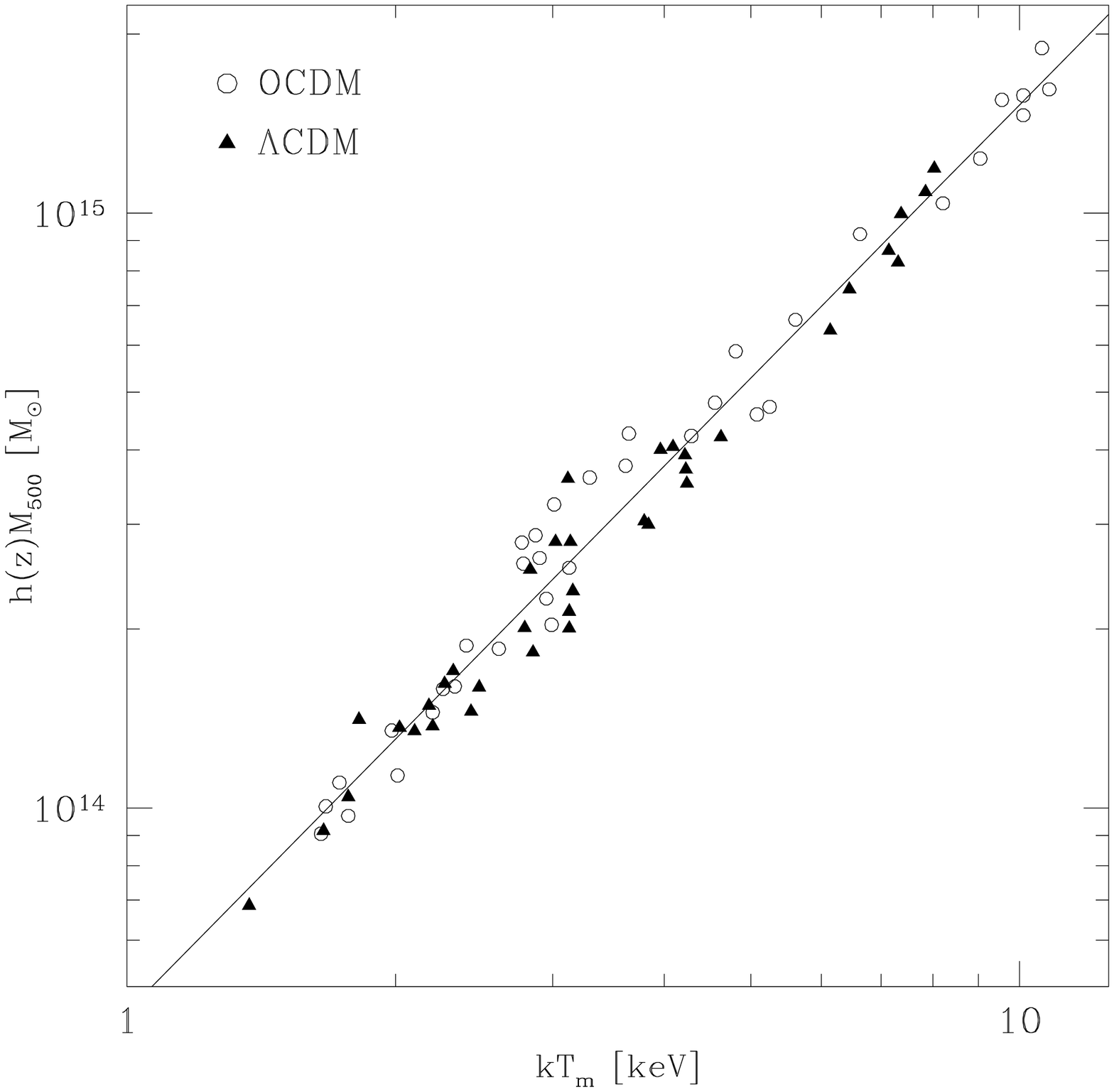}      
\end{minipage}
\begin{minipage}[t]{0.48\textwidth}
 \caption[ ]{Universal cluster density profiles obtained from numerical simulations of structure formation.  The radius is scaled to the virial radius, corresponding to a density contrast of 200 as compared to the mean density of the Universe in which the cluster is embedded.   The Dark matter and gas densities are normalized to this density. Figure from \cite{navarro95}.}
 \label{fig:NFW}
\end{minipage}
\hspace{+0.02\textwidth}
\begin{minipage}[t]{0.48\textwidth}
 \caption[ ]{Virial mass-temperature relationship for simulated clusters. When the mass is scaled by $h(z)$, it varies as $T^{3/2}$ independent of redshift and cosmology. Figure from \cite{mathiesen01}.}
 \label{fig:MT}
\end{minipage}
 \end{figure}

\noindent $\bullet$ The gas internal structure is universal, as this is the case for the dark matter (Fig.~\ref{fig:NFW}).

\noindent $\bullet$  The gas, within the virial radius,  is roughly in hydrostatic equilibrium in the potential of the dark matter. The virial theorem  then gives:
\begin{equation}
\frac{G \mu {\rm m_{p}}\ \MV }{2\ \RV}  = \beta_{T}\ \kT
\label{eq:viriel}
\end{equation}
where  $T$ is the gas mean temperature and $\beta_{\rm T}$ is a normalization factor, which depends on the cluster internal structure. Since this structure  is universal,  $\beta_{\rm T}$ is a constant,  independent on $z$ and cluster mass.

\noindent $\bullet$ The gas mass fraction $\fgas$ reflects the Universe value, since the gas 'follows' the collapse of the dark matter. It is thus constant:
\begin{equation}
\fgas =  \frac{\Mgas}{\MV} = cst
\label{eq:fgas}
\end{equation}

%%%%%%%%%%%%%%%%%%%%%%%%%%%% 
%%%%%%%%%%%%%%%%%%%%%%%%%%%% 

Therefore,  X--ray clusters are expected to exhibit self-similarity:

\noindent $\bullet$  Each cluster is defined by two parameters only:  its mass and its redshift. Because masses are difficult to measure, it is traditional to rather use the cluster temperature. From the basic equations Eq.~\ref{eq:rho}-\ref{eq:fgas},  one can derive a scaling law for each physical property, $Q$, of the form $Q\propto A(z)T^{\alpha}$, that relates it to the redshift and  temperature. The evolution factor,  $A(z)$, in the scaling relations is due to the evolution of the mean dark matter (and thus gas) density which varies as the critical density of the Universe:
\begin{equation}
  \overline{\rhog} \propto \overline{\rhoDM}= \delta \rhocz \propto h^{2}(z)
  \label{eq:rho2}
  \end{equation}
For instance, the total and gas  mass scales as $\Mgas \propto \MV \propto h^{-1}(z)  T^{3/2}$ (Fig.~\ref{fig:MT}), the virial radius as $\RV  \propto   h^{-1}(z) T^{1/2}$. The X-ray luminosity is $\Lx = \int \rhog^{2} dV \Lambda(T)$, where $\Lambda(T) \propto T^{1/2}$, assuming bremsstrahlung emission. It thus scales as $\Lx \propto h(z) T^{2}$.

\noindent $\bullet$
The radial profile of any physical quantity (e.g gas or dark matter density, gas temperature)
can be expressed in scaled
coordinates. The considered quantity is normalized according
to the scaling relation estimated at the cluster temperature and
redshift and the radius is expressed in units of the virial radius.  Then, the scaled radial profiles are the same for all clusters, whatever their redshift or  their temperature (see Fig.~\ref{fig:NFW}).

The self-similar model has been validated  with  numerical simulations of large scale structure formation that include the gas dynamics but only gravitation \cite{navarro95,evrard96,bryan98,eke98} and by similar semi-analytical hierarchical models \cite{cavaliere99}.  However, there is  some ambiguity in the definition of the  'virial' mass of a cluster, or equivalently in the value of the density contrast $\delta$ used in Eq.~\ref{eq:rho}.   This is fully discussed in \cite{voit04} (see also \cite{white01}). In the spherical top-hat model, where a cluster is considered as a spherical perturbation which has just collapsed,  the boundary of a cluster is well defined. It corresponds to  $\delta = 18\pi^{2} \sim
200$ in a critical density Universe (SCDM model), while $\delta$ is a function $\Delta(z,\Omm,\Oml)$ of  redshift and  $\Omm,\Oml$ in a $\Lambda$CDM cosmology \cite{bryan98}.  In reality, there is no such thing as a strict boundary between a relaxed part of a cluster and an infall region. Recent numerical simulations  \cite{mathiesen01,evrard02r} indicate that tighter scaling laws are obtained if one uses a common value, $\delta \sim200$,  for all redshift  and cosmologies (Fig.~\ref{fig:MT}).  Nevertheless, the spherical model definition is often used in the literature.  In that case,  note that 
the expected evolution factor of the scaling laws is different:  $h^{2}(z)$ is  replaced by $\Delta(z,\Omm,\Oml)h^{2}(z)$ (Eq.~\ref{eq:rho2}).

Since the first X-ray observations of clusters, the statistical properties of the observed cluster population
have been compared with these theoretical
predictions. This has provided valuable insight into the physics that governs the large
scale structure formation and evolution of both the Baryonic and the Dark
Matter components. 

\subsection{The Dark matter in local clusters}
\label{sec:statdm}

 \subsubsection{Theoretical predictions}

%The Dark Matter  distribution in clusters is a sensitive test of current scenarios of structure formation and of the
%nature of the Dark Matter itself.
Recent high resolution simulations predict
that Cold Dark Matter profiles are cusped in the center
\cite{navarro97,moore99,navarro04,diemand04}. An example is  the NFW profile \cite{navarro97} given by
\begin{equation}
\rhoDM(r) = \frac{\rhocz \dc}{(c r/\Rv) (1+ c r/\Rv)^2}
\label{eq:nfw}
\end{equation}
\noindent where $\rhoDM(r)$ is the mass density, $\Rv$ is the radius corresponding to a density contrast of $200$ (roughly the virial radius) and $c$ is a concentration parameter. $\dc$ is the characteristic dimensionless density, related to the concentration parameter by  $\dc = (200/3) c^3/[ln (1+c) - c/(1+c)]$.  If $c$ was a constant, the profiles of all clusters would be perfectly self-similar.  Actually, it is expected to vary slightly with $z$ and system mass \cite{navarro97,dolag04}.  The corresponding integrated mass profile is of the form: $M (r) = \Mv m(cr/\Rv)/m(c) $, where $\Mv$ is the mass enclosed  within $\Rv$ (the 'virial mass'). This mass profile  can be directly compared with observations.

The NFW density profile varies from $\rhoDM \propto r^{-1}$
at small radii to $\rhoDM  \propto r^{-3}$ at large radii. The exact slope of the dark matter profile in the center is still debated (see  \cite{diemand04,navarro04} for latest results).  For instance, the simulations in  \cite{moore99} give a steeper slope  $\rhoDM \propto r^{-1.5}$.
\begin{figure}[t]
\begin{minipage}[c]{0.48\textwidth}
\centering
\includegraphics[width=\textwidth]{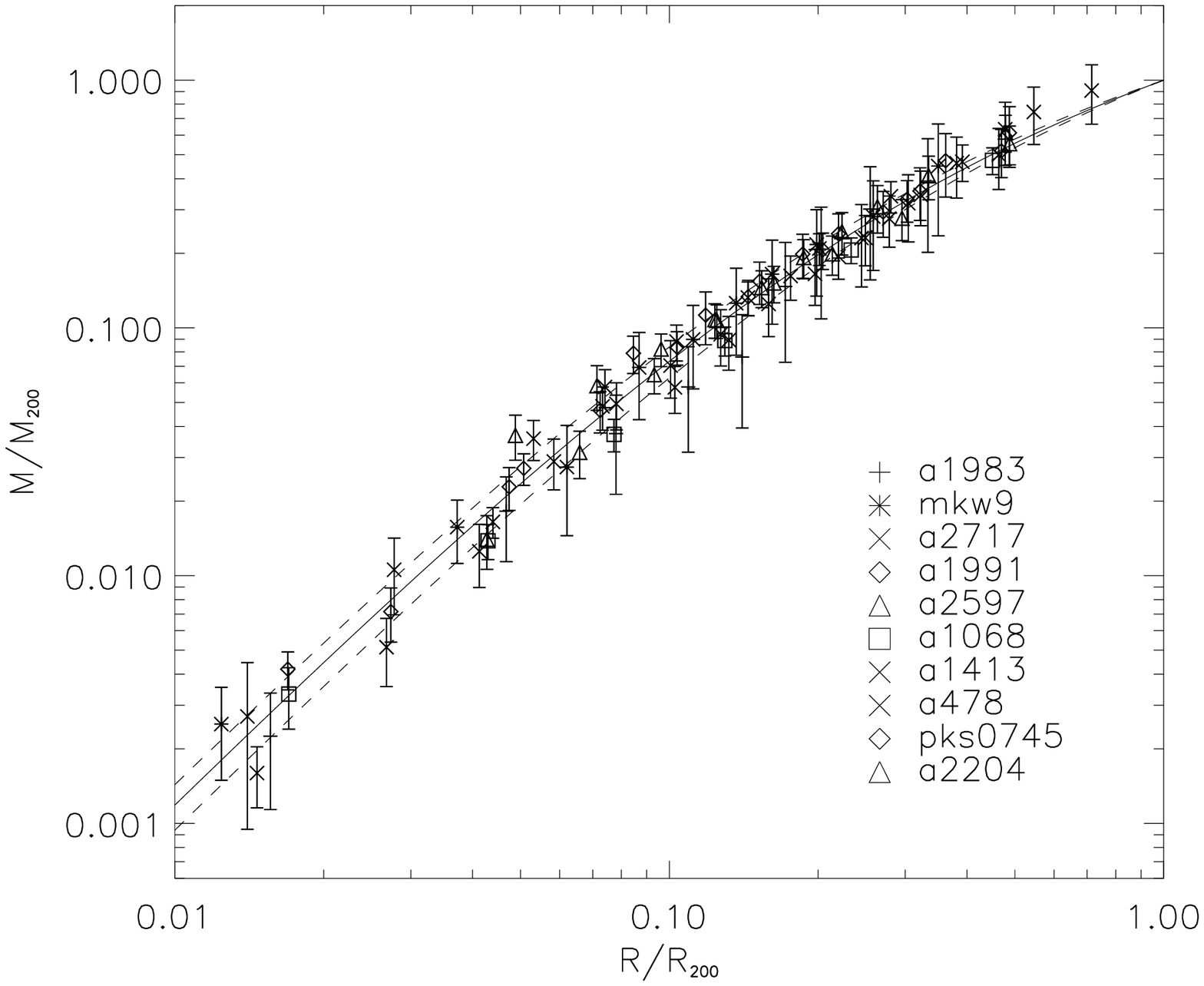}     % includes figure foo.eps
 \end{minipage}
\hspace{+0.02\textwidth}
\begin{minipage}[c]{0.48\textwidth}
\centering \includegraphics[width=\textwidth]{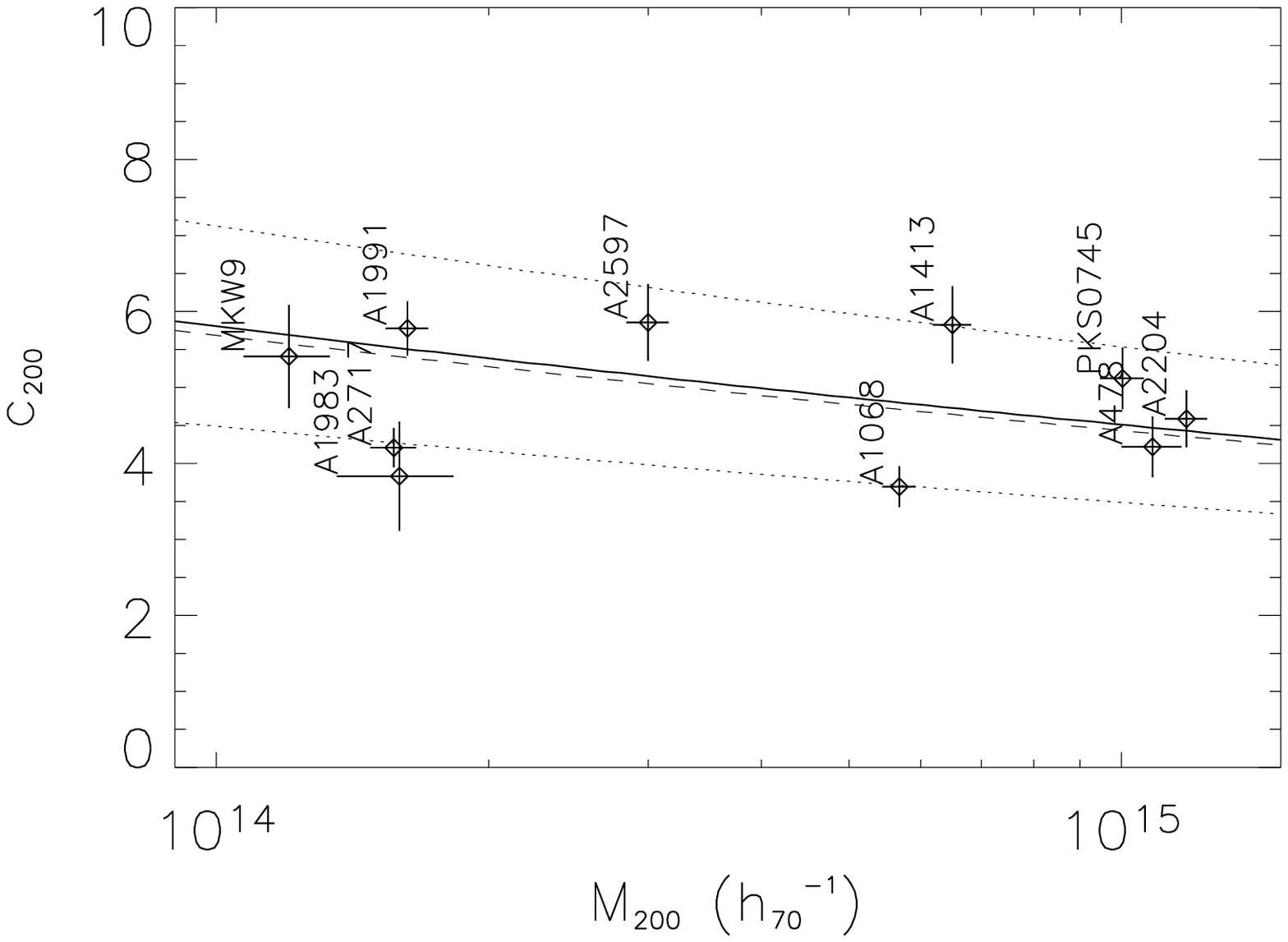}     % includes figure foo.eps
\end{minipage}
\begin{minipage}[t]{0.48\textwidth}
 \caption[ ]{Integrated total mass profiles for a sample of clusters in the temperature range $2-9~\keV$ measured with \xmm.  The mass is scaled to
$\Mv$, and the radius to $\Rv$, both values being derived from the
best fitting NFW model. The solid line corresponds to the mean
scaled NFW profile and the two dashed lines are the associated standard
deviation. Figure from \cite{pointecouteau05}.}
 \label{fig:msc}
\end{minipage}
\hspace{+0.02\textwidth}
\begin{minipage}[t]{0.48\textwidth}
 \caption[ ]{Concentration parameter $c_{200}$ versus the
    cluster mass
  $M_{200}$ (derived from fitting a NFW model to the data presented Fig.~\ref{fig:msc}).  The solid line represents the variation of $c_{200}$
  for clusters at $z=0$ from numerical simulations
  \cite{dolag04}. The dotted lines are the standard deviation
  associated with this relation. The dashed line represents the same
  relation at a redshift of $z=0.15$( the maximum redshift for
 the sample). Figure from \cite{pointecouteau05}.}
 \label{fig:cm}
\end{minipage}
 \end{figure}

\begin{table}[t]
\begin{minipage}[c]{0.48\textwidth}
\begin{tabular}{lclc}
\hline
Relation & Slope & Reference & Note \\
	\hline
\LxT & $2.64\pm 0.27$ &\cite{markevitch98a} & a\\
     & $2.88\pm 0.15$ & \cite{arnaud99} & b  \\
	\hline
\MgT &$1.98 \pm 0.18$ & \cite{mohr99} & c\\
 &$1.71 \pm 0.13$ & \cite{vikhlinin99}& \\
 	&   $1.89 \pm 0.20$ & \cite{ettori02} & c \\
	&$1.80 \pm 0.16$ &  \cite{castillo03}& c \\
	\hline
\EMT & 1.38 &  \cite{neumann01} &   \\
       \hline
\ST & $0.65 \pm 0.05$  &\cite{ponman03} & d   \\
\hline
\fgT &$0.34 \pm 0.22$ & \cite{mohr99} & c\\
 & $0.66 \pm 0.34$  & \cite{ettori02}  & e  \\
\hline
\end{tabular}
\end{minipage}
\hspace{+0.02\textwidth}
\begin{minipage}[t]{0.48\textwidth}
 \caption{ Logarithmic slope of local scaling laws from the literature (not exhaustive).    In the standard self-similar model: $\Lx \propto  T^{2}$,  $\Mgas \propto T^{1.5}$, $EM \propto T^{0.5}$ and $S \propto T$. {\it Notes}: a: Corrected for cooling flow contribution; b: non cooling-flow clusters; c: Gas mass integrated within $R_{500}$;  d: entropy estimated at $0.1 \Rv$; e: $\fgas$ integrated within $R_{1000}$. }
\label{tab:tab1}
\end{minipage}
 \end{table}
\subsubsection{Observed mass profiles}

With \chandra\ and \xmm\, we can now measure precisely the  total mass
distribution in clusters (from Eq.~\ref{eq:HE}) over a wide range of radius, from $\sim 0.001~\Rv$ \cite{lewis03}  up to $\sim 0.7~\Rv$ \cite{pratt02}.    The validity of the X-ray method to derive masses from  the hydrostatic equilibrium  equation  has been validated for relaxed clusters, by comparing \chandra\ mass estimates with independent lensing mass estimate \cite[and reference therein]{allen01}.  The observed mass profiles are well fitted using a NFW model. This is true not only for massive clusters  \cite{david01,allen01,arabadjis02,pratt02,lewis03,buote04,pointecouteau04}, but also for low mass systems  \cite{pratt03,pratt05}.   A profile with a flat core is generally rejected with a  high confidence level.  In a few cases \cite{lewis03, buote04, pointecouteau04}, the inner slope has even been measured precisely enough\footnote{Note that such measurements are not limited by the instrument capabilities but by  the number of suitable targets. In most clusters the very center is disturbed (see below), invalidating the HE approach to compute the central mass profile.}  to distinguish between various predictions.   They  favor a slope of $\alpha\sim -1$. This steep density profile in the center of clusters is incompatible with the flattened core DM profiles predicted  by self-interacting Dark Matter models \cite{arabadjis02,buote04}.  In one case, the validity of the NFW model was tested up to the virial radius \cite{pratt02}.

Recently, the first quantitative check of the universality of the mass profile was performed  with \xmm\ \cite{pratt05,pointecouteau05}.
As shown Fig.~\ref{fig:msc}, the mass profiles scaled in units of $\Rv$ and $\Mv$ nearly coincide, with a dispersion of less than
$15\%$ at $0.1~\Rv$.  Furthermore, the shape is quantitatively consistent with the predictions. The derived concentration parameters  are consistent with the  $c$--$\Mv$ relation derived from numerical simulations for a $\Lambda$CDM cosmology (Fig.~\ref{fig:cm}).

This excellent {\it quantitative} agreement with theoretical predictions  provides strong evidence in favor of the Cold Dark Matter cosmological scenario, and shows that the physics of the Dark Matter collapse is well understood, at least down to the cluster scale.

Finally, it is worth mentioning again the observation of 1E0657-56, which yields very interesting, independent constraints on the Dark Matter. A comparison of the gas, galaxies and weak-lensing maps demonstrates the presence and dominance of non-baryonic Dark Matter in this cluster and shows that the cross-section of the dark matter particles is low, excluding again most of the self-interacting dark matter models \cite{clowe04,markevitch04}.

\subsection{Gas properties in local clusters}
\label{sec:statgas}

It has been known for nearly 20 years  that gas properties deviate  from the standard self-similar model predictions: the $\LxT$ relation is steeper than expected.  This was the first  indication that the gas physics should be looked at more closely and that non-gravitational processes could play a role (e.g. \cite{evrard91}).  We have now a much more precise view of cluster properties.  The new picture that emerged from recent observations is that local
clusters, down to remarkably low mass ($\kT \sim 2 \keV$), do obey self-similarity. However, the scaling laws differ
from simple expectations.   In this part, I summarize these observations, focusing on clusters above a typical temperature of $2~\keV$.  Below that temperature, one typically finds groups. The latest results indicate that their properties may follow the trends observed in clusters, albeit with a significantly increased dispersion \cite{osmond04}. For a recent review on group properties, the reader may refer to \cite{mulchaey04r}.

%%%%%%%%%%%%%%%%%%%%%%%%%%%%%%%%%%%%%
\begin{figure}[t]
\begin{minipage}[c]{0.48\textwidth}
\centering
\includegraphics[width=\textwidth]{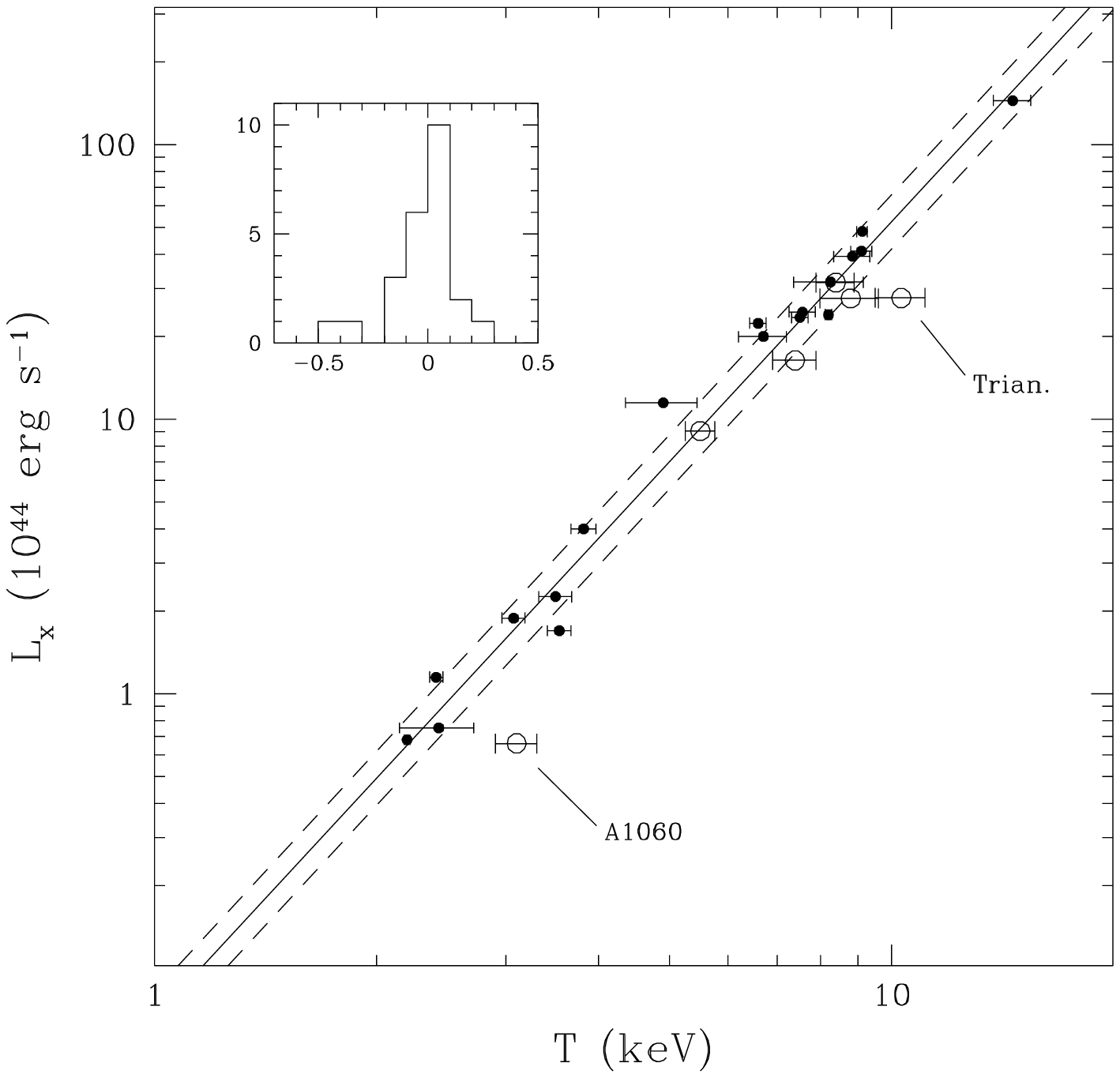}    
 \end{minipage}
\hspace{+0.02\textwidth}
\begin{minipage}[c]{0.48\textwidth}
\includegraphics[width=\textwidth,clip]{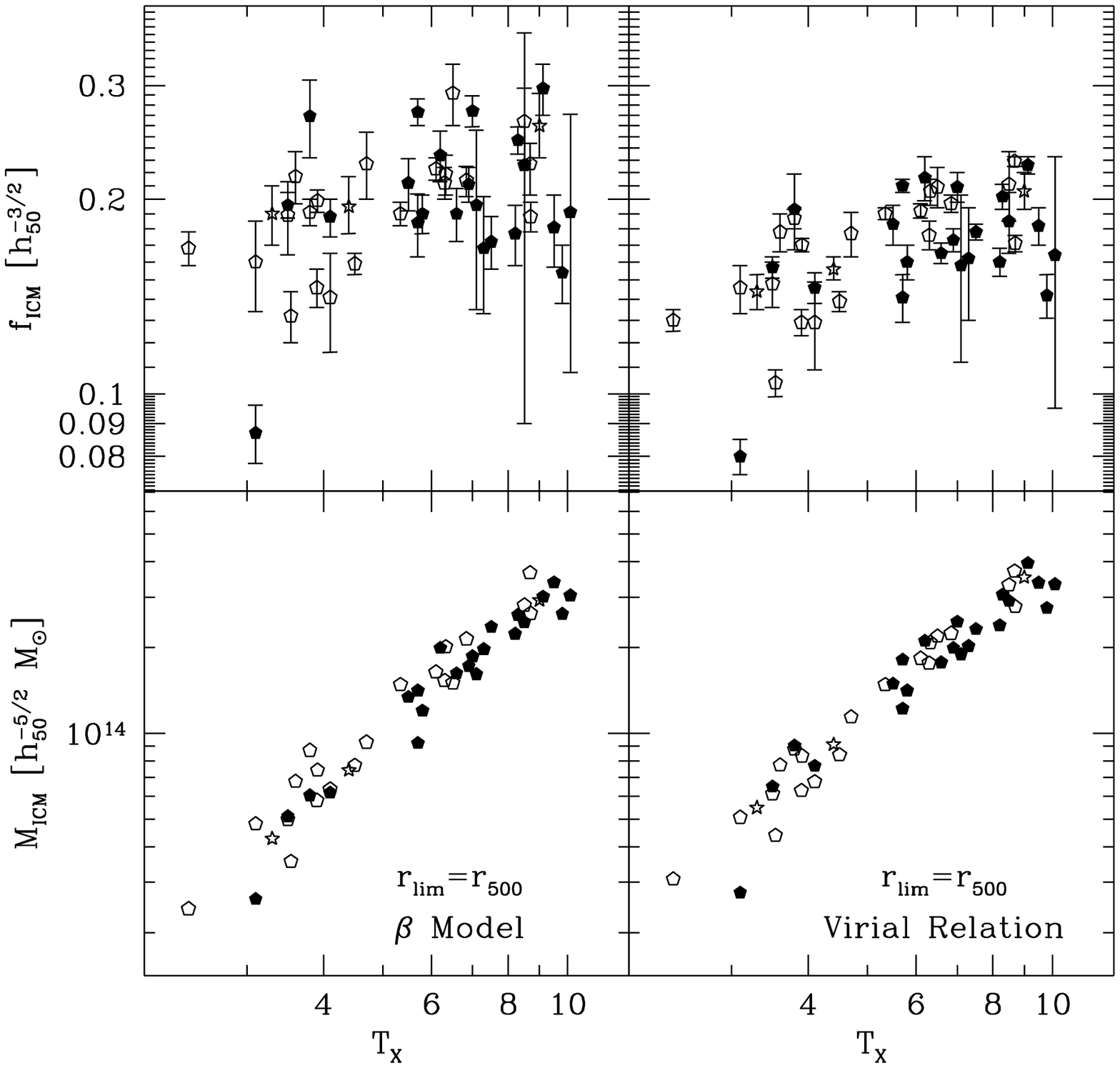}     
\end{minipage}
\begin{minipage}[t]{0.48\textwidth}
 \caption[ ]{The relation between the bolometric luminosity  and the temperature.  Only clusters without strong cooling in the center are considered. The data are mostly from \ginga\ observations. The relation is steeper than expected in the standard self-similar model: $\Lx \propto T^{2.88\pm0.15}$. Figure from \cite{arnaud99}.}
 \label{fig:LxT}
\end{minipage}
\hspace{+0.02\textwidth}
\begin{minipage}[t]{0.48\textwidth}
\caption[ ]{The relation between the gas mass and the temperature. The gas mass was measured with \rosat\ and estimated within $R_{500}$.  The temperature was measured with the \asca\ or earlier missions.  The relation is steeper than expected in the standard self-similar model: $\Mgas \propto T^{1.98\pm0.18}$. Figure from \cite{mohr99}.}
 \label{fig:MgasT}
\end{minipage}
 \end{figure}
%%%%%%%%%%%%%%%%%%%%%%%%%%%%%%%%%%%%%

\subsubsection{Local scaling laws}

Scaling laws relate various physical properties  with  $T$.  This includes global gas properties like the X-ray luminosity $\Lx$ \cite{markevitch98a,arnaud99}, the gas mass $\Mgas$  \cite{mohr99,vikhlinin99,ettori02,castillo03} or the gas mass fraction $\fgas$ \cite{mohr99,ettori02}.  The corresponding scaling relations are steeper than expected (Fig.~\ref{fig:LxT},\ref{fig:MgasT}). However, an unambiguous picture can be obtained only by studying the internal structure of clusters. For instance, a steepening of the \LxT\ relation could be due to a systematic increase of  the mean gas density  with $T$ (a simple modification of the scaling laws) or to a variation of cluster shape with $T$ (a break of self-similarity).  By looking at radially averaged profiles of interesting quantities, particularly the density and the entropy,   two issues can be addressed at the same time. The first is to see whether the profiles agree in {\it shape\/}; the second is to investigate the {\it scaling\/} of the profiles.  The various scaling relations  are summarized in Table~\ref{tab:tab1}.

\subsubsection{Gas density  profiles}
\label{sec:ng}

\begin{figure}[t]
\includegraphics[width=\textwidth]{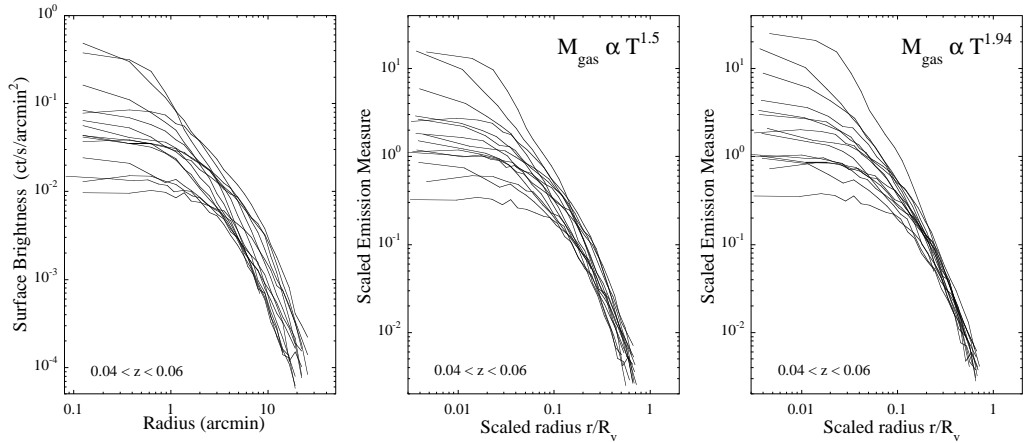}     % includes figure foo.eps
 \caption[ ]{Left: Surface
brightness profile of 15 nearby ($0.04<z<0.06$) and hot clusters ($\kT>3.5~\keV)$ observed with
\rosat. Middle: Scaled emission measure profiles.
The radius is normalized to the virial radius and $EM$ is scaled
according to the classical laws ($EM \propto T^{1/2}$ or equivalently
$\Mgas \propto T^{3/2}$).  Right: Same assuming $\Mgas \propto
T^{1.94}$, or $EM \propto T^{1.38}$. Figure adapted from \cite{neumann99,neumann01}.}
 \label{fig:SxEM}
 \end{figure}

The gas content and density distribution can been  studied through the emission measure along the line of sight $EM(r) = \int_{r}^{\Rv} \ne^{2}~dl$, which is easily derived from the X-ray surface brightness profile (Sec.~\ref{sec:obsphys}, Fig.~\ref{fig:SxEM}). The scaled $EM$ profiles of {\it hot} clusters measured with \rosat\ were found to be similar in shape outside typically $0.1-0.2\Rv$   \cite{neumann99,vikhlinin99,neumann01}.  There is a large dispersion in the central region, linked with the presence of a cooling core (see Sec.~\ref{sec:cool}).  Outside that  region the universal profile is well fitted by a $\beta$--model, with $\beta=2/3$ and $\rc \sim 0.12\Rv$ \cite{arnaud02a}. Note however that more precise \xmm\ data on hot clusters gives  a steeper density profile in the center, even for weak cooling flow clusters \cite{pratt02}.   This is a consequence of the cusped nature of the dark matter profile.

In the standard self-similar framework, the mean gas density does not depend on the temperature (Eq.~\ref{eq:rho2}). The emission measure is thus expected to  scale as $\Rv$, i.e $EM \propto T^{0.5}$. The scatter in the scaled $EM$ profiles was found to be considerably reduced if a much steeper $EM$---$T$ relation, $EM \propto T^{1.38}$, was used to scale the profiles \cite[see Fig.~\ref{fig:SxEM}]{neumann01}. This explains the steepening of the \LxT\ relation and translates into $\Mgas \propto T^{1.94}$, which is consistent with the observed steepening of the \MgT\ relation (Table~\ref{tab:tab1}) as compared to the standard $\Mgas \propto T^{1.5}$ scaling. A test case  study  with \xmm\ of a cool cluster (A1983) suggests that clusters follow the universal scaled $EM$ profile down to temperature as low as $\kT \sim 2~\keV$ \cite{pratt03}.

There is converging evidence that the gas distribution in clusters is more inflated than the dark matter distribution. The integrated  gas mass fraction increases with scaled radius \cite{sanderson03}. The increase, measured with \sax\,  is about $25\%$ between a density contrast of $\delta=2500$ (about $1/3$ of the virial radius) and $\delta=500$  \cite{ettori03}.  The results at $\delta=500$ were largely based on extrapolations and could be biased. However, recent \xmm\  and \chandra\ results seem to confirm this trend. From a sample of  six massive clusters observed by \chandra,  an average $\fgas$ value of $0.113\pm0.013$ was derived at $\delta =2500$ \cite{allen02} . At this density contrast, the value  derived from the \xmm\ analysis  of A1413 is  $\fgas=0.11$, perfectly consistent with Chandra results \cite{pratt02}. The \xmm\ data  extends up to $\delta=500$ and show that $\fgas$ keeps increasing with radius to reach $\fgas\sim0.14$ at $\delta =500$ ($25\%$  increase).

\subsubsection{Temperature profiles}

There is also a  similarity in the temperature
profiles of hot clusters observed with \asca\ and \sax\ beyond the cooling core region
\cite{markevitch98b,irwin00,degrandi02}.  In relaxed clusters, there is usually a drop  of temperature towards the center ($r \lesssim 0.1\Rv$). This corresponds  to the cooling core region (see Sec.~\ref{sec:cool}).  There is also a tendency for clusters with cooling cores
to have flatter temperature profiles at large scale than non cooling core clusters, suggesting that the profile shape depends on the cluster dynamical state \cite{degrandi02}.    

 A \xmm\ study of an unbiased sample of clusters shows a variety of shapes, probably linked to various dynamical states \cite{zhang04}. The self-similarity of shape seem to be  confirmed by \chandra\ \cite{allen01,vikhlinin04} and \xmm\ data \cite{arnaud04,piffaretti04} for relaxed clusters.   However, no consensus has been reached yet on the exact shape of the profiles. This was already the case for \asca\ and \sax\ studies and this is still the case with \xmm\ and \chandra.  Some studies find relatively flat profiles (within $\sim 20\%$),  beyond the cooling core  region, up to $0.3\Rv$ \cite{allen01} or even to $0.5\Rv$ \cite{arnaud04}.   Other studies found steadily decreasing profiles, by $30\%$ between $0.1\Rv$ and $0.5\Rv$  \cite{piffaretti04} or even by $50\%$ \cite{vikhlinin04}.

\subsubsection{Gas entropy}

The studies of the gas density profiles suggest that the departures of the gas scaling laws from the standard self-similar model are due to a non-standard scaling of the mean gas density with temperature and not to a break of self-similarity.  To understand the physical origin of these deviations, one must consider the gas entropy rather than the density.   The `entropy' is traditionally defined as $S = \kT/\ne^{2/3}$ and is related to the true thermodynamic entropy via a logarithm and an additive constant.   It  is a fundamental characteristic of the ICM, because it is a probe of the thermodynamic history of the gas \cite{voit02,voit04}.  The entropy profile of the gas  and the shape of the potential well, in which it lies, fully define  the X-ray properties of a relaxed cluster \footnote{The gas density and temperature profiles can be determined from  the entropy profile and the hydrostatic equation (Eq.~\ref{eq:HE}) with a boundary condition at the virial radius}.  In the standard self-similar picture, the entropy, at any scaled radius, should scale simply as $S \propto h(z)^{4/3}T$.

\begin{figure}[t]
\begin{minipage}[c]{0.48\textwidth}
\centering
\includegraphics[width=\textwidth]{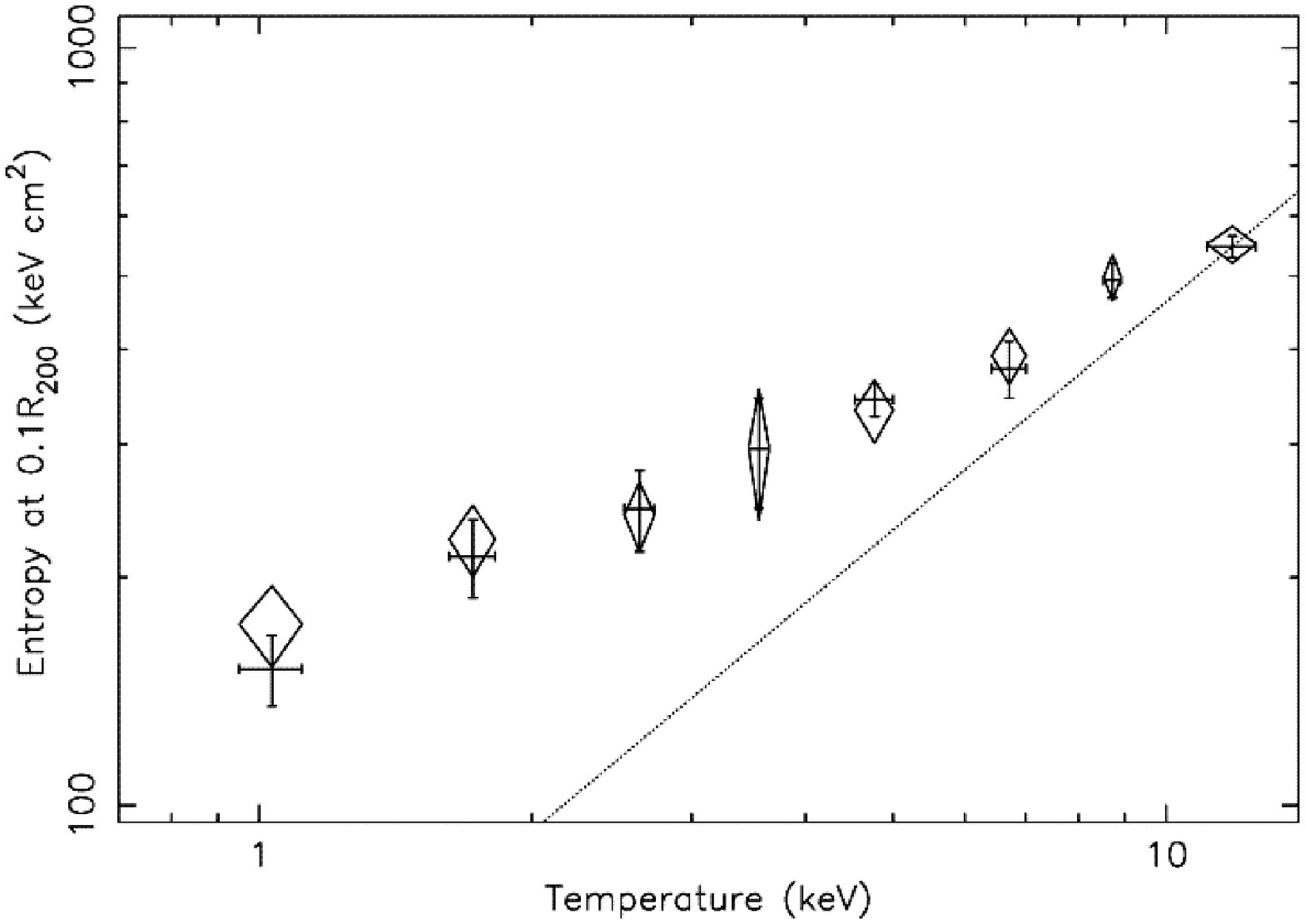}     % includes figure foo.eps
 \end{minipage}
\hspace{+0.02\textwidth}
\begin{minipage}[c]{0.48\textwidth}
\centering \includegraphics[width=\textwidth]{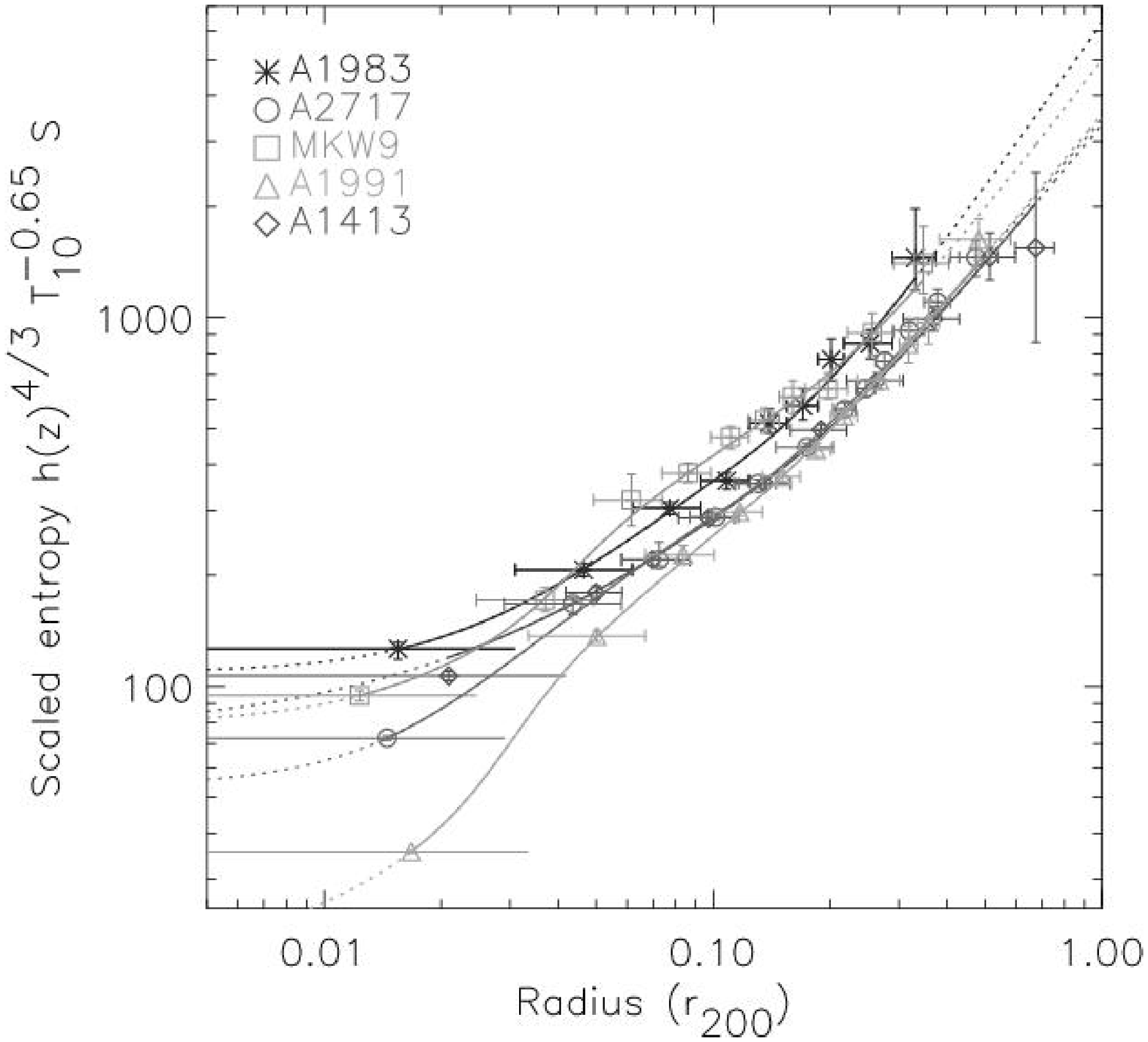}     % includes figure foo.eps
\end{minipage}
\begin{minipage}[t]{0.48\textwidth}
 \caption[ ]{Gas entropy at $0.1~\Rv$ versus cluster temperature. Data  of individual clusters, derived from \rosat\ and \asca\ observations, have been grouped by temperature bins. The dotted line shows the standard self-similar prediction, $S \propto T$. The relation is shallower than expected in the standard self-similar model: $S \propto T^{0.65\pm0.05}$. Figure from \cite{ponman03}.}
 \label{fig:ST}
\end{minipage}
\hspace{+0.02\textwidth}
\begin{minipage}[t]{0.48\textwidth}
 \caption[ ]{\xmm\ scaled entropy profiles. The temperature ranges from $2~\keV$ to $6.5\keV$. The entropy has been scaled following the empirical relation, $S \propto h(z)^{-4/3}T^{0.65}$. The radius is scaled to the measured virial radius, $\Rv$. Figure from \cite{pratt05}. }
 \label{fig:Sprof}
\end{minipage}
 \end{figure}

Since  the pioneering work  of \cite{ponman99}, it is known that the entropy measured at $0.1 \Rv$ exceeds the value attainable
through gravitational heating alone, an effect that is especially noticeable in low mass systems. Various
non-gravitational processes have been proposed to explain this entropy excess, such as heating before or after collapse (from SNs or AGNs) or radiative cooling.  A recent study of 66 nearby systems observed by \asca\ and \rosat\  \cite{ponman03} shows that the \ST\ relation follows a power law but with a smaller slope than expected: the entropy measured at $0.1 \Rv$ scales as $S \propto T^{0.65}$ (see Fig.~\ref{fig:ST}).  High quality  \xmm\ observations \cite[see Fig.~\ref{fig:Sprof}]{pratt03,pratt05} show a remarkable self-similarity in the shape of the entropy profiles down to low mass ($\kT \sim 2~ \keV$).  Stacking analysis of \rosat\  data gives the same results \cite{ponman03}.  Except in the very centre, the \xmm\ entropy profiles are self-similar in shape, with close to power law behavior in the $0.05~\Rv < r < 0.5~\Rv$ range. The slope is slightly shallower than predicted by shock heating models, $S(r)\propto r^{0.94\pm0.14}$. The normalization of the profiles  is consistent with the $S \propto T^{0.65}$  scaling.  Similar results were obtained more recently on a larger sample observed with \xmm\ \cite{piffaretti04}.

The self-similarity of shape of the entropy profile is a strong constraint for models.  Simple pre-heating models, which predict large isentropic cores, must be ruled out. In addition to the gravitational effect, the gas history probably depends on the interplay between cooling and various galaxy feedback mechanisms (see \cite{voit04} and other reviews in this book). 

\subsubsection{The complex physics in cluster core}
\label{sec:cool}

As mentioned above, there is a very large dispersion in the core properties, within typically $0.1\Rv$.  This is linked to the complex physics at play in the cluster center.  In the center of clusters the  gas density is high.  The cooling time, which scales as $t_{\rm cool}  \propto T^{1/2}/\ne$ can be shorter than the 'age' of the cluster, the time since the last major merger event. We thus expect the temperature to decrease due to radiative cooling and the density to increase so that the gas stays in quasi hydrostatic equilibrium. Both \chandra\ \cite{allen01} and \xmm\  \cite{kaastra04} observations clearly confirm this temperature drop  in cooling clusters.  These new observations have also dramatically changed our vision of cooling cores in clusters.  This topic has been the subject of a recent conference \cite{reiprich04r} and is only briefly discussed  here.

A major surprise is the lack of very cool gas, inconsistent with standard isobaric cooling flow models. The \xmm/RGS high resolution spectra exhibit strong emission from cool plasma at just below the ambient temperature, $T$, down to $T/2$, but also exhibit a severe deficit of emission compared to the predictions at lower temperatures \cite{peterson03}. The standard model is also inconsistent with \xmm/EPIC data \cite{molendi01,bohringer02,kaastra04}. In parallel, arc-second spatial imaging with \chandra\ have revealed complex interaction between AGN activity in the cluster center and the intra-cluster medium \cite[for a review]{blanton04}.  One observes X-ray cavities or 'bubbles', created by the central AGN radio lobes  as they displace the X-ray gas. They are usually surrounded by cool rims and not by shocked gas.  However, a shock was recently discovered at the boundary of the cavities in  MS0735.6+7421 \cite{macnamara05}. There are also 'ghost' cavities, probably  bubbles that have buoyantly risen away from the cluster center.   Spectacular examples of such phenomena are observed in the center of the Perseus cluster \cite{fabian00,fabian03}. In this cluster ripples are also observed in the X-ray surface brightness  of the gas surrounding the central bubble.  This was interpreted as resulting from  the propagation of weak shocks and viscously dissipating sound waves due to repeated outbursts of the central AGN  \cite{fabian03}.  Cold fronts and  sloshing gas are also observed with \chandra\ in the center of relaxed clusters \cite[for a review]{markevitch02r}.

Whether and how both phenomena, the absence of very cool gas and AGN/ICM interaction,  are connected is still unclear \cite[for reviews]{voit04,bregman04,fabian03r}.  For instance, AGN heating may limit cooling but  conduction could also play a role in heating the central region.  A better understanding of cooling and AGN heating in the central part of clusters has further implications because both phenomena  play a role at larger scales in clusters and during galaxy formation. Finally, the core properties have also a substantial impact on the  \LxT\ relation. When a cooling core is present, the luminosity is boosted (due to the peaked density profiles) and the mean temperature is decreased (due to the temperature drop in the center). This results in a large dispersion in the \LxT\ relation. The dispersion is decreased when these effects are corrected for \cite{markevitch98a}. Then the  \LxT\ relation is the same as for  clusters without strongly cooling cores  \cite{arnaud99}. 

  \subsubsection{The $M$--$T$ relation}

The \MT\ relation is a fundamental scaling relation. Since other scaling relations are expressed in terms of the temperature $T$,  the \MT\ relation provides the missing link between the gas properties and the mass. Furthermore, estimations of the cosmological
parameters from cluster abundances in clusters, or from the spatial distribution of clusters, heavily rely on this relation to
relate the mass to the observables available from X-ray surveys (see Sec.\ref{sec:cosmoab}).
A sustained observational effort to measure the local \MT\ relation has been undertaken using
\rosat, \asca\ and \sax, but no definitive picture  had emerged. Does the mass scale as $T^{3/2}$ as expected
\cite{horner99,ettori02,castillo03}? Is this true only in the high mass range ($\kT\gtrsim4~\keV$), with a steepening at lower mass
\cite{nevalainen00,xu01,finoguenov01}?  Is the slope higher than expected over the entire mass range \cite{sanderson03}? This was unclear.  The normalizations of the \MT\ relation derived from \asca\ data are generally lower than predicted by adiabatic numerical simulations \cite{evrard96}, by typically $40\%$ \cite{nevalainen00,finoguenov01}.  On the other hand, using \sax\ data, a normalisation consistent with the predictions was obtained, albeit with large error \cite{ettori02}.

%%%%%%%%%%%%%%%%%%%%
\begin{figure}[t]
\begin{minipage}[c]{0.40\textwidth}
\centering
\includegraphics[width=\textwidth, angle=-90.]{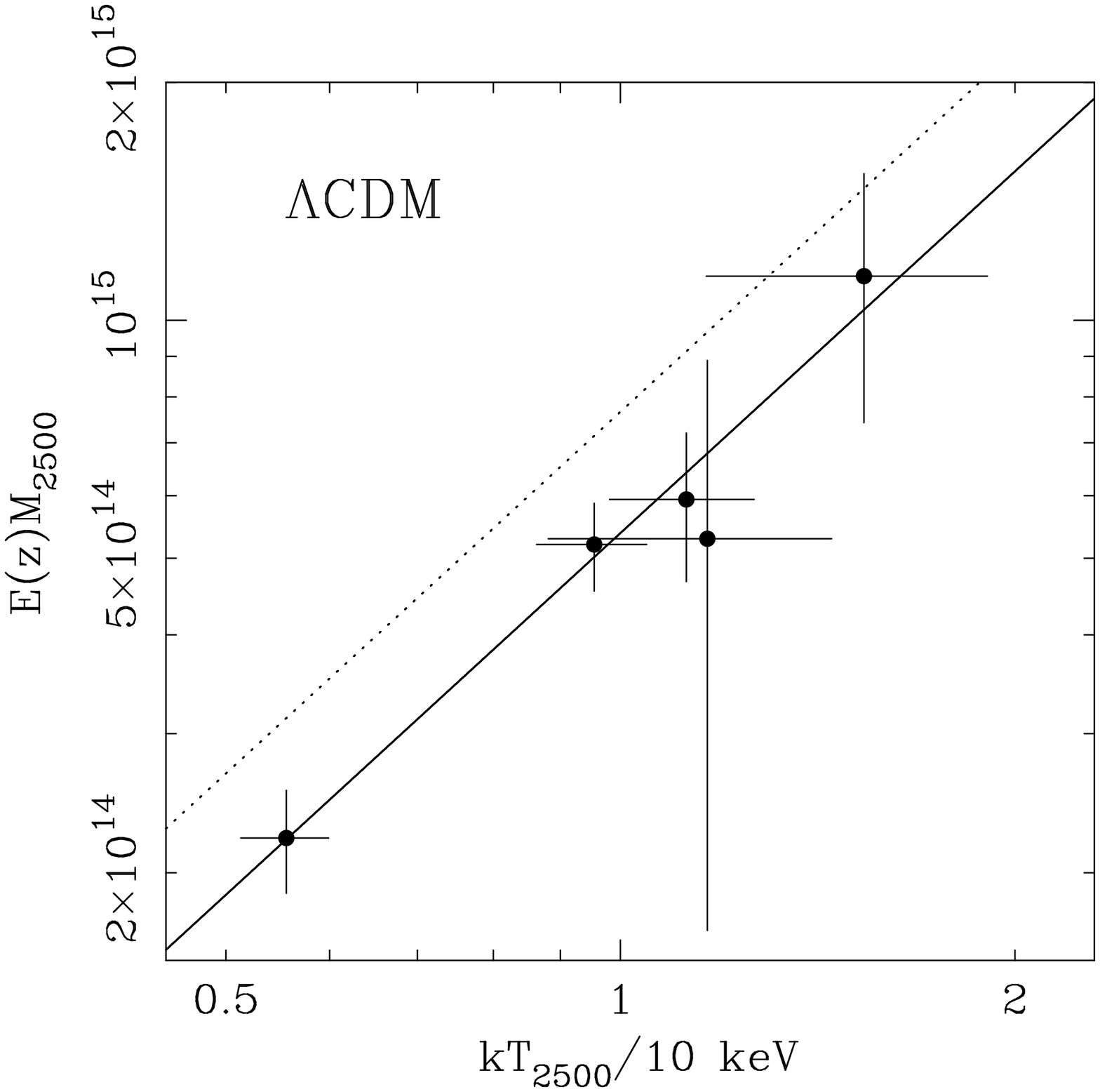}      
 \end{minipage}
\hspace{+0.02\textwidth}
\begin{minipage}[c]{0.56\textwidth}
\centering \includegraphics[width=\textwidth, clip]{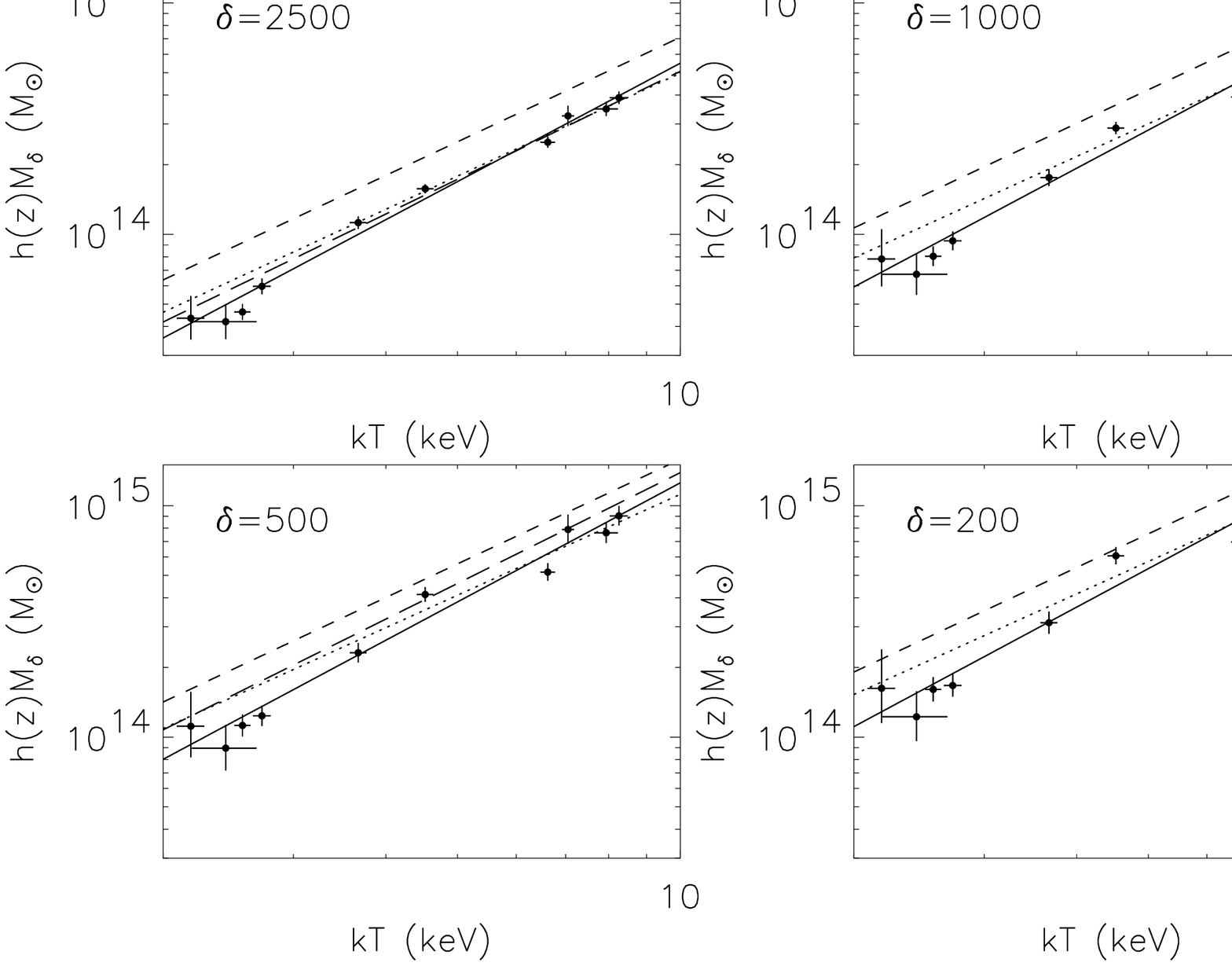}     
 \end{minipage}
\begin{minipage}[t]{0.40\textwidth}
 \caption[ ]{The $M_{2500}$--$T$ relation observed with \chandra\ for hot clusters. Solid line: The best fit power law, $M_{2500} \propto T_{2500}^{1.51\pm0.27}$. Dashed line: the predicted relation from adiabatic numerical simulations \cite{evrard96}. Figure from \cite{allen01}.}
 \label{fig:MT2500}
\end{minipage}
\hspace{+0.02\textwidth}
\begin{minipage}[t]{0.56\textwidth}
 \caption[ ]{The $M_{500}$--$T$ relation observed with \xmm.  Solid line: The best fit power law for the whole sample. Dotted line: same  for the hot cluster subsample.  Dashed line: the predicted relation from adiabatic
numerical simulations \cite{evrard96}. Long-dashed line:  
the relation derived from a numerical simulation including radiative
cooling, star formation and SN feedback \cite{borgani04}.
   Figure from \cite{arnaud05}. }
 \label{fig:MT500}
\end{minipage}
 \end{figure}
%%%%%%%%%%%%%%%%%%%%

These studies had to rely largely on extrapolation to deduce the virial mass, and they were limited by the low resolution and the statistical quality of the temperature profiles. As shown above, significant progress on mass estimates has been made with \chandra\ and \xmm.  A  first  \chandra\  study  \cite{allen01} of  five hot clusters ($kT>5.5$~keV) derived a \MT\ relation slope of $1.51\pm0.27$, consistent with the self-similar model (see Fig.~\ref{fig:MT2500}). It confirmed the offset in normalisation. However, due to the
relatively small \chandra\ field of view, the \MT\ relation was established at $R_{2500}$, i.e., about $\sim 0.3\Rv$. More recently, the \MT\ relation was established  down to lower  density contrasts ($\delta=2500$ to $\delta=200$)  from a sample of ten nearby relaxed
galaxy clusters covering a wider temperature range, $[2-9]~\keV$ \cite{arnaud05}. The masses were derived from precise mass profiles
measured with \xmm\ at least down to $\delta=1000$ and extrapolated beyond that radius using the best fitting NFW model.  The $M_{2500}$--$T$ for hot clusters is perfectly consistent with the \chandra\ results. 
The logarithmic slope of the $M$--$T$ relation was well constrained. It is the same at all $\delta$, reflecting the self-similarity of the mass profiles. At $\delta=500$ (see Fig.~\ref{fig:MT500}), the slope of the relation for the sub-sample of hot clusters ($\kT>3.5~\keV$) is consistent with the standard self-similar expectation: $\alpha=1.49\pm0.15$. The relation steepens when the whole sample is considered, but the effect is fairly small: $\alpha=1.71\pm0.09$. The normalisation of the relation differs, at all density contrasts from the prediction of gravitation based models (by $\sim30\%$).  Models that take into account radiative cooling and galaxy feedback \cite{borgani04,kay04} are generally in better agreement with the data \cite[for full discussion]{arnaud05}.

\subsection{Evolution of cluster properties}
\label{sec:stat_evol}

The standard self-similar model makes strong predictions for the evolution of cluster properties. Distant clusters should have the same internal structure as nearby clusters, but they  should be denser, smaller and more luminous (Sec.~\ref{sec:statstandard}).  Physical properties derived from X--ray observations, as well as theoretical predictions, depend on the assumed cosmology. Fortunately,  we have now concordant constraints (Sec.~\ref{sec:cosmo}) on cosmological parameters, showing that we live in a flat low density Universe ($\Omm \sim 0.3, \Oml \sim 0.7$). This greatly simplifies  the issue of cluster formation and evolution.

The \rosat\  and \asca\  observations gave the first indication that the self-similarity does hold at $z>0$.  The universal emission measure  profile appears to extend to $z\sim0.8$, with a redshift scaling consistent with the expectation for a $\Lambda$CDM cosmology  \cite{arnaud02a}. A significant evolution in the normalisation of the \LxT\ relation was obtained, consistent with the self-similar model. This was also the case  in other studies made assuming this cosmology\footnote{Previous studies assumed a SCDM cosmology and found no evolution of the \LxT\ relation.  As discussed in \cite{arnaud02a}, this is an artifact due to the choice of a 'wrong' cosmology:  the distance and therefore the luminosity are underestimated}  \cite{reichart99,novicki02}, although more recently no significant evolution was detected  in a large sample of 79 clusters \cite{ota04}.  However, all these observations  were highly biased towards massive systems, mostly clusters discovered by the EMSS, and their statistical quality was poor.

%%%%%%%%%%%%%%%%%%%%
 \begin{figure}[t]
\begin{minipage}[c]{0.48\textwidth}
\centering
\includegraphics[width=\textwidth]{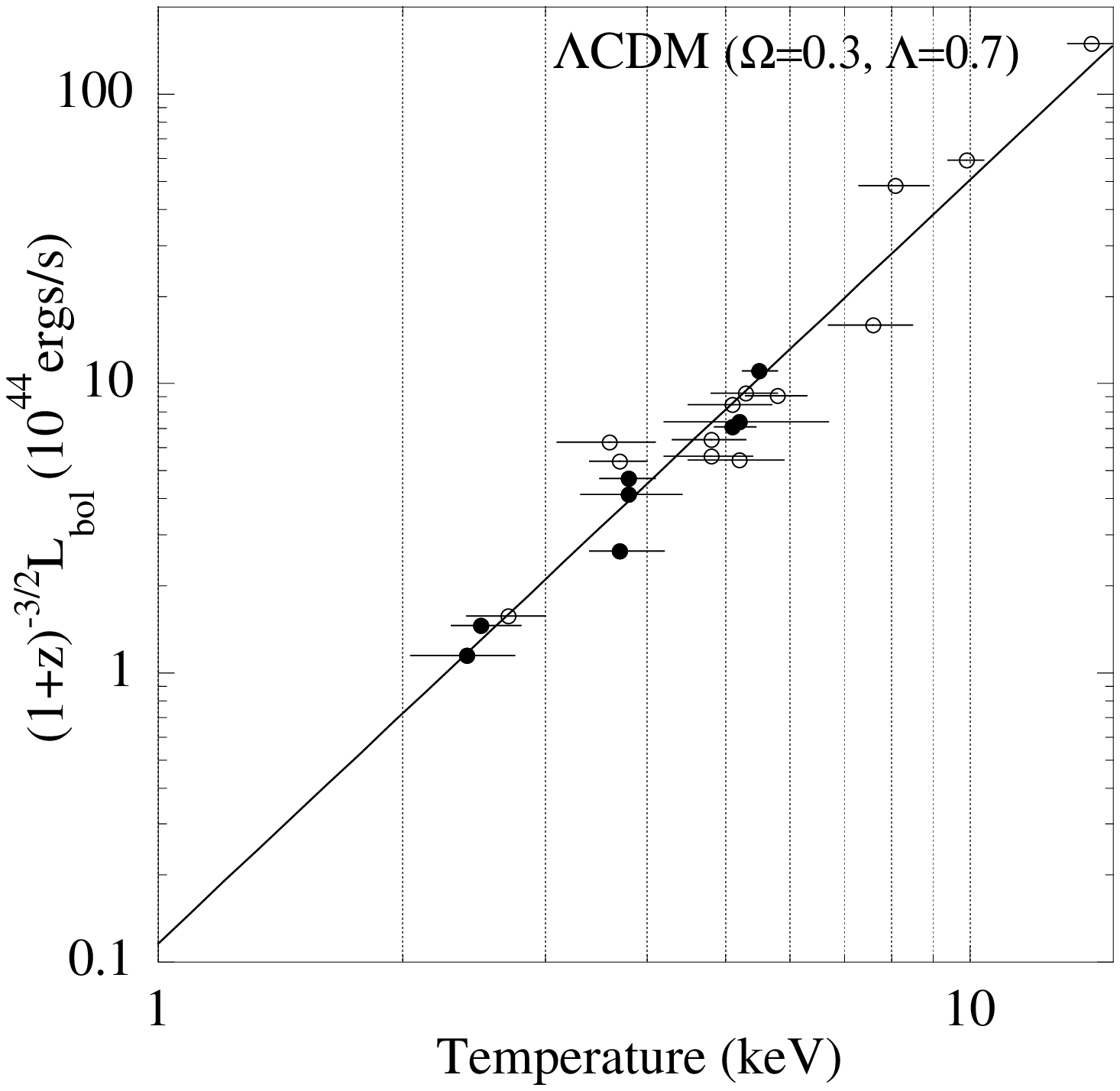}
 \end{minipage}
\hspace{+0.02\textwidth}
\begin{minipage}[c]{0.48\textwidth}
\centering
\includegraphics[width=\textwidth]{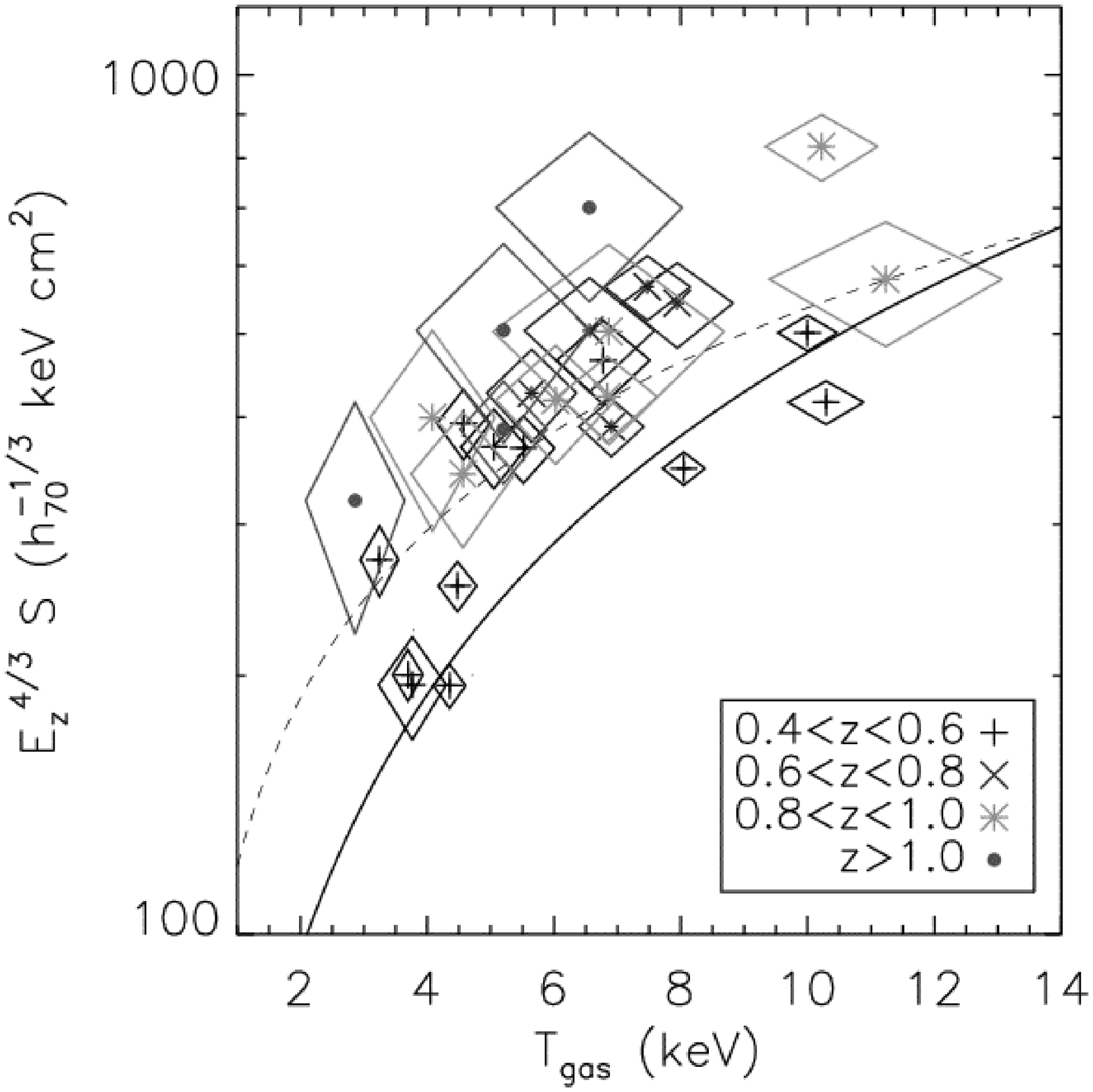}
 \end{minipage}
\begin{minipage}[t]{0.48\textwidth}
 \caption[ ]{Correlation between bolometric luminosity
of distant clusters and temperature.  Filled circles: \xmm\ data for
SHARC clusters \cite{lumb04}; Open circles: \chandra\ data \cite{vikhlinin02}.  When the luminosity
is scaled by $(1+z)^{3/2}$, the data points are consistent with the
local \LxT\ relation (solid line), indicating evolution.  Figure adapted from \cite{lumb04,vikhlinin02}}
 \label{fig:LxTz}
\end{minipage}
\hspace{+0.02\textwidth}
\begin{minipage}[t]{0.48\textwidth}
 \caption[ ]{Entropy measured at $0.1\Rv$ versus temperature for a sample of distant clusters observed with \chandra. Data for clusters in various $z$ bins are indicated by different symbols. The entropy is rescaled according to the expected evolution. Dashed line: local \ST\ relation as measured by \cite{ponman03}. Figure from \cite{ettori04}.}
 \label{fig:STz}
\end{minipage}
 \end{figure}
%%%%%%%%%%%%%%%%%%%%

With \xmm\ and \chandra, we can now make high quality studies of distant  clusters from X--ray samples assembled using \rosat\ observations (Sec.~\ref{sec:survey}).  These  new samples  cover a much wider   mass range than the EMSS survey. Recent \xmm\ and \chandra\   evolution studies do confirm that  clusters follow scaling laws up to high z \cite{vikhlinin02,lumb04,ettori04}.  The self-similarity of shape was also confirmed  on a test case at $z=0.6$ \cite{arnaud02b}. However, the amount   of evolution remains uncertain.  For instance, depending on the data considered and the analysis    procedure, the normalization the \LxT\ relation has been found to  evolve more than expected in the standard model, as expected or less than expected. The standard self-similar model predicts that the  normalization A(z) varies as h(z),   $A(z) \equiv h(z) \sim  (1+z)^{0.6-0.9}$ for the favored $\Lambda$CDM cosmology. The first  studies comparing  respectively \chandra\  \cite{vikhlinin02} and \xmm\  \cite{lumb04} data with the local relation  measured with \asca\  \cite{markevitch98a}   gave similar evolution factors:  $A(z) \equiv  (1+z)^{1.52\pm0.26}$ and  $A(z)\equiv(1+z)^{1.5\pm0.30}$ respectively.  This is  larger  than expected. A more recent study \cite{ettori04}, using a larger set of \chandra\  data and the same
local reference but a different procedure, gives  $A(z)\equiv(1+z)^{0.62\pm0.28}$,  perfectly consistent with the
expectation. In contrast, a smaller evolution   than expected was obtained when using the local relation measured
with \sax: $A(z) \equiv h(z) (1+z)^{-1.04\pm0.32}$.

The evolution of the other  scaling relations are still poorly  constrained.  The evolution of the \MT\ relation is consistent with the expectation \cite{ettori04}. There is some hints that the evolution of the \MgT\ relation  is smaller than expected \cite{vikhlinin02,ettori04}, while the \ST\ relation would be higher than expected \cite[see Fig.~\ref{fig:STz}]{ettori04}. However, the effects are not very significant.

This illustrates the difficulty in studying cluster evolution. High precision is required because the  evolution expected in the
'reference' standard model is small. The normalization  of the key   \LxT, \MT\ and \ST\ relations evolves as $h(z)$, $h(z)^{-1}$ and $h(z)^{-4/3}$ respectively, where $h(z)$ is the Hubble constant.  The  evolution factor, $h(z)$, varies by $30\%$ between $z=0$ and $z=0.5$  ($56\%$  at $z=0.8$). To distinguish between various models, statistical and systematic errors have to be well below
these figures.  This requires the use of large unbiased cluster samples, covering  a wide mass range,  in both the local Universe and at high redshift.  Biased results on the  evolution could be obtained, for instance, by comparing the \LxT\ relations at various $z$ with non representative sub-samples of cooling flow clusters (or by using not fully consistent treatment of the cooling center). Ideally, one also wants to compare data obtained with the same instrument, in order to minimize systematic uncertainties. For instance, a calibration error of $10\%$ in the kT measurements is equivalent to a $30\%$ systematic error on the luminosity ($L_{\rm bol}  \propto T^{3}$). Such an error could introduce a bias equivalent to the expected evolution.   Ongoing \xmm\ and \chandra\ large projects,  aiming at studying the properties of large unbiased distant and local samples, will fulfill the above requirements.  Significant progress on the evolution of cluster properties are expected from these projects.

\section{Constraining cosmological parameters with X-ray observations of clusters}
\label{sec:cosmo}

Several independent methods can be used to constrain the cosmological parameters from cluster X-ray observations.  This includes the baryon fraction in clusters, the cluster abundance and its evolution, and the cluster spatial distribution. All methods rely in principle or in practice on specific assumptions on the scaling and structural properties of clusters.  

\subsection{The baryon fraction}
In the simplest model of cluster formation, the contents of clusters is a fair sample of the Universe as a whole (Sec.~\ref{sec:statstandard}). The baryon mass fraction in clusters is then $\fb = \Omb/\Omm$, where $\Omb$ and $\Omm$ are the mean baryon density and the total matter density  of the Universe.   $\fb$ is the sum of the gas and galaxy mass fractions: $\fb = \fgas +\fgal$.   Combined with the $\Omb$ value estimated from Big Bang nucleo-synthesis or CMB measurements, $\fb$ in clusters can be used to measure $\Omm$  \cite{white93}. The method requires an independent knowledge of  the Hubble constant $h$, which  enters in the determination of $\Omb$ ($\Omb \propto h^{-2}$), of  $\fgas$ ($\fgas\propto h^{-3/2}$) and of $\fgal$ ($\fgal \propto h^{-1}$). Note that the  baryonic mass in rich clusters is dominated by the X-ray gas. The gas mass fraction alone sets an upper limit on $\Omm$. 

Cluster sample studies that rely on  measurements of both $\fgas$ and $\fgal$ are rare \cite{lin03}. $\Omm$  is most often constrained from $\fgas$ only, assuming a constant $\fgal/\fgas$ ratio,  taken from other cluster studies \cite{ettori03,allen04}.  A further difficulty is that $\fgas$ increases with the integration radius (Sec.~\ref{sec:ng}) and  numerical simulations indicate that $\fgas$ within the virial radius is slightly smaller than  the Universe's  value \cite{eke98}. Observed  values must be corrected for these effects.  The correction is about $20\%$ for $\fb$ values estimated within $1/3$ of the virial radius \cite{allen04}.   Corrections factors are deduced from adiabatic numerical simulations \cite{lin03,allen04} and may not be perfectly adequate since  these simulations do  not include galaxy formation and fail to account for the observed ICM properties (Sec.\ref{sec:statgas}).  One also observes a significant increase of  $\fb$  with system mass, mainly resulting from the increase of $\fgas$ \cite{mohr99,lin03}. This increase of $\fgas$ is likely due to non-gravitational processes, less important in high mass systems (Sec.\ref{sec:statgas}). Therefore, most studies are restricted to massive clusters to minimize systematic errors. To firmly establish which cluster populations are fair samples of the Universe, the variation of $\fgas$ with system mass must be fully understood, which remains to be done.

 Present $\fb$ data provides a tight constraint on $\Omm$. All recent studies favor a low $\Omm$ Universe and are in excellent agreement, e.g.  $\Omm = 0.37 \pm 0.08$  from \sax\ data \cite{ettori03},  $\Omm = 0.28 \pm 0.03$  from \rosat/\asca\ data \cite{lin03} and 
$\Omm = 0.30 \pm 0.04$ from \chandra\ data \cite{allen02}.  The variation of $\fb$ with mass and radius, mentioned above,  are the main source of systematic uncertainties for high precision cosmology.

%%%%%%%%%%%%%%%%%%%%%%%%%%%%%%%%%%%%%
\begin{figure}[t]
\begin{minipage}[c]{0.48\textwidth}
\centering 
\includegraphics[height=\textwidth, angle=-90]{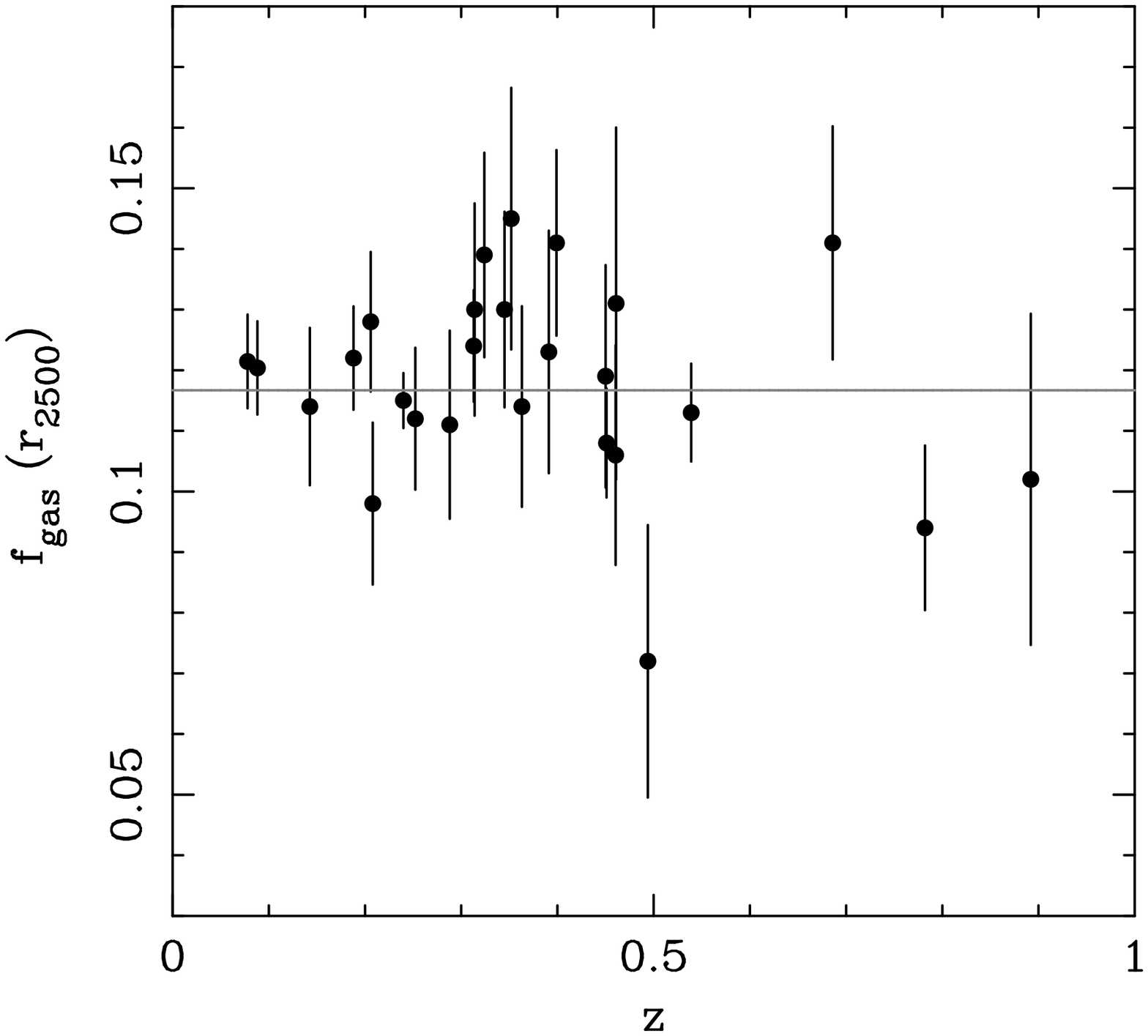}     
 \end{minipage}
\hspace{+0.02\textwidth}
\begin{minipage}[c]{0.48\textwidth}
\centering \includegraphics[width=\textwidth]{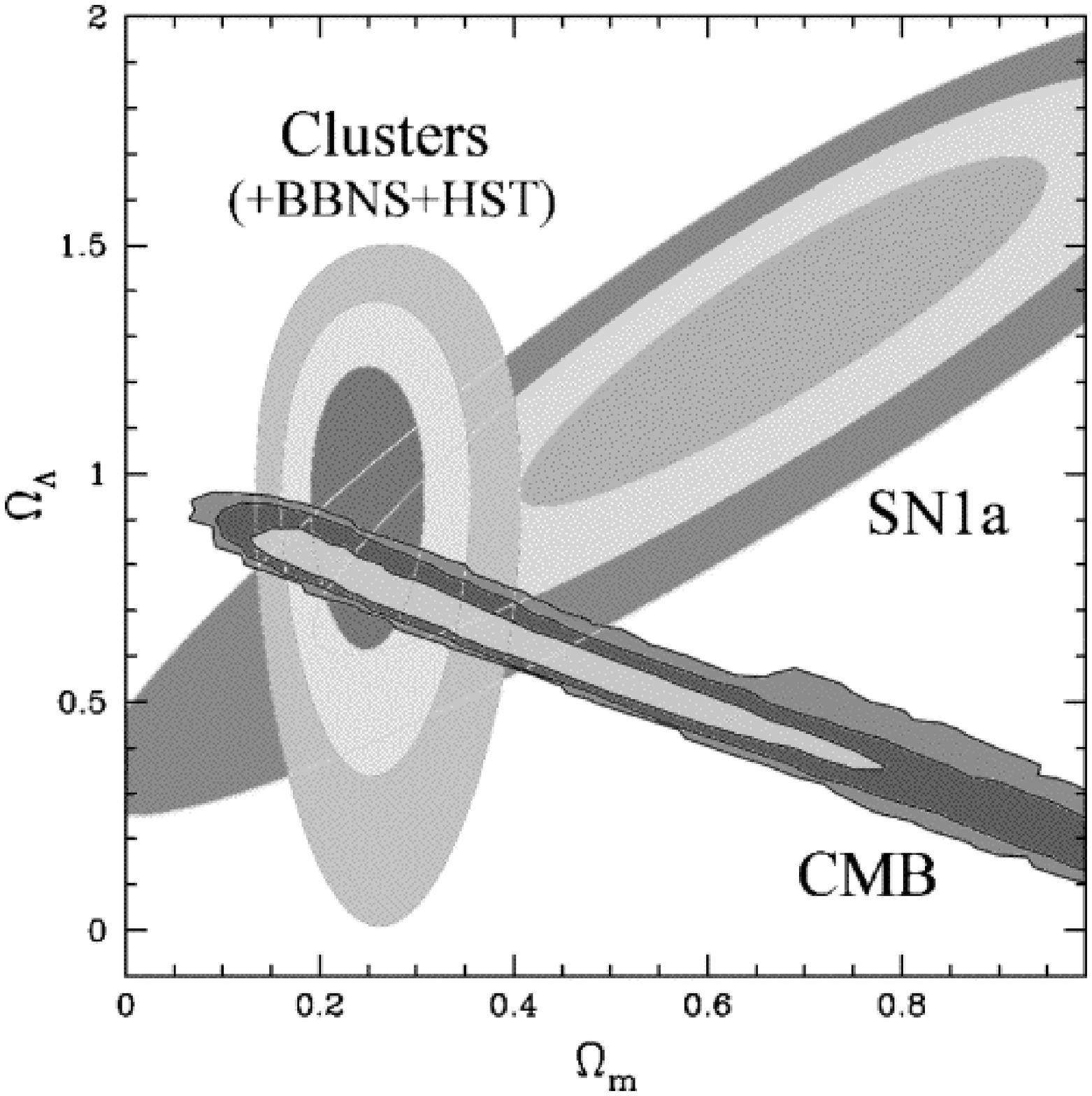}     
\end{minipage}
\begin{minipage}[t]{0.48\textwidth}
 \caption[ ]{The X-ray gas mass fraction ($1\sigma$ errors)  versus $z$ for a $\Lambda$CDM cosmology with  $\Omm=0.25,\Oml=0.96$.   In this 'best fit' cosmology $\fgas$ is constant.  Figure from \cite{allen04}.}
 \label{fig:fgasz}
\end{minipage}
\hspace{+0.02\textwidth}
\begin{minipage}[t]{0.48\textwidth}
 \caption[ ]{The $1$, $2$ and $3\sigma$ confidence constraints in the $\Omm,\Oml$ plane  obtained from the analysis of the cluster $\fgas$ data. Also shown are the independent results obtained from CMB and SNI data. Figure from \cite{allen04} }
 \label{fig:fgascosmo}
\end{minipage}
 \end{figure}
%%%%%%%%%%%%%%%%%%%%%%%%%%%%%%%%%%%%%

\subsection{The gas fraction as distance indicator}
Assuming again that $\fgas$ is universal and thus constant with $z$,  $\fgas(z)$ can be used  as distance indicator  \cite{sasaki96}. The values derived from observations depend on the angular distance as $\fgas \propto (D_{\rm A}(z))^{3/2}$. It  will be constant only for the correct underlying cosmology (Fig.~\ref{fig:fgasz})  \footnote{More generally, non evolving cluster properties can be used as distance indicators. This includes the gas mass fraction, but also the isophotal radius \cite{mohr00}, or properties corrected for evolution, like the scaled emission measure profiles \cite{arnaud02a}}. Because the variation of  $D_{\rm A}$ with redshift is controlled by the expansion rate of the Universe, $H(z)$, $\fgas(z)$  thus provides constraints  on $\Omm$ and $\Oml$ for a $\Lambda$ CDM cosmology, (or, more generally, on the Dark Energy present density and  equation of state). Note, however, that  there is a large degeneracy between  the  constraints on $\Omm$ and $\Oml$.  

Because it is  a pure geometrical test, the method is straightforward. However, it requires high quality data (precise mass estimates) on a sample of clusters at various redshifts and it is not free of possible systematical errors. $\fgas$ varies with radius and mass at $z=0$. Therefore, to compare the gas fraction at various $z$, one estimates $\fgas$ at the same fraction of the virial radius and for clusters of similar mass (preferentially massive clusters) \cite{ettori03,allen02}. This is sufficient to avoid biases only if cluster mass profiles are similar at all $z$ and if  $\fgas$  at a given mass does not evolve with $z$. These assumptions must be verified from detailed observations at high $z$ (in particular observations of the structural properties of clusters), compared with numerical simulations. 

The principle of this  test is similar to that of  the cosmological test  using SNI as  'standard' candles \cite[and references therein]{riess98,perlmutter99,riess04}.  Therefore, it  probes the same domain of cosmological parameters as SNI experiments. However,  there are also  systematic uncertainties inherent to the use of SNI (dust extinction, possible evolution of the luminosity).  Much safer constraints can thus be obtained by cross-checking the results obtained with SNI and with the gas fraction.   The constraints provided by $\fgas(z)$ (and SNI)  complement the observations of the CMB anisotropies \cite[for WMAP results]{spergel03}, which do not probe the same domain of cosmological parameters. 

The recent results obtained from \chandra\ observations are very encouraging \cite{allen04} (see also \cite{allen02,ettori03}). The whole information provided by $\fgas$ was used: absolute value and apparent evolution with $z$. For a $\Lambda$CDM cosmology, this yields: $\Omm = 0.245 \pm 0.04$ and $\Oml = 0.96 \pm 0.2$ ($68\%$ confidence level), with  $\Omb h^2 = 0.0214\pm0.02$ and  $h = 0.72\pm0.08$ as only priors (Fig.~\ref{fig:fgascosmo}). $\Oml$ is positive at the $3\sigma$ level. The complementarity with CMB observations  is obvious in Fig.~\ref{fig:fgascosmo}.   Note also the complementarity of $\fgas$ and SNI measurements:  the ($\Omm$, $\Oml$) degeneracy intrinsic to pure distance measurement techniques (SNI or $\fgas(z)$ observations) has been  broken because the local $\fgas$ value provides independent constraints on $\Omm$. On the other hand, the constraints from $\fgas(z)$ on $\Oml$  at a given $\Omm$ value are still less stringent than those provided by SNIs (Fig.~\ref{fig:fgascosmo}). This can be improved using larger samples or more precise estimates of $\fgas$. 
Finally the consistency between$\fgas$, SNI and CMB constraints is reassuring: their respective confidence contours in the   $\Omm$,  $\Oml$ plane intersect in a common domain. 

\subsection{Cosmological parameters from  cluster abundance and evolution}
\label{sec:cosmoab}

An independent and powerful method to constrain the cosmological parameters is to measure the growth rate of linear density perturbations, as reflected in the evolution of the galaxy cluster mass distribution function, $n(M,z)$, i.e.,  the comoving number density of clusters of mass $M$ at redshift $z$.  This cosmological test is not as direct as the previous one, since the mass function also depends on the spectrum of the initial density fluctuations. 

\begin{figure}[t]
\begin{minipage}[b]{0.50\textwidth}
\centering 
\includegraphics[width=\textwidth]{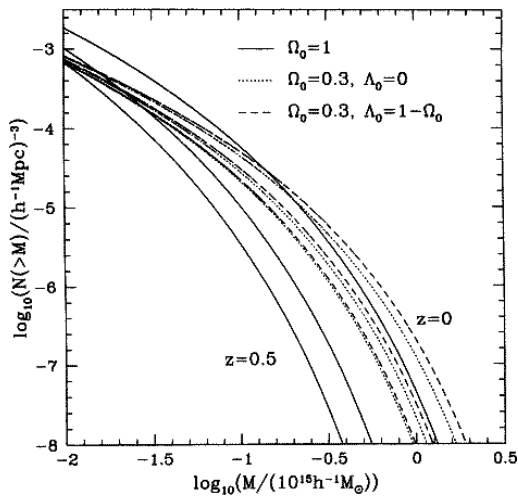}     
 \end{minipage}
\hspace{+0.02\textwidth}
\begin{minipage}[b]{0.48\textwidth}
 \caption[ ]{Predicted evolution of the cluster mass function. The comoving number density of clusters with mass larger than M, is plotted as a function of M, for three cosmological models:  $\Omm=1.$ (solid lines), an open model with $\Omm=0.3$  (dotted lines) and a flat model with $\Omm=0.3,\Oml=0.7$ (dashed lines).  In each case, the predictions for $z=0.5,0.33,0$ are shown.  Figure from \cite{eke96}.  }
\label{fig:XMF}
\end{minipage}
\end{figure}

From large N-body simulations \cite{evrard02} we can now accurately trace the formation of dark matter halos.  Such simulations have been used to establish universal parametric formulae for the cluster mass function in various cosmological models  \cite{sheth01,jenkins01}, improving over the original formula provided by \cite{press74}.  The mass function depends mostly on $\Omm$ and $\sigma_{8}$ (i.e. the normalisation of the fluctuation power spectrum $P(k)$), but also on $h$,  $\Omb$, $\Oml$ and $n$ (the shape of $P(k)$). The {\it evolution} of the mass function, specially of the number of high mass clusters, is extremely sensitive to $\Omm$ \cite{oukbir92,eke96}, as illustrated Fig.~\ref{fig:XMF} (see also Fig.6 in \cite{borgani01a}).  It was first proposed in \cite{perrenod80} to use the evolution of X--ray cluster abundance to constrain $\Omm$. 

%%%%%%%%%%%%%%%%%%%%%%%%%%%%%%%%%%%%%
%%%%%%%%%%%%%%%%%%%%%%%%%%%%%%%%%%%%%

In practice, this method requires to build large, well controlled, samples of local and distant clusters. Furthermore, the mass can never be directly determined and other(s) observable(s) have to be used as surrogates.  I review below the available data and current constraints on the cosmological parameters.  For more detailed reviews, the reader may refer to \cite{rosati02,henry03}.

\subsubsection{The X-ray cluster surveys}
\label{sec:survey}

Beyond the galactic plane, the X-ray sky is essentially populated with AGNs (about $90\%$ of the sources at the typical ROSAT detection limit)  and galaxy clusters. There is a diffuse X-ray background emission, due mostly  to the local galactic emission (in the soft energy band  used normally to detect clusters).

Detecting clusters in X-ray has several advantages \cite{rosati02}. First, the detection of X-rays is an unambiguous signature of the presence of a true potential well.  Second, clusters are high contrast objects in the X-ray sky (the X-ray flux depends on the square of the density) and can be distinguished from other sources (AGNs) by their spatial extent (since ROSAT pointed observations). By contrast, the  field galaxy population sets a serious problem to optical surveys, as it rapidly overwhelms the galaxy overdensity associated with clusters as $z$ increases. X-ray surveys are also less subject to projection effects. Third, quantitative selection criteria can be defined: one searches for extended sources above a given flux limit. X-ray cluster samples are thus flux-limited samples.  This allows for precise estimates of the survey volume and thus of space density of clusters. 

%%%%%%%%%%%%%%%%%%%%%%%%%%%%%%%%%%%%%
\begin{figure}[t]
\begin{minipage}[c]{0.48\textwidth}
\centering 
\includegraphics[width=\textwidth]{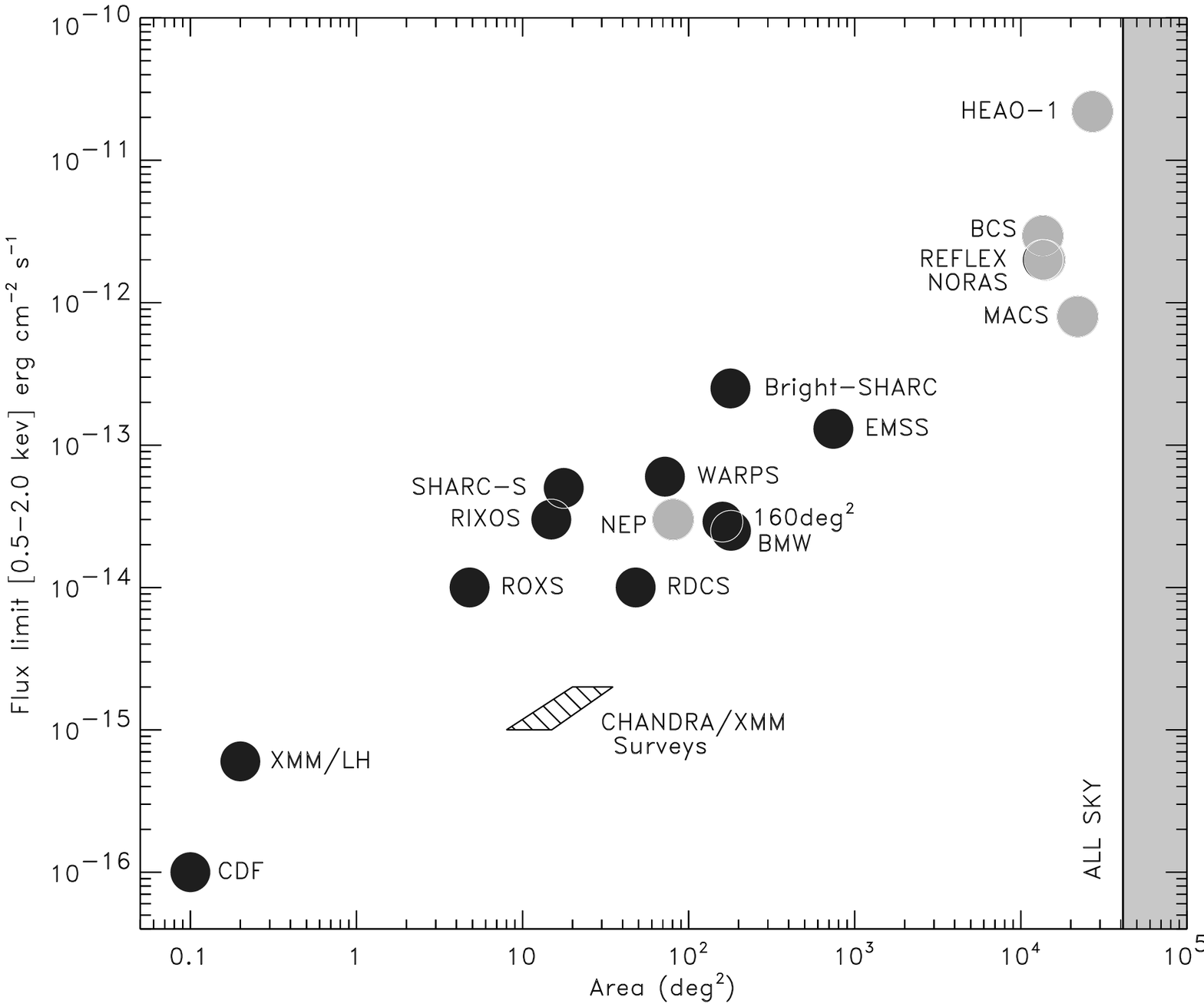}     % includes figure foo.eps
 \end{minipage}
\hspace{+0.02\textwidth}
\begin{minipage}[c]{0.48\textwidth}
\centering 
\includegraphics[width=\textwidth]{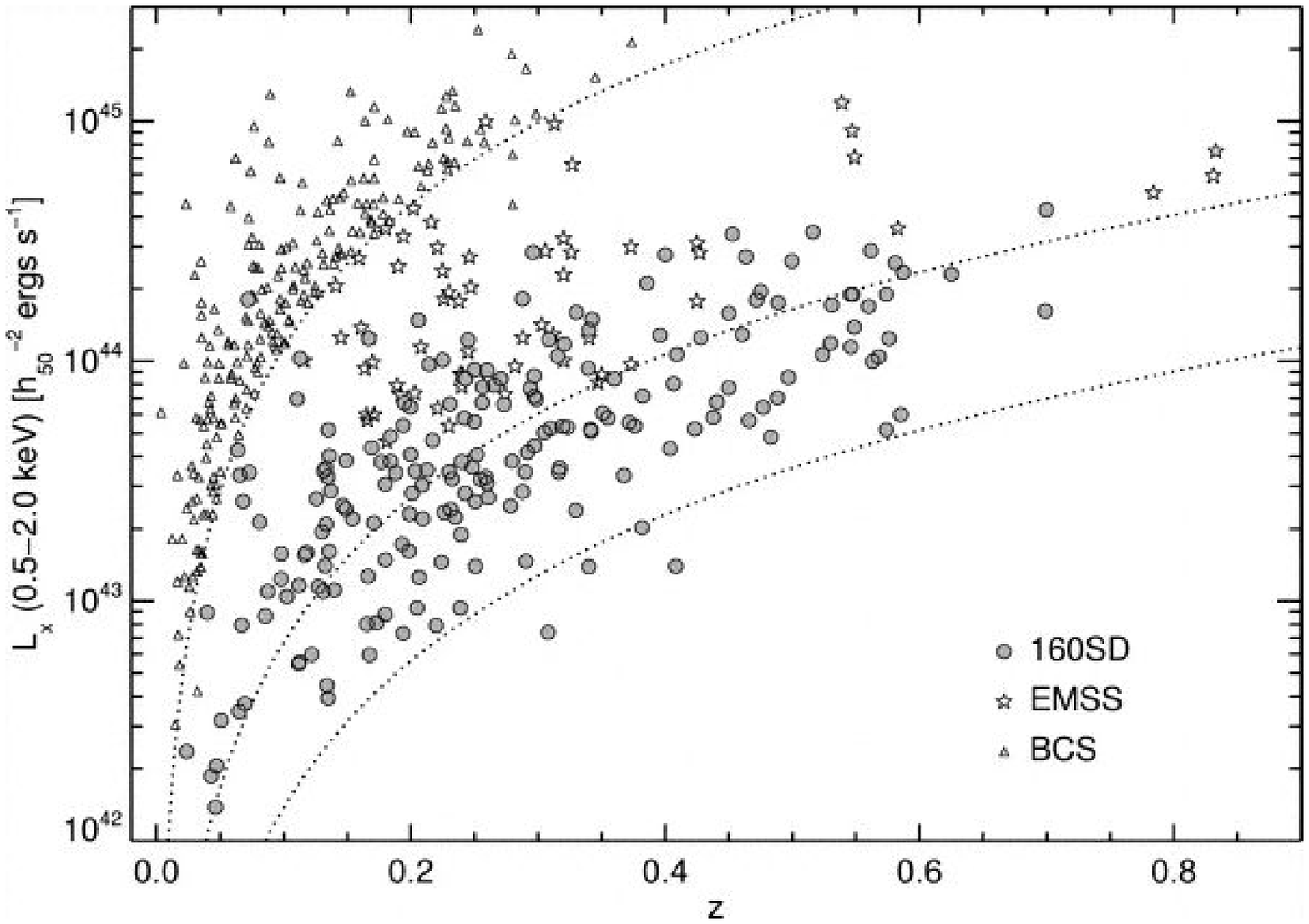}     % includes figure foo.eps
\end{minipage}
\begin{minipage}[t]{0.48\textwidth}
 \caption[ ]{Solid angles and flux limits of various X-ray cluster surveys (references in text and in \cite{rosati02}). Dark filled circles: serendipitous surveys constructed from a collection of pointed observations. Light shaded circles: surveys covering contiguous areas. Figure from \cite{rosati02}.}
 \label{fig:survey}
\end{minipage}
\hspace{+0.02\textwidth}
\begin{minipage}[t]{0.48\textwidth}
 \caption[ ]{ X--ray luminosity and redshift distribution of the 160SD \cite{mullis03}, EMSS \cite{henry04} and BCS \cite{ebeling00} cluster samples.  The dotted curves (left to right) are indicative flux limit of the surveys:  $2.7 10^{-12}$,  $1.5 10^{-13}$, $3. 10^{-14}~{\rm  ergs/cm^{2}/s}$  in the $(0.5-2)~\keV$ energy band.  Figure from \cite{mullis04} }
 \label{fig:Lxz}
\end{minipage}
 \end{figure}
%%%%%%%%%%%%%%%%%%%%%%%%%%%%%%%%%%%%%

However, there are several difficulties in constructing well controlled samples of X-ray clusters. In particular, it  requires intensive optical follow-up to confirm the identification and to measure redshifts.  Another  critical point is to precisely estimate the 'selection function', that is, how the survey samples the true population of clusters. This includes the computation of the sky coverage $\Omega(S_{\rm X})$: the effective area covered by the survey as a function of X-ray flux. This is not trivial because the exposure time, background, PSF, efficiency etc.. vary across the telescope field of view and from pointing to pointing and so does the detection probability. The detection probability also depends on the cluster morphology. In view of these difficulties, one usually relies on Monte-Carlo simulations to study the survey selection function  \cite{vikhlinin98,adami00}. 

Several cluster surveys have been conducted since the early HEAO-1 survey (see \cite{rosati02} for a complete list and descriptions of all X-ray cluster surveys). The RASS (Rosat All Sky Survey) was the first (and still unique) X-ray imaging Survey of the whole sky. Several cluster catalogs were or are still being constructed from the RASS. Published catalogs of X-ray selected clusters include the BCS \cite{ebeling00}, the NORAS \cite{bohringer00} and the REFLEX catalogs \cite{bohringer04}. More than $1000$ clusters have been discovered,  mostly at low redshift.  For instance, the REFLEX cluster catalog contains $\sim 450$ clusters  out to $z\sim0.3$.  Serendipitous search of clusters is performed using archival pointed observations:  first with the EINSTEIN satellite (EMSS survey \cite{gioia90}), then with ROSAT, and now with XMM \cite{romer01} and Chandra.  The EMSS cluster catalog \cite[and references therein]{henry04} is still largely used for cosmological studies  and several ROSAT serendipitous cluster catalogs are now available in the literature:  the Bright SHARC \cite{romer00}, the WARPS-I \cite{perlman02}, the Southern SHARC \cite{burke03} and the 160SD \cite{mullis03} catalogs. The sky coverage and flux limit  of various surveys are given in Fig.~\ref{fig:survey}.  Compared to the RASS, ROSAT serendipitous surveys go  much deeper and thus cover  a larger $z$ range (up to $z\sim 1.3$). However, their sky coverage is much smaller, so that serendipitous cluster catalogs include  from $12$ to $100$ clusters. 

\begin{figure}[t]
\begin{minipage}[c]{0.48\textwidth}
\centering 
\includegraphics[width=\textwidth]{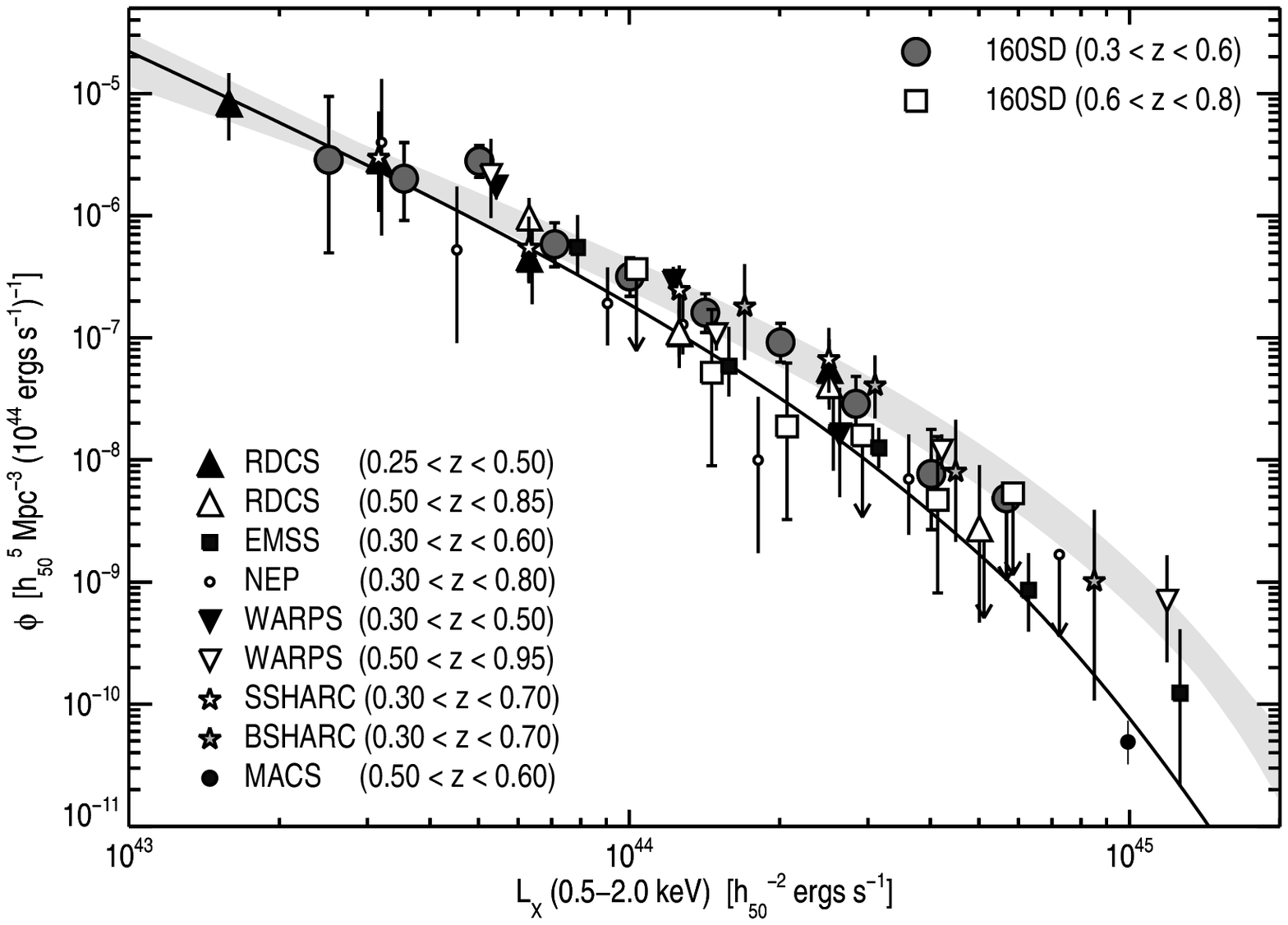}     % includes figure foo.eps
 \end{minipage}
\hspace{+0.02\textwidth}
\begin{minipage}[c]{0.48\textwidth}
\centering 
\includegraphics[width=\textwidth]{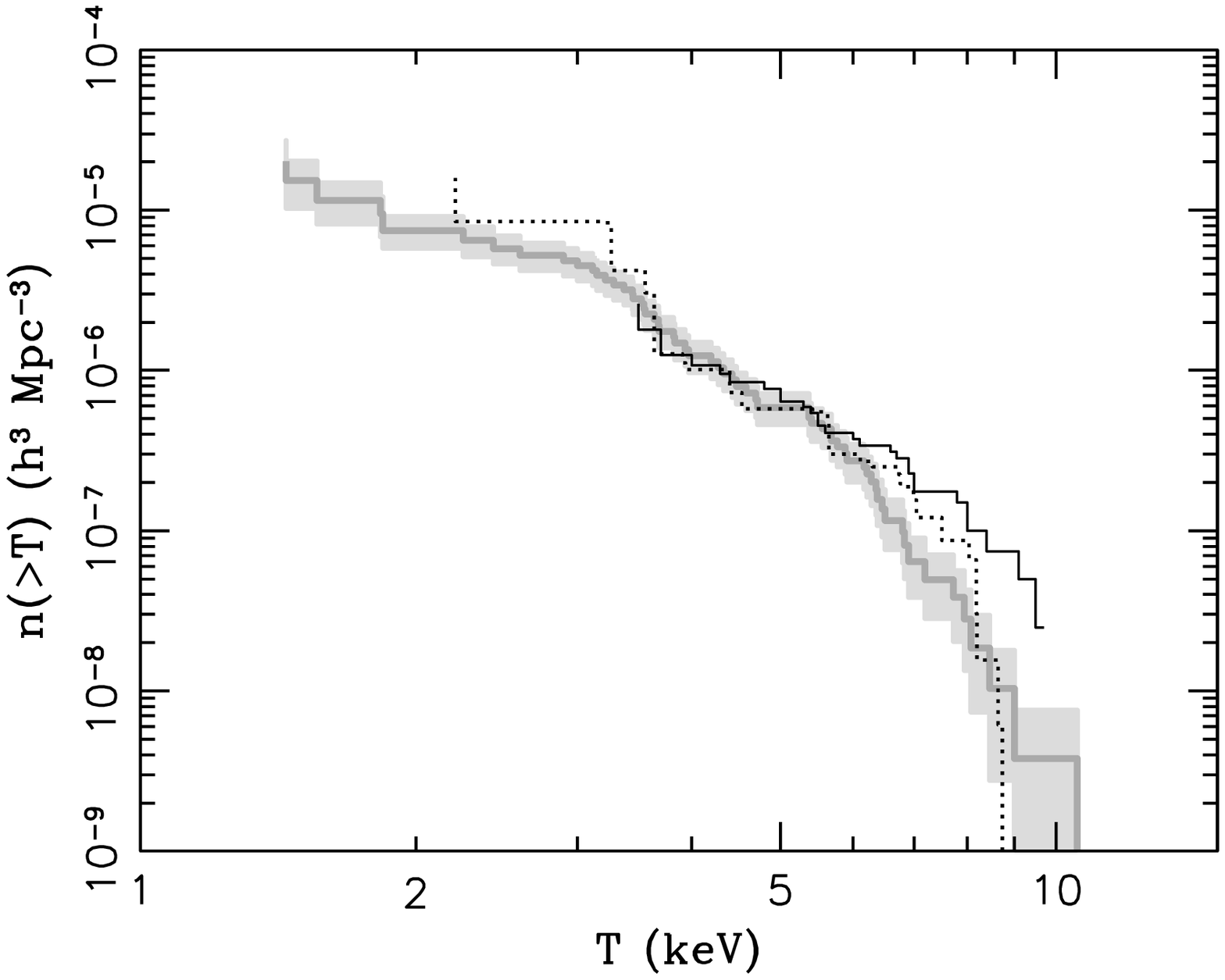}     % includes figure foo.eps
\end{minipage}
\begin{minipage}[t]{0.48\textwidth}
 \caption[ ]{Compilation of high-redshift XLFs as measured by eight independent X-ray flux--limited surveys (Einstein de Sitter universe), compared with the local XLF (shaded area). Figure from \cite{mullis04}.}
 \label{fig:XLF}
\end{minipage}
\hspace{+0.02\textwidth}
\begin{minipage}[t]{0.48\textwidth}
 \caption[ ]{Cumulative temperature function (grey solid line with its $68\%$ error band) established  from the HIFLUCGS cluster sample \cite{ikebe02}. Previous results from  \cite{markevitch98a} and \cite{henry00} are shown by the solid and dotted black lines. Figure from  \cite{ikebe02}.}
 \label{fig:XTF}
\end{minipage}
 \end{figure}

The mass and redshift coverages depend on the survey area and flux limit in a complex way.   The lower luminosity limit (and thus mass limit)  increases with $z$ (Fig.~\ref{fig:Lxz}), since X--ray surveys are flux-limited, at variance with SZ surveys.  The abundance of clusters rapidly decreases with luminosity (Fig.~\ref{fig:XLF}). To detect massive (and thus rare) clusters, the sky coverage must be wide. Going deeper  (decreasing the flux limit of a survey) extends the low mass coverage. However, this does not necessarily augment  the $z$ coverage.  Indeed, the survey area may  become  too small to detect clusters above the luminosity limit at high $z$. These points are illustrated Fig.~\ref{fig:Lxz}. 

Cluster catalogs derived from X--ray surveys contain only  basic information: X--ray luminosity and redshift. To measure the temperature (and possibly the mass) X-ray follow-up is needed, usually with the next generation satellite(s).  This may change with XMM serendipitous surveys, where we expect that a fraction of the detected clusters will be bright enough to allow for the measurement of  temperature \cite{romer01}.  

%%%%%%%%%%%%%%%%%%%%%%%%%%%%%%%%%%%%%
%%%%%%%%%%%%%%%%%%%%%%%%%%%%%%%%%%%%%

\subsubsection{Measures of cluster space density}
 
The X-ray Luminosity Function (XLF) in various $z$ bins, is a direct  product of X-ray cluster surveys. Estimating the selection function in terms of luminosity is simple: one has $\Omega(L_{\rm X},z)\equiv \Omega(S_{\rm X})$ where $S_{\rm X} = L_{\rm X}/D_{\rm l}(z)^{2}$.   
There is an excellent agreement between the local XLFs derived from various surveys \cite{rosati02}, and  between the various XLFs at high redshifts derived from \rosat\ Surveys \cite{mullis04}, as shown Fig.~\ref{fig:XLF}.  A significant evolution of the XLF is found only at the bright end ($L_{\rm X} \gtrsim 5~10^{44} ergs/s$) and above $z\sim 0.5$ \cite{mullis04}.  

The drawback of using the XLF for cosmology  is that the luminosity is very sensitive to the detailed gas properties: density distribution in the core, thermodynamic history, dynamical state. The XLF is thus not easily related to the mass function. The most common method is to combine the empirical \LxT\ and the \MT\ relations. The scatter of the \LxT\ relation and the uncertainty on its evolution are the major sources of systematic uncertainties. 

In principle, the temperature is  more directly related to the mass than the luminosity, and the X-ray temperature function (XTF) is a better substitute for the mass function.  However, because the XTF is derived from cluster samples selected according to their flux,  a knowledge of the \LxT\ relation (and its scatter) is still required to compute the temperature selection function from $\Omega(S_{\rm X})$. Furthermore,  the XTF  has  generally a lower statistical quality and mass coverage than the XLF, because the temperature is only measured for sub-samples of available  X-ray cluster catalogs (specially at low mass and high z).  

The agreement among various local XTFs  is less good than for the local XLFs. This is illustrated in Fig.~\ref{fig:XTF}. The XTF recently derived \cite{ikebe02} from the HIFLUGCS cluster sample \cite{reiprich02}, which contains the 63 brightest clusters from the RASS,   is compared to earlier XTFs \cite{markevitch98a} (30 clusters) and \cite{henry00} (25 clusters).  The XTFs agree within $20\%$ in the $[3-6]~\keV$ range but differ by as much as a factor $3$ above $\sim 6-7~\keV$ (see also discussion in  \cite{blanchard00} and their Fig 4 and 5).  The only presently available  XTF  at high $z$ (Fig.~\ref{fig:XTFz}) is derived from ASCA follow-up of the EMSS Survey \cite{henry04}.

The HIFLUGCS cluster sample  is currently the largest local cluster sample that incorporates temperature measurements and also imagery data (from ROSAT pointed observations).
Using these data, the first (and still unique) cluster mass function, XMF,  for X-ray selected clusters  was established \cite{reiprich02}.   It can be directly compared to the theoretical predictions, but this test is not free of possible systematic errors. The masses had to be estimated using a simple isothermal $\beta$--model for the gas distribution (and the HE equation).  Moreover,  the determination of the selection function is not straightforward: the \LxM\  relation (and its scatter)  must be known to deduce the mass selection function from $\Omega(S_{\rm X})$. This relation was estimated using an extended sample of $106$ clusters.  

Constraining cosmological parameters from the XTF or the XLF requires a good calibration of the \MT\ and/or \LxM\  relations, both at low and high $z$, This is difficult to achieve, because it requires precise total mass measurements. To avoid this problem, it was recently proposed to use the baryon  mass $M_{\rm b}$, which is easy to measure, as a proxy of the total mass \cite{vikhlinin03,voevodkin04}.  The Baryon mass function BMF is directly related to the theoretical mass function under the assumption that the baryon fraction in clusters  is universal and equal to $\Omb/\Omm$.  In practice, the small variation with mass is taken into account.  The $L_{\rm X}$--$M_{\rm b}$ and its evolution must still be known to compute the selection function in terms of  $M_{\rm b}$, but this is relatively easily measured. The local  BMF, derived from a sub-sample of 52 clusters from the HIFLUCGS sample, is plotted in Fig.~\ref{fig:BMF}, together with the high redshift BMF derived from \chandra\ follow-up of bright 160SD clusters \cite{vikhlinin03,voevodkin04}.    

%%%%%%%%%%%%%%%%%%%%%%%%%%%%%%%%%%%%%
\begin{figure}[t]
\begin{minipage}[c]{0.48\textwidth}
\centering 
\includegraphics[width=\textwidth]{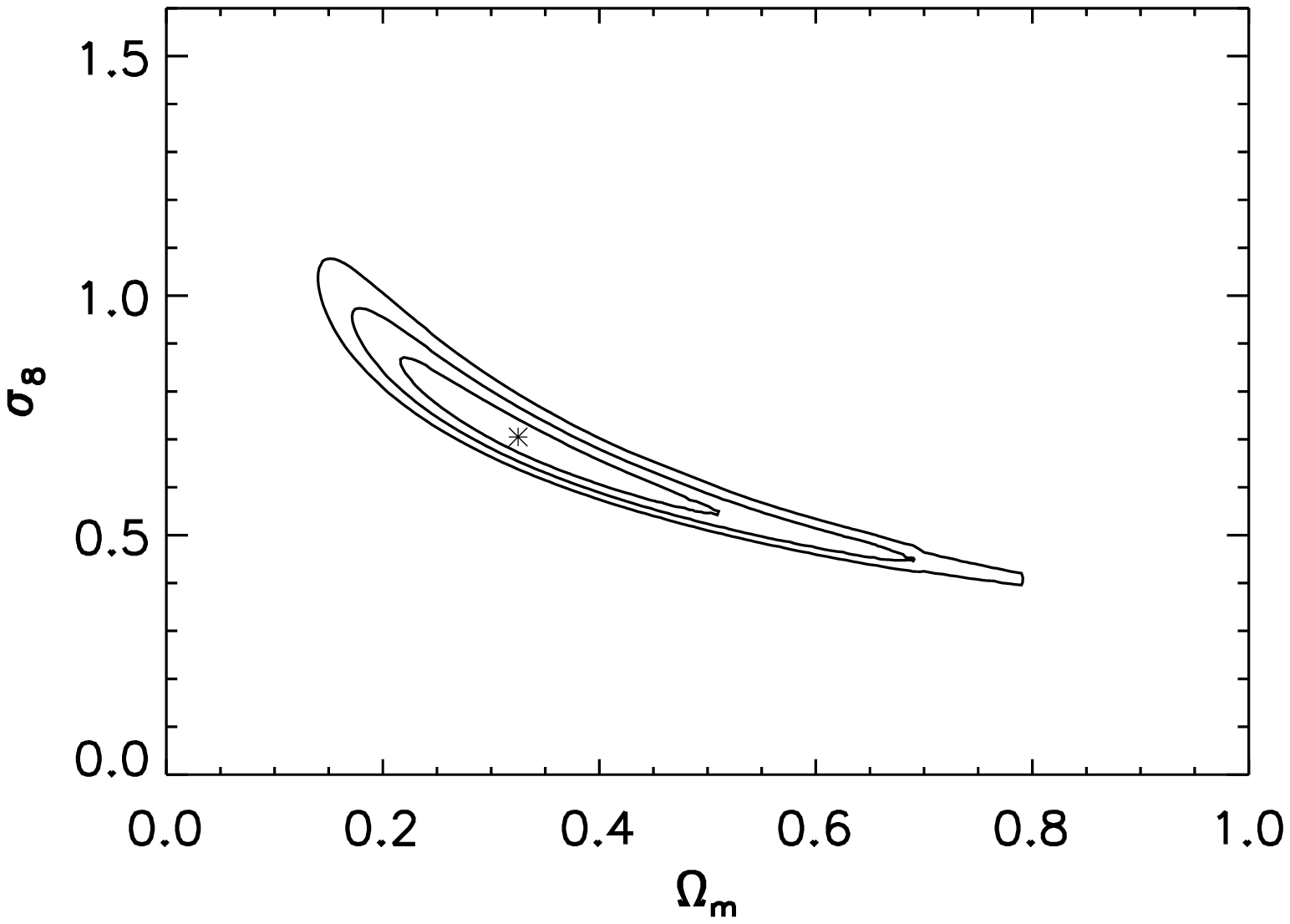}     
 \end{minipage}
\hspace{+0.02\textwidth}
\begin{minipage}[c]{0.48\textwidth}
\centering 
\includegraphics[width=\textwidth]{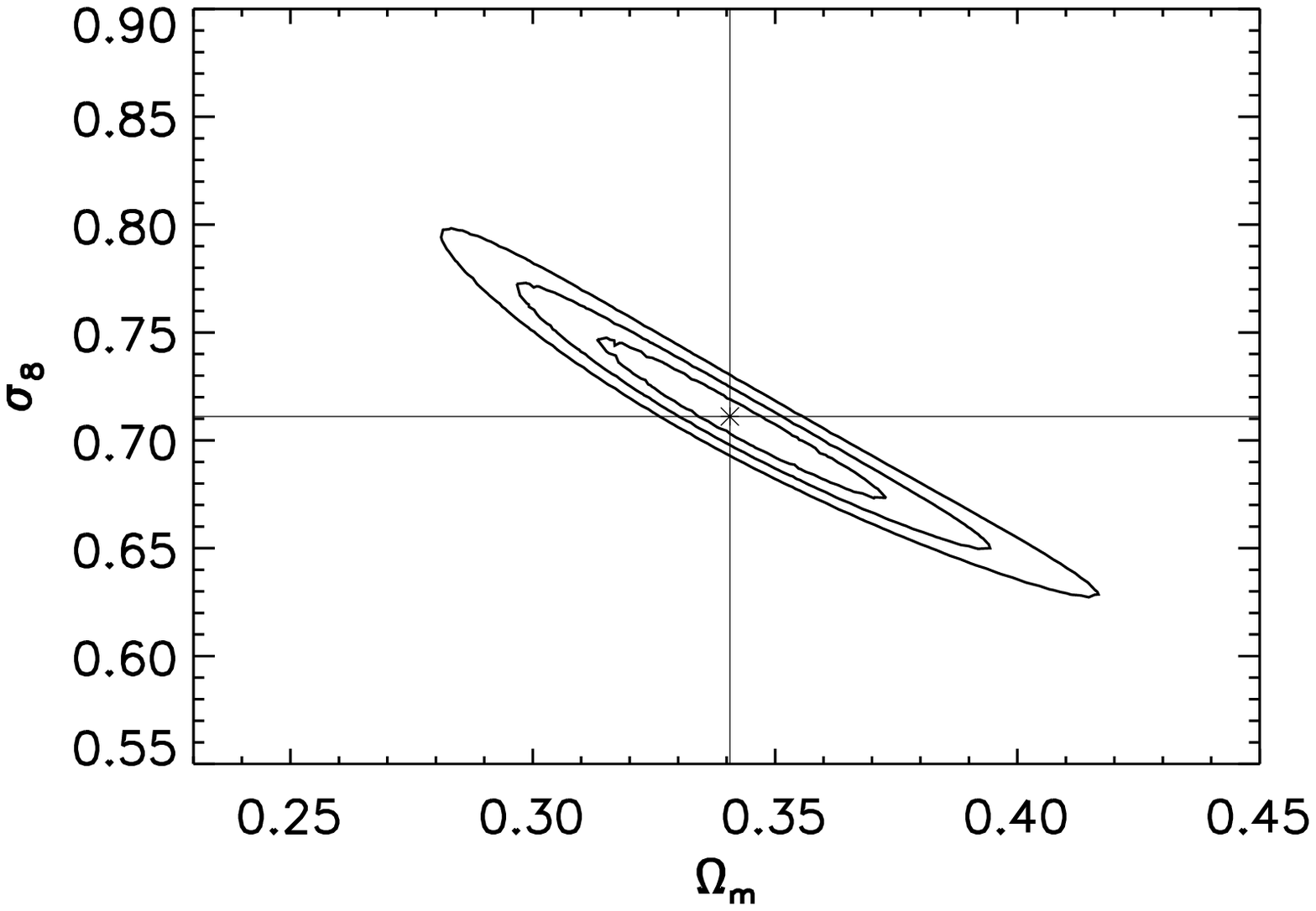}     
\end{minipage}
\begin{minipage}[t]{0.48\textwidth}
 \caption[ ]{Likelihood contours ($1$--$3\sigma$ levels) in the $\sigma_{8}$--$\Omm$ plane obtained from the REFLEX cluster abundances. Note the degeneracy between $\sigma_{8}$ and $\Omm$. Figure from \cite{schucker03}}
 \label{fig:XLFcosmo}
\end{minipage}
\hspace{+0.02\textwidth}
\begin{minipage}[t]{0.48\textwidth}
 \caption[ ]{Same as Fig.~\ref{fig:XLFcosmo} but obtained  using both cluster abundance and large-scale distribution (scale is not the same).  Note the degeneracy breaking. Figure from \cite{schucker03}}
 \label{fig:XLFPkcosmo}
\end{minipage}
 \end{figure}
%%%%%%%%%%%%%%%%%%%%%%%%%%%%%%%%%%%%%
 
\subsubsection{Constraints from local abundances}

Recent  studies of the local distribution functions  \cite{blanchard00,ikebe02,reiprich02,viana02,allen03,pierpaoli03,schucker03}  indicate that  $\sigma_{8} \Omm^{0.4-0.6} = 0.4-0.6$, or $\sigma_{8}\sim 0.8$ for $\Omm=0.3$ \cite{rosati02}.  
For instance, Fig.~\ref{fig:XLFcosmo} shows the constraint derived from the Reflex XLF \cite{schucker03}, using the \LxM\ relation from \cite{reiprich02}.  Note the well-known degeneracy between $\sigma_{8}$ and $\Omm$ (see \cite{rosati02} for explanation).   Constraints from the XTF and the XLF from RASS data are consistent, but the XLF gives stricter constraints due to  the wider mass coverage \cite{pierpaoli03}. 

Errors on ($\sigma_{8}$,$\Omm$) are currently dominated by systematic uncertainties.  The major source of errors on $\sigma_{8}$ (at fixed $\Omm$) is the uncertainty on the normalization of the \MT\ relation \cite{pierpaoli03,viana03,henry04}.  If one assumes a higher mass for a given temperature, the amplitude of the mass function corresponding to an observed XTF is higher  and a higher $\sigma_{8}$ value is obtained. The discrepancies amoung various studies on $\sigma_{8}$ are largely due to the different  normalizations used \cite{henry04}.   The use  of different cosmological priors
and statistical methods also contributes to the observed differences \cite{pierpaoli03}.  Furthermore, a precise knowledge and proper treatment of the intrinsic scatter of the scaling relations (\MT\ and  \LxT) is critical. For instance, neglecting the scatter in the \LxM\ relation biases $\sigma_{8}$ towards high values \cite{pierpaoli03}. 

\subsubsection{Breaking the degeneracy using local cluster clustering}

Galaxy clusters can also be used to trace the large-scale structure of the Universe, which depends on the cosmology and the initial density spectrum. The spatial distribution of clusters thus provide cosmological constraints, complementary to those obtained from cluster abundance.  This requires to survey large contiguous regions of the sky. 

The  large-scale clustering and  abundance of clusters was measured with unprecedented accuracy  with the REFLEX survey. Recently, these data were analyzed simultaneously \cite{schucker03}. As shown in Fig.~\ref{fig:XLFPkcosmo}, this largely breaks the degeneracy between $\sigma_{8}$  and  $\Omm$ observed when only the local XLF is used.  The REFLEX sample gives  $\Omm = 0.34\pm 0.03$ and $\sigma_{8} = 0.71 \pm 0.04$ ($1\sigma$ statistical errors). 

 %%%%%%%%%%%%%%%%%%%%%%%%%%%%%%%%%%%%%
 \begin{figure}[t]
\begin{minipage}[c]{0.48\textwidth}
\centering 
\includegraphics[width=\textwidth]{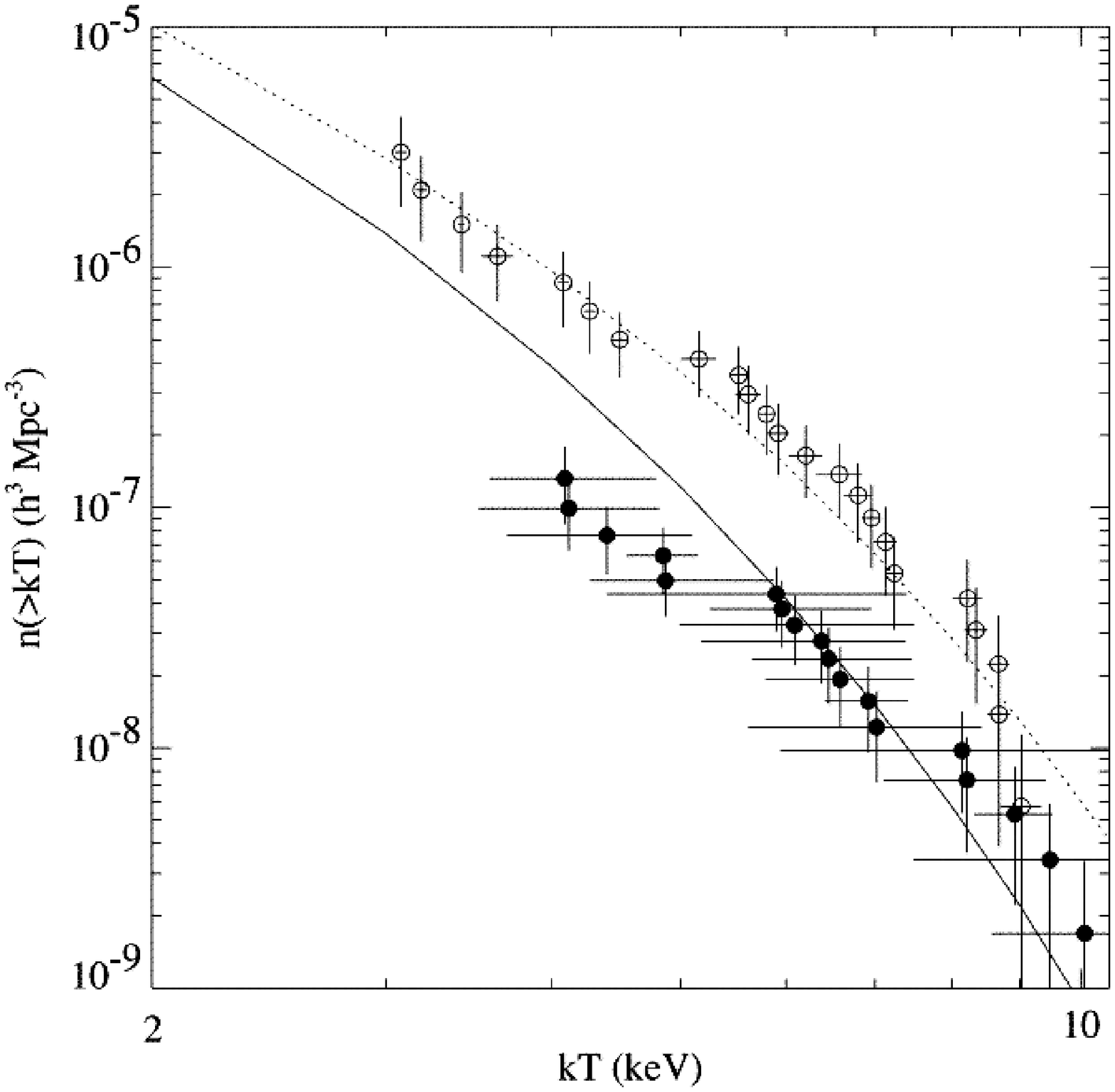}     
 \end{minipage}
\hspace{+0.02\textwidth}
\begin{minipage}[c]{0.48\textwidth}
\centering 
\includegraphics[width=\textwidth]{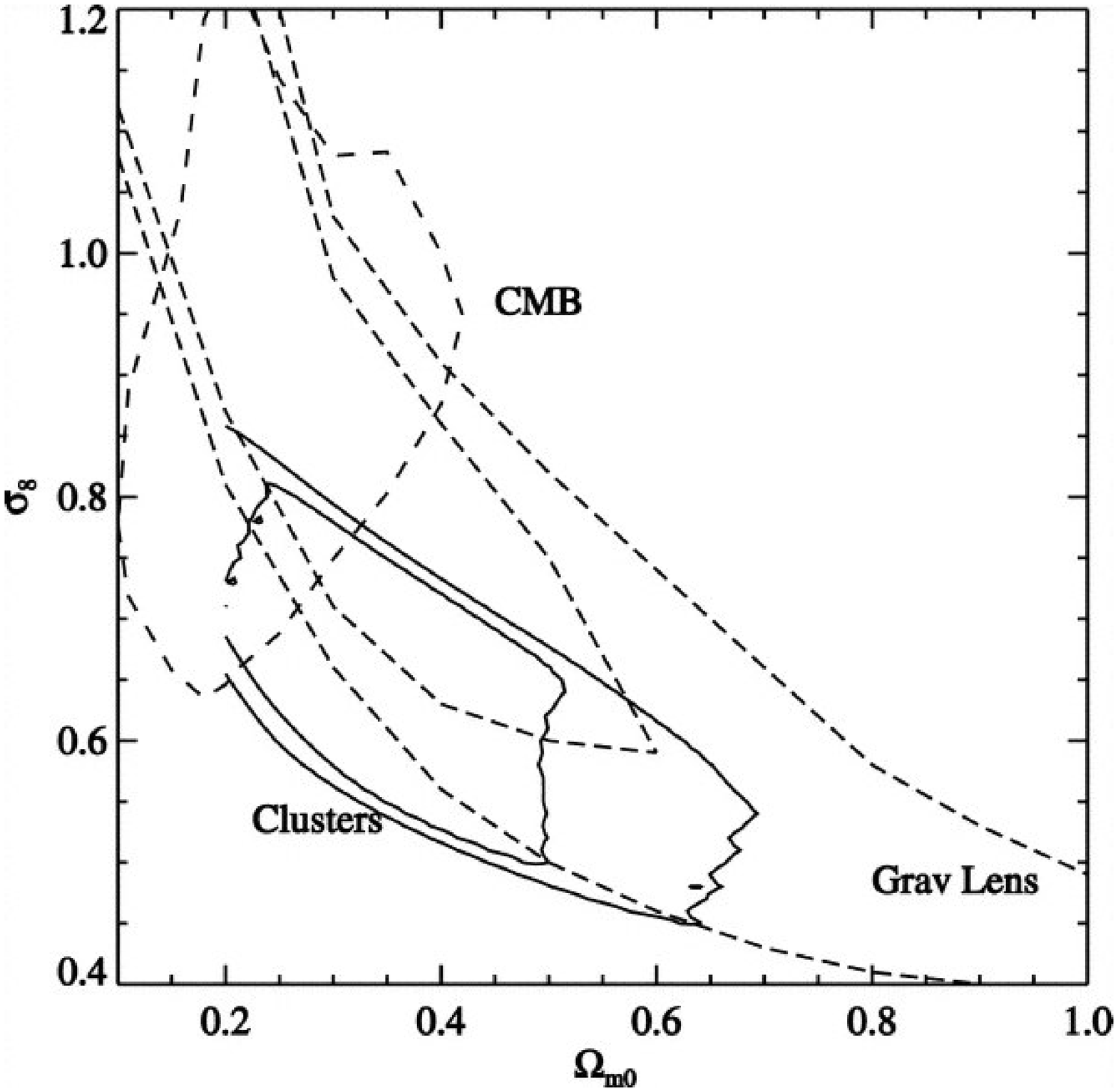}     
\end{minipage}
\begin{minipage}[t]{0.48\textwidth}
 \caption[ ]{Cumulative temperature function established by \cite{henry04} at low (open circle) and high (filled circles) redshift.  The high redshift XTF is derived from ASCA follow-up of EMSS cluster sample. The lines are the best fit models ($\Omm=0.28$, $\Oml=0.94$). Figure from \cite{henry04}}. 
  \label{fig:XTFz}
\end{minipage}
\hspace{+0.02\textwidth}
\begin{minipage}[t]{0.48\textwidth}
 \caption[ ]{Constraints on $\sigma_{8}$ and $\Omm$ ($1$ and $2\sigma$ confidence contours) derived from the temperature function and its evolution, compared to constraints from CMB and weak lensing observations. Figure from   \cite{henry04} }
 \label{fig:XTFzcosmo}
\end{minipage}
 \end{figure}
%%%%%%%%%%%%%%%%%%%%%%%%%%%%%%%%%%%%%

%%%%%%%%%%%%%%%%%%%%%%%%%%%%%%%%%%%%%
 \begin{figure}[t]
\begin{minipage}[c]{0.48\textwidth}
\centering 
\includegraphics[width=1.0\textwidth]{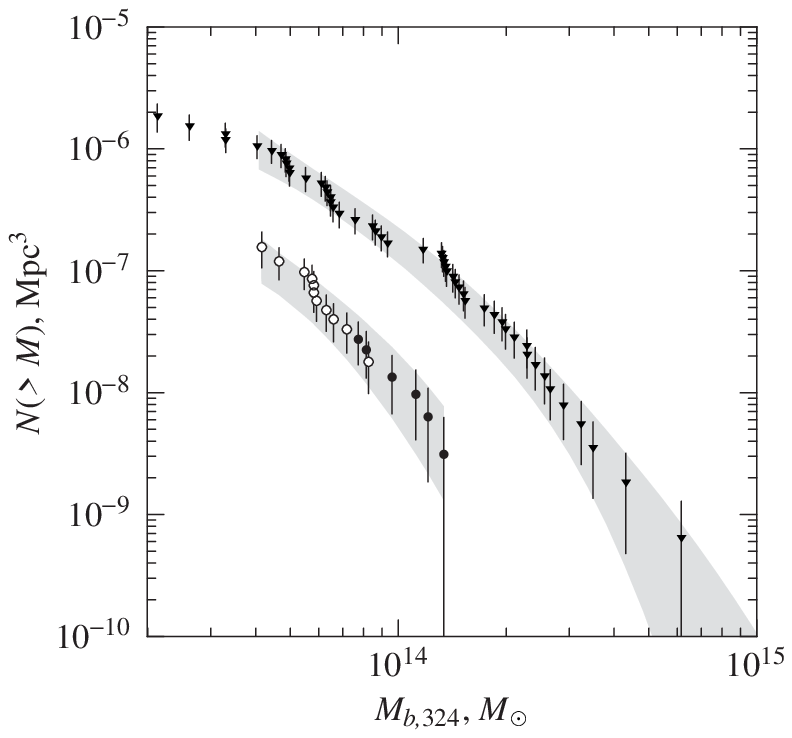}      
 \end{minipage}
\hspace{+0.02\textwidth}
\begin{minipage}[c]{0.48\textwidth}
\centering 
\includegraphics[width=1.0\textwidth]{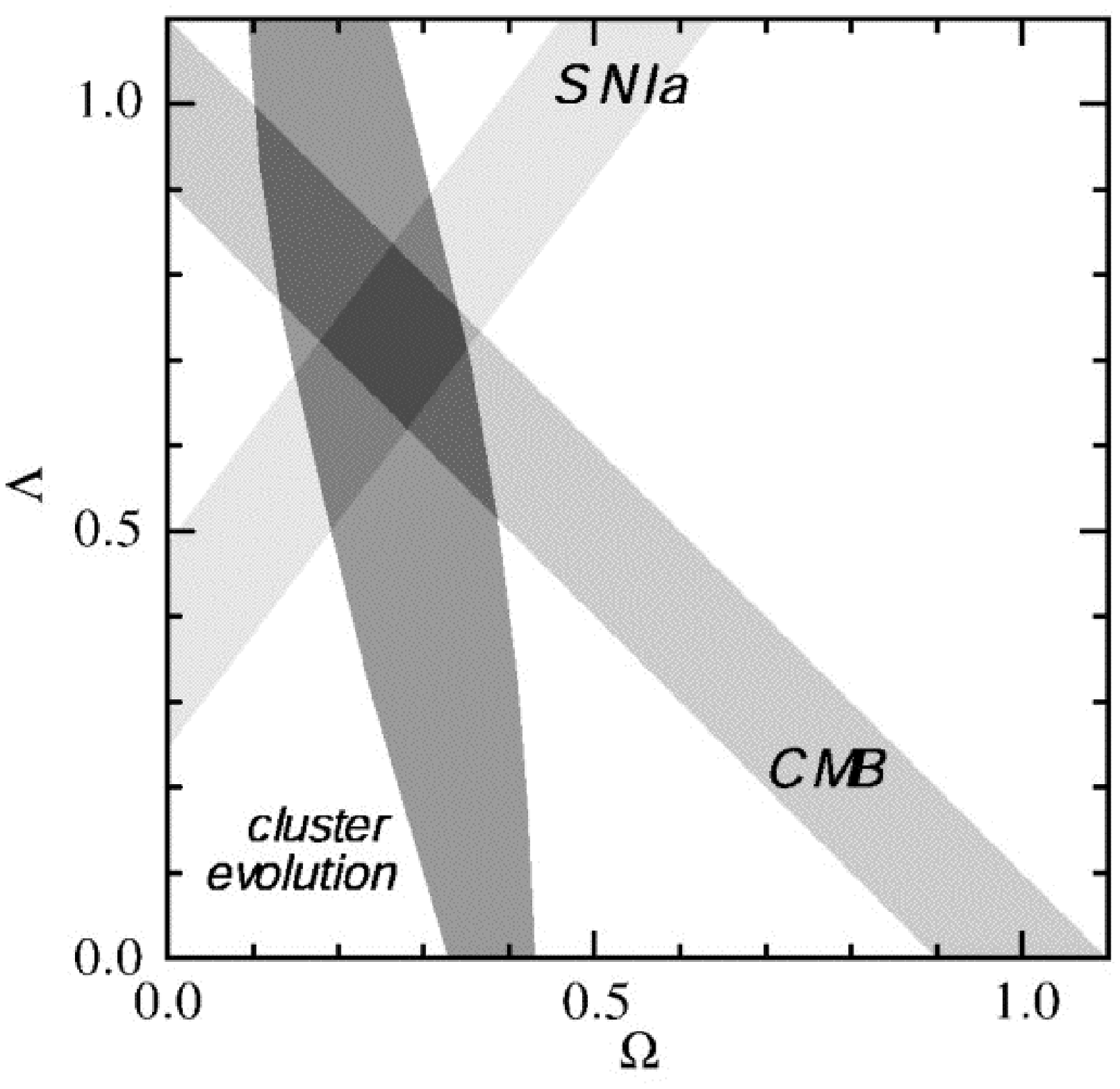}     
\end{minipage}
\begin{minipage}[t]{0.48\textwidth}
 \caption[ ]{Baryon mass function for the160SD survey sample in the redshift interval $0.4 < z < 0.8$ (filled circles)  assuming $\Omm=0.3$ and $\Oml=0.7$ \cite{vikhlinin03}. The local mass function from \cite{voevodkin04}  is shown by triangles.  Figure from \cite{vikhlinin03}}
 \label{fig:BMF}
\end{minipage}
\hspace{+0.02\textwidth}
\begin{minipage}[t]{0.48\textwidth}
 \caption[ ]{Comparison of constraints derived from the baryon mass function (Fig.~\ref{fig:BMF}) with those from distant supernovae and the CMB ($68\%$ confidence regions).  Figure from \cite{vikhlinin03} }
 \label{fig:BMFcosmo}
\end{minipage}
 \end{figure}
%%%%%%%%%%%%%%%%%%%%%%%%%%%%%%%%%%%%%

\subsubsection{Constraints from evolution }

As mentioned above, the evolution of the mass function is very sensitive to $\Omm$. The degeneracy between $\sigma_{8}$ and  $\Omm$, obtained when using local cluster abundance, can be broken if  the cluster distribution functions (XLF, XTF..) at higher redshift is known. 

Till recently, cosmological tests based on the evolution of the XLF gave ambiguous results due to the large uncertainty on the evolution  of the \LxT\  relation \cite{rosati02}.   This evolution is better constrained with \xmm\ and \chandra.  However, a consensus has not yet been reached.  An analysis of the X-ray luminosity distribution in the RDCS sample (103 clusters out to $z\sim0.85$) gives $\Omm=0.35^{+0.13}_{-0.1}$ and $\sigma_{8} = 0.66 \pm 0.06$ \cite{borgani01b}. A SCDM  Universe ($\Omm=1$) is excluded with a high confidence level: $\Omm=0.1-0.6$ at the $3\sigma$ level  when  systematic uncertainties are taken into account. In this study, the normalisation of the \LxT\ relation was assumed to vary as $(1+z)^{A}$, with $A=0-1$ in agreement with  \chandra\ observations. 
On the other hand, a study of the evolution of cluster number counts \cite{vauclair03}, as measured in various surveys, using as local reference the local XTF \cite{blanchard00}, gives $\Omm=0.85-1$.  As in the previous study  \cite{borgani01b},  a standard evolution of the \MT\ relation was assumed, and a similar evolution of the \LxT\ relation was used  as constrained from \xmm\ and \chandra\ data.   

In contrast, concordant constraints are obtained from the evolution of the XTF \cite[see Fig~\ref{fig:XTFzcosmo}]{henry04}  or the BMF \cite[see Fig~\ref{fig:BMFcosmo}]{vikhlinin03}, consistent with the value derived from the XLF RDCS study \cite{borgani01b}. For instance, the BMF evolution \cite{vikhlinin03}  gives  $\Omm=0.24\pm0.12$ for a flat Universe ($68\%$ confidence level). The XTF evolution \cite{henry04} gives a best fit  $\Omm=0.33$ value. It  was also used to constrain the equation of state of the Dark Energy: $w = -0.42\pm0.21$ ($68\%$ confidence level). 

\subsubsection{Conclusion}

Clusters of galaxies are powerful cosmological tools. Several {\it independent} methods can be used to constrain the cosmological parameters  from cluster X-ray (or S-Z) observations.  They are complementary to constraints provided by CMB, SNI and weak lensing observations.  Nearly all studies indicate a low $\Omm$ universe. $\sigma_{8}$ and $\Omm$  are typically determined with $\pm20\%$ accuracy, whereas   $\Oml$ (or the equation of state of the Dark Energy) is not yet well constrained. High precision cosmology with clusters is a priori  possible.  However, the key issue is to control and decrease the systematic uncertainties due to our imperfect  knowledge of the  physics that govern cluster formation and evolution.  This includes a better knowledge of the intrinsic scatter and evolution of the scaling laws.  A better precision of the abundance of massive clusters at high $z$ should also constrain much more tightly $\Oml$ and $\Oml$.  This requires a survey with a very large sky coverage. 

\section{Perspectives}
	
The \xmm\ and \chandra\ observatories are designed to operate until at least 2010.  In the coming years, large efforts  will be devoted on statistical analysis of known cluster samples, using archival data or  large projects.   Several Large Projects are in progress, to follow-up unbiased samples of local or distant clusters discovered by \rosat.  In parallel, the \xmm\ (and to a lesser extent \chandra) satellites will provide new cluster samples for cosmology, extending to lower mass and possibly higher $z$. This includes serendipitous surveys, like the XCS \cite{land04}, or contiguous surveys, like the XMM-LSS \cite{pierre04}. 

For understanding the physics of  the intra-cluster medium, key information is still missing. We cannot measure the velocity structure of the gas.  Moreover, the high resolution and spatially resolved spectroscopic data  required to investigate the complex 
cluster core  is not yet available. This information can only be provided by bolometer array type instruments, as will be on board Astro-E2, planned to be launched in a few months \cite{inoue04}.  

Understanding non thermal processes in clusters (e.g electron acceleration by shocks or turbulence during  merger events, effect on the cluster evolution) is another open issue.  Astro-E2 will greatly improve our capability to observe high energy emission from clusters. Further progresses  will require spatially resolved spectroscopy  at high energies (up to E $\sim80$ keV). When  combined with radio data, this will allow  us to map both the magnetic field and the non thermal particle population.  Several projects are under study, like  {\it SIMBOL-X}  \cite{ferrando04} and {\it NeXT} \cite{inoue04}. 

Planck (to be launched in 2007) should detect about $10000$ clusters via the SZ effect.  The statistical study of this sample, unique by its size, depth and sky coverage, will constrain cosmological parameters  and provide information on the physics of structure formation.  It will be extremely useful (and sometimes mandatory) to combine the SZ data with X-ray data, like those obtained by \xmm.  Planck and \xmm\ surveys have not the same sky, mass and redshift coverage and it is unrealistic to think of a complete X-ray follow-up of the Planck sample. However, sub-samples can still be used,  to calibrate for instance the $Y$--$M$ relation. This relation must be known to constrain cosmology on the basis of SZ cluster abundances. 

The next generation of X--ray observatories, Constellation-X and XEUS, is already on the horizon for the years $2010+$.  They should allow us to probe the hot Universe at much higher $z$ than presently. We will  study early Black Holes, groups of galaxies at $z\sim 2$  and their evolution to the massive clusters of today,  and investigate nucleosynthesis down to the present epoch. These satellites  will  perfectly complement observatories like ALMA, JWST that will look at the 'cool' component of the Universe.

\acknowledgments

\end{document}